\RequirePackage{amsmath}
\documentclass[runningheads]{llncs}
\usepackage{longtable}

\usepackage{todonotes}
\usepackage[english]{babel}

\usepackage{xcolor}
\usepackage{float}
\usepackage{comment}
\colorlet{punct}{red!60!black}
\definecolor{background}{HTML}{EEEEEE}
\definecolor{delim}{RGB}{20,105,176}
\colorlet{numb}{magenta!60!black}
\usepackage{subfig}
\usepackage{multirow}
\usepackage{booktabs}
\usepackage{pgfplots}
\usepackage{pgfplotstable}
\usepackage[font=small,labelfont=bf,tableposition=top,labelsep=period]{caption}
\usepgfplotslibrary{fillbetween}
\usetikzlibrary{calc}
\usepgfplotslibrary{groupplots}
\pgfplotsset{compat=newest}

\usepackage{graphicx}

\usepackage{algorithm}
\usepackage{algpseudocode}
\usepackage{pseudocode}

\usepackage{tikz}
\usetikzlibrary{positioning}

\usepackage{cprotect}
\usepackage{multirow}
\usepackage{graphicx}
\usepackage[T1]{fontenc}
\usepackage{varioref}
\usepackage{xspace}

\usepackage{paralist}
\usepackage{fancybox}
\usepackage{xcolor}
\usepackage{calc}
\usepackage{verbatim}
\usepackage{todonotes}

\usepackage{booktabs}

\usepackage{listingsVDM}

\newcommand{\keywords}[1]{\par\addvspace\baselineskip
  \noindent\keywordname\enspace\ignorespaces#1}
\lstdefinelanguage{VDM++}
  {morekeywords={act, active, fin, req, waiting, abs, all, allsuper, always, and, answer, 
     assumption, async, atomic, be, bool, by, card, cases, char, class, comp, compose, conc, cycles,
     dcl, def, definitions, del, dinter, div, dlmodule, do, dom, dunion, duration, effect, elems, else, elseif, end,
     error, errs, exists, exists1, exit, exports, ext, floor, for, forall, from, functions, 
     general, hd, if, imports, in, inds, infer, init, inmap, input, instance, int, inter, inv, inverse, iota, is, 
     isofbaseclass, isofclass, inv, inverse, lambda, len, let, map, measure, mu,
     mutex, mod, module, nat, nat1, new, merge, 
     munion, not, of, operations, or, others, per, periodic, post, power, pre, pref, 
     private, protected, public, qsync, rd, responsibility, return, reverse,  
     sameclass, parameters, psubset, pure, rem, renamed, rng, sel, self, seq, seq1, set, skip, specified, st, 
     start, startlist, state, static, stop, stoplist, sporadic, subclass, subset, subtrace, sync, system, then, thread, 
     threadid, time, tixe, tl, to, token, traces, trap, types, undefined,
     union, uselib, using, values, 
     variables, while, with, wr, yet, RESULT, false, true, nil, periodic pref, rat, real},
   sensitive,
   frame=trBL,
   morecomment=[l]--,
   morestring=[b]",
   morestring=[b]',
   frameround=fttt
  }[keywords,comments,strings]
\lstdefinelanguage{JavaCC}
  {morekeywords={options, PARSER\_BEGIN, PARSER\_END, SKIP, TOKEN},
   sensitive=false,
  }[keywords]

\usepackage{listings}
\usepackage{xcolor}

\colorlet{punct}{red!60!black}
\definecolor{background}{HTML}{EEEEEE}
\definecolor{delim}{RGB}{20,105,176}
\colorlet{numb}{magenta!60!black}

\lstdefinelanguage{json}{
	basicstyle=\scriptsize\ttfamily,
	numbers=left,
	numberstyle=\scriptsize,
	stepnumber=1,
	numbersep=8pt,
	showstringspaces=false,
	breaklines=true,
	frame=lines,
	backgroundcolor=\color{white},
	literate=
	*{0}{{{\color{numb}0}}}{1}
	{1}{{{\color{numb}1}}}{1}
	{2}{{{\color{numb}2}}}{1}
	{3}{{{\color{numb}3}}}{1}
	{4}{{{\color{numb}4}}}{1}
	{5}{{{\color{numb}5}}}{1}
	{6}{{{\color{numb}6}}}{1}
	{7}{{{\color{numb}7}}}{1}
	{8}{{{\color{numb}8}}}{1}
	{9}{{{\color{numb}9}}}{1}
	{:}{{{\color{punct}{:}}}}{1}
	{,}{{{\color{punct}{,}}}}{1}
	{\{}{{{\color{delim}{\{}}}}{1}
	{\}}{{{\color{delim}{\}}}}}{1}
	{[}{{{\color{delim}{[}}}}{1}
	{]}{{{\color{delim}{]}}}}{1},
}

\usepackage{graphicx}
\usepackage{vdmlisting}
\usepackage{url}
\usepackage{tcolorbox}

\lstdefinelanguage{html}{ 
  backgroundcolor={\color[gray]{1}},
  basicstyle=\small\ttfamily,
  morekeywords={code}, 
  sensitive=true, 
  morestring=[s]{"}{"}, 
  style=HtmlStyle 
} 
\lstdefinestyle{HtmlStyle}{ 
}

\usepackage{graphicx}
\usepackage{todonotes}
\usepackage{pgfplotstable}
\usepackage{pgfplots}
\usepackage{booktabs}
\usepackage{tikz}
\usetikzlibrary{trees,shapes,arrows}
\usetikzlibrary{shapes.multipart}
\usetikzlibrary{positioning}
\usepackage{amssymb}
\usepackage{amsmath}
\usepackage{multirow}
\usepackage{xcolor}

\usepackage{graphicx}
\usepackage{array}
\usepackage{hyperref}
\usepackage{siunitx}
\usepackage{amsmath}
\usepackage{fancyvrb}
\usepackage{listings}
\newcolumntype{L}[1]{>{\raggedright\let\newline\\\arraybackslash\hspace{0pt}}m{#1}}
\newcommand{\xfnm}[1][]{\ifx!#1!\else\unskip,\space#1\fi}

\usepackage{graphicx}
\usepackage{amsmath}
\usepackage{esdiff} 
\usepackage{url}
\usepackage{soul}
 \usepackage{comment} 

\lstdefinestyle{VDM}
{
  frame=single,
  basicstyle=\small\ttfamily,
  escapechar=!,
  breaklines=true,
  frameround=false,
  linewidth=\columnwidth,
  morekeywords={atomic,is,inv,values,dcl,forall,in,set,nil,and,let,be,st,set1,pure,nat,pre,post,map,to,of,true,false},
  moredelim={[is][keywordstyle]{@}{@}},
}

\newtheorem{prop}{Property}

\begin{document}

\let\origref\ref
\let\origpageref\pageref
\let\origlabel\label
\newcommand\locallabels[1]{%
  \renewcommand\label[1]{\origlabel{#1##1}}%
  \renewcommand\ref[1]{\origref{#1##1}}%
  \renewcommand\pageref[1]{\origpageref{#1##1}}%
}

\title{Proceedings of the 19$^{th}$ International Overture Workshop}
\institute{}
\author{Hugo Daniel Macedo \and Casper Thule \and Ken Pierce (Editors)}
\authorrunning{Hugo D. Macedo et al.}   
%

\maketitle

\setcounter{page}{1}

\chapter*{Preface}
\markboth{Preface}{Preface}

The 19th in the ''Overture'' series of workshops on the Vienna Development Method (VDM), associated tools and applications was held as a hybrid event both online at in person at Aarhus University on October 22, 2021.
VDM is one of the longest established formal methods, and yet has a lively community
of researchers and practitioners in academia and industry grown around the modelling
languages (VDM-SL, VDM++, VDM-RT) and tools (VDM VSCode, VDMTools, VDMJ, ViennaTalk,
Overture, Crescendo, Symphony, and the INTO-CPS chain). Together, these provide a
platform for work on modelling and analysis technology that includes static and dynamic
analysis, test generation, execution support, and model checking.

Research in VDM is driven by the need to precisely describe systems. In order to do so, it is also necessary for the associated tooling to serve the current needs of researchers and practitioners and therefore remain up to date. The 19th Workshop reflected the breadth and depth of work supporting and applying VDM. This technical report includes first a paper on industrial usage. This is followed by a session on VDM related papers concerning techniques for developing models and applying VDM for modelling systems. The last sessions is related to Cyber-Physical Systems and simulation. 

As the pandemic is still in effect across the world this workshop was held both online and in person. We applaud the possibility of meeting some collegues in person and still remaining in contact with others online. It is our sincere hope that the next Overture Workshop will see an increase in personal attendance.

We would like to thank the authors, PC members, reviewers and participants for
their help in making this a valuable and successful workshop, and we look forward
together to meeting once more in 2022.

\medskip
\begin{flushright}\noindent
Hugo Daniel Macedo, Aarhus
Casper Thule, Aarhus\\
Ken Pierce, Newcastle\\
\end{flushright}

\tableofcontents

\chapter*{Organization}

\section*{Programme Committee}
\begin{longtable}{p{0.3\textwidth}p{0.7\textwidth}}

Tomo Oda & Software Research Associate Incorporated, Japan\\[12pt]

Marcel Verhoef & European Space Agency, The Netherlands\\[12pt]

Paolo Masci & National Institute of Aerospace (NIA), USA\\[12pt]

Peter Gorm Larsen & Aarhus University, Denmark\\[12pt]

Nick Battle & Newcastle University, UK\\[12pt]

Fuyuki Ishikawa & National Institute of Informatics, Japan\\[12pt]

Keijiro Araki & National Institute of Technology, Kumamoto College, Japan\\[12pt]

Sam Hall & Newcastle University, UK\\[12pt]

Hugo Daniel Macedo & Aarhus University, Denmark\\[12pt]

Ken Pierce & Newcastle University, UK\\[12pt]

Marcel Verhoef & European Space Agency, The Netherlands\\[12pt]

Kenneth G. Lausdahl & AGROCorp International and Aarhus University, Denmark
\end{longtable}

\mainmatter              

\titlerunning{19th Overture Workshop, 2021}

\setcounter{page}{5}

\clearpage

\begingroup
\renewcommand\theHchapter{5-Fraser:\thechapter}
\renewcommand\theHsection{5-Fraser:\thesection}
\locallabels{5-Fraser:}
\setcounter{footnote}{0}
\setcounter{chapter}{0}
\setcounter{lstlisting}{0}
\fontfamily{ptm}\selectfont

\makeatletter
\def\input@path{{5-Fraser/}}
\makeatother

\graphicspath{{5-Fraser}}

\lstset{
    basicstyle=\scriptsize\ttfamily,
    captionpos=b
}

\title{Behaviour driven specification}
\subtitle{A case study in combining formal and agile methods}

\author{Simon Fraser, Alessandro Pezzoni}
\authorrunning{\protect\raggedright S. Fraser, A Pezzoni}
\titlerunning{Behaviour driven specification}

\institute{Anaplan Limited, York, UK\\
    \email{\{simon.fraser, alessandro.pezzoni\}@anaplan.com\\}
    \url{http://www.anaplan.com}
}

\maketitle              

\begin{abstract}
    The use of formal methods in industry may be resisted by developers who have embraced agile methods.
There is often a perception that formal methods require much upfront work and are unable to adapt quickly to a changing landscape.
This leads to a view that formal methods are a barrier to embracing change and are unsuitable for projects involving the iterative evolution of requirements.
In this paper, we use a project at Anaplan -- a company where agile methods are the norm -- to illustrate how formal methods can be integrated into an agile development process.
Specifically, we introduce the concept of behaviour driven specification as an agile-compatible method of constructing a formal specification and show how the acceptance criteria produced by this process not only provide an accessible form of requirements capture, but provide a means for validation of both the specification and a target implementation.

\keywords{Software Validation \and Behavior Driven Development (BDD) \and Agile Methods \and Formal Methods}

\end{abstract}

\section{Introduction}\label{sec:introduction}

The \gls{vdm} is one of the approaches to follow, when applying formal methods during the development of computer systems.
The method prescribes the development of digital/computer models of the system under development in one of the the \gls{vdm} specification languages and dialects. If the models are described in an executable subset it is then possible to execute and analyse them to reach a high-fidelity system description, which is then subsequently used to produce code in the  programming languages used to operate the system implementation.

To support all the steps involved in the development of a \gls{vdm} model, there are several tools and \glspl{ide} supporting the variety of \gls{vdm} specification languages and dialects to different levels \cite{Larsen&16}.
VDMTools were the first available commercial tool developed in the  mid-1990s \cite{3-Rask:Larsen01}. Then Overture \cite{3-Rask:Larsen&10a} brought free and open-source support to \gls{vdm} in 2005, and many others are now available \cite{Oda&15,Tran&19}.

The Overture tool is one of the most complete and popular \glspl{ide} for \gls{vdm}. It consists of multiple plugins that extend the Eclipse \gls{ide}, which appeared in the 2000s.
Eclipse is the most popular\footnote{See \url{https://pypl.github.io/IDE.html}} in its class, but it is being challenged by a new contender, the \gls{vscode}\footnote{See \url{https://code.visualstudio.com/}} editor, which is based on Electron\footnote{See \url{https://www.electronjs.org/}.} and fully leverages web technologies.
%
%
Although not an \gls{ide}, \gls{vscode} modernized the \gls{ide} world, with  the introduction of the \gls{lsp}\footnote{See \url{https://microsoft.github.io/language-server-protocol/}.} and the \gls{dap}\footnote{See \url{https://microsoft.github.io/debug-adapter-protocol/}.}, which enable the development of \gls{ide} features in a general manner. With the two protocols the editor becomes indistinguishable of an \gls{ide}, and the implementation code becomes reusable and better maintainable.
However, the Overture language core does not implement the LSP and DAP protocols, so the effort required to move towards \gls{vscode} is significant, but we expect it to pay off in the long run.

In this paper we investigate the efforts needed to support \gls{vdm} in \gls{vscode} with as much as possible of the support using the standardised protocols \gls{lsp} and \gls{dap}.
To support specification lanuage features not found in programming languages we propose an extension to \gls{lsp}, \gls{slsp}.
As part of the investigation we have extended VDMJ such that it can be used as a language server that supports both these protocols. 
The server supports syntax-checking, type-checking and go-to functionality using the \gls{lsp} protocol, debugging by using the \gls{dap} protocol, and \gls{pog} and \gls{ct} using the \gls{slsp} protocol.
We have also developed a \gls{vscode} extension which connects to the language server in order to provide the language features in the \gls{ide}.
In addition, we have investigated which features are required to provide full language support for specification languages and which of these are supported by the protocols.

Our work departs from a previous proof of concept, and adds further support for the LSP protocol, support for all the VDM dialects, support for the DAP protocol and support for the \gls{slsp} protocol. 
In addition, this is the first research paper on the topic, and we foresee more publications to emerge from the works towards fully supporting VDM development using \gls{vscode}.
We believe the resulting extensions will become the next in line supporting \gls{vdm} development. A modern and robust IDE, which development starts now. 

The remaining parts of this paper starts with an overview of the background necessary to understand this paper in \Cref{3-Rask:sec:background}. Afterwards \Cref{sec:Implementation} explains how the standard protocols \gls{lsp} and \gls{dap} have been used to implement the core of the \gls{vscode} support for \gls{vdm}. This is followed by \Cref{sec:Evaluation} which is evaluating the efforts that has been conducted. 
In \Cref{sec:relatedWork} we describe related work.
Finally, \Cref{3-Rask:sec:conclusion} provides concluding remarks and future work.

\section{Background}\label{sec:background}
In this section, we introduce VDMJ which is a tool that provides  language features for \gls{vdm}. 
We further describe the development tool \gls{vscode}, which is used as the \gls{gui} for integration with the language server.
Finally, we describe the standardised protocols \gls{lsp} and \gls{dap} used to decouple the language features from the development tool.

\subsection{VDMJ}

VDMJ \cite{Battle09} is a command-line tool written in Java, that provides basic language support for the \gls{vdm} dialects VDM-SL, VDM++ and VDM-RT. 
It includes a parser, a type checker, an interpreter, a debugger, a proof obligation generator and a combinatorial test generator with coverage recording, as well as VDMUnit support for automatic testing and user definable annotations.
These are implemented using an extensible \gls{ast} analysed using a visitor framework\cite{Gamma&95}.

Using VDMJ for the language server allows parts of the language support to be reused. 
However, many features of VDMJ are not supported by standardised protocols, thus only a subset of the functionality is used in the language server.


\subsection{Visual Studio Code}

\gls{vscode}\footnote{See \url{https://code.visualstudio.com/}} is a free source code editor. 
It has built-in support for the programming languages JavaScript and TypeScript and further enables support for other languages through a rich ecosystem of extensions.
\gls{vscode} uses a folder or workspace system for interacting with a project and a document system for handling the source code files in the project.
This allows \gls{vscode} to be language-agnostic and delegate language specific functionality to an extension. 
Given this design reasoning, a need for the standardisation of decoupling between editor and extension has been identified.
At the time of writing, three different protocols have been developed for this purpose. 
Namely \gls{lsp}\footnote{See \url{https://microsoft.github.io/language-server-protocol/}} (described in \Cref{subsec:lsp}), \gls{dap}\footnote{See \url{https://microsoft.github.io/debug-adapter-protocol/}} (described in \Cref{subsec:dap}) and Language Server Index Format (LSIF)\footnote{See \url{https://microsoft.github.io/language-server-protocol/overviews/lsif/overview/}}.

\subsection{Language Server Protocol}
\label{subsec:lsp}

Since most \glspl{ide} support a variety of programming languages, \gls{ide} developers face the problem of keeping up with ongoing programming language evolution \cite{Bunder19a}. 
At the same time language providers are interested in providing as many \gls{ide} integrations as possible to serve a broad audience. 
Consequently, integrating every language, $m$, in every \gls{ide}, $n$, leads to a $m \times n$ complexity.

\begin{figure}[htb]
    \centering
    \includegraphics[width=0.75\textwidth]{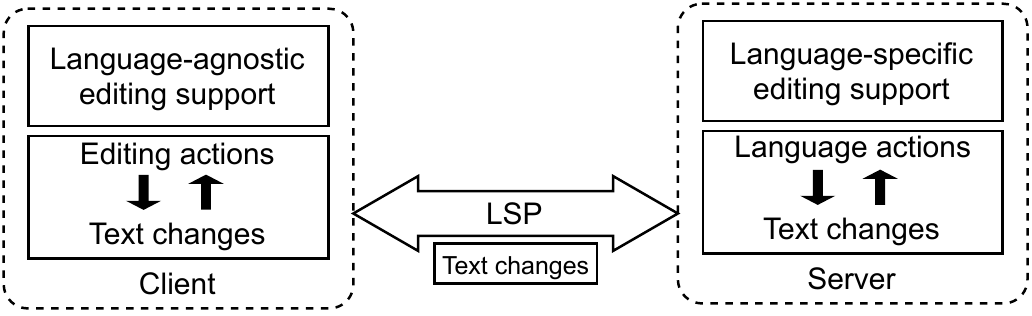}
    \caption[\gls{lsp} approach to language support.]
        {\gls{lsp} approach to language support. Borrowed from \cite{Rodriguez&18}.}
    \label{fig:LSPApproach}
\end{figure}

The \gls{lsp} protocol defines a standardised protocol to be used to decouple a language-agnostic development tool (client) and a language-specific server that provides language features like syntax-checking, hover information and code completion.
This is illustrated in \Cref{fig:LSPApproach}.
The client is responsible for managing editing actions without any knowledge of the language and the server validates the correctness of the source code and reports issues and language-specific information to the client.
To facilitate this the \gls{lsp} protocol communicates using language neutral data types such as document references and document positions.

Many tools support the \gls{lsp} protocol which reduces the time needed to create a client implementation\footnote{See \url{https://microsoft.github.io/language-server-protocol/implementors/tools/}}.
Furthermore, new development tools only have to support the protocol, which can be done with little effort compared to native integration \cite{Bunder19b}. 
Additionally, as the server is separated from the development tool it can be used for multiple tools.
This allows tools to easily support multiple languages and features. 
Thus, by decoupling the language implementation from the editor integration the complexity is reduced to $m + n$.

\subsection{Debug Adapter Protocol}
\label{subsec:dap}
The \gls{dap} protocol is a standardised protocol for decoupling \glspl{ide}, editors and other development tools from the implementation of a language-specific debugger. 
The \gls{dap} protocol uses language neutral data types, which makes the protocol possible to use for any text-based language.
The debug features supported by the protocol includes: different types of breakpoints, variable values, multi-process and thread support, navigation through data structures and more.

To be compatible with existing debugger components, the protocol relies on an intermediary debug adapter component.
It is used to wrap one or multiple debuggers, to allow communication using the \gls{dap} protocol.
The adapter is then part of a two-way communication with a generic debugger component, which is integrated in a given development environment as illustrated in \Cref{fig:DAPArchitecture}. 
Thus, the protocol reduces a $m \times n$ problem of implementing each language debugger for each development tool into a $m + n$ problem.

\begin{figure}[htb]
    \centering
        \includegraphics[width=0.8\textwidth]{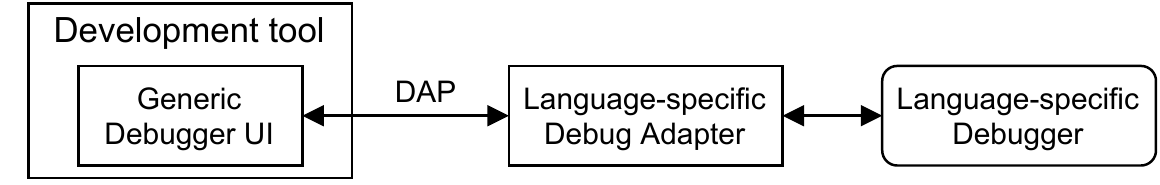}
    \caption[The decoupled architecture where the \gls{dap} protocol is used.]
        {The decoupled architecture where the \gls{dap} protocol is used. }
    \label{fig:DAPArchitecture}
\end{figure}


\section{Behaviour driven specification}\label{sec:behaviourDrivenSpec}
When using agile methods, work is broken into small chunks that each provide some value to a stakeholder in the system.
These chunks are often framed as user stories~\cite{cohn2004user} which describe the work to be done from the perspective of the stakeholder.
Part of the user story is a set of acceptance criteria (AC) that give a checklist of what must be achieved in order to consider the work complete.

Test driven development (TDD)~\cite{beck2003test} is a process frequently used by teams employing agile methods.
Whenever the developer wishes to change the code, failing tests are created before writing just enough code to make those tests pass.
The developer then repeats the process until they have made all the changes required.
TDD does not align neatly with user stories as the granularity of the changes tested is usually much finer than that of the user story.

Behaviour driven development (BDD)~\cite{wynne2017cucumber} is a process that can be used in addition to TDD and which operates at the level of a user story.
BDD involves the creation of scenarios written from the point of view of the stakeholder, and which exemplify their requirements.
Each scenario is a form of Hoare Triple~\cite{hoare1969axiomatic} where we capture that:
\begin{itemize}
    \item {\em given} an initial context (precondition)
    \item {\em when} the stakeholder performs an action (command)
    \item {\em then} the outcome is as expected (postcondition)
\end{itemize}
These scenarios are written using a domain-specific language (DSL) for which the developer creates an adapter that enables the scenario to be executed and verified.

We believed that we could adapt BDD to form a behaviour driven specification process which would enable us to create a formal specification following agile principles.
Our principal aim was to be able to comprehensively specify small chunks of the system on a regular cadence, which would enable developers to leverage the specification to validate the corresponding chunks of implementation, providing constant feedback on its correctness and quickly identifying where differences had arisen -- whether intentional or not.
We also wished to provide the developers with a set of requirements for the system in a form with which they were familiar -- acceptance criteria.
By doing this, we also aimed to significantly reduce the need for natural language commentary on the specification, replacing it with scenarios written in a DSL which would be both less ambiguous and more maintainable.

\subsection{What does BDS involve?}\label{subsec:what-is-bds?}

The BDS process is much like BDD, and it too requires a DSL in which to write scenarios.
However, unlike BDD, where the inputs are ACs and the output is code, with BDS we look to establish ACs and use those to drive the creation of the specification, with both being valuable artifacts of the process.

For each behaviour, the first step is to write a natural language description of the AC.
A scenario is then created to exemplify the AC; this may involve the extension of the DSL if new language is needed to express the criterion.
This scenario now forms an executable acceptance criterion (EAC) -- we shall discuss how these are run in section~\ref{subsec:executable-acceptance-criteria} -- that we can use to validate the behaviour of both the specification and the target implementation.
When first run this EAC should fail, otherwise it is likely that the AC is redundant or that the scenario does not correctly capture the unicity of the AC.
The specifier then makes just enough changes to the specification to satisfy the EAC without invalidating any others.
This process is repeated until all of the ACs of the behaviour have been established and agreed with the stakeholder.
At this point the EACs can be handed to a developer to implement the behaviour.
The developer can now start their work with a set of ACs, including examples scenarios that fail against the target implementation --- an ideal place to start BDD.

We can rapidly create scenarios to exemplify behaviour questions that arise during implementation, and we can run them against the specification to determine the expected outcome.
If the developer is not satisfied, we can take that example and the specified results, and present to the stakeholder.
If the stakeholder believes this scenario is distinct from our existing EACs and wants a change of behaviour, we make the scenario a new EAC and change just enough of the specification to satisfy it.
At this point, the specifier may discover that it is not possible to make a change that is consistent with the existing EACs.
If so, the conflicting criteria are presented to the stakeholder and a resolution agreed.
The use of the DSL allows us to do this in the language of the stakeholder, but without the ambiguity of natural language.

Similarly, the developer may discover efficiency or performance issues with implementing a particular EAC, or in fact, any member of the team may spot some behaviour they disagree with.
Once again, these can be discussed with the stakeholder, the outcome of the EAC changed, and the requirements of the behaviour, the specification, and the implementation iteratively refined.

The use of EACs not only encourages a more agile approach to specification, but enables its continuous validation through DSL-based requirements which are accessible to everyone involved in the project.

\subsection{Running executable acceptance criteria}\label{subsec:executable-acceptance-criteria}

Running scenarios requires a DSL in which to express the actions and checks of the system in the stakeholder's language, and means by which to translate that language into commands and queries that the implementation can execute.
Cucumber~\cite{wynne2017cucumber} is synonymous with BDD, but we found it unsuitable for use with BDS.
Although Cucumber accepts scenarios in the given-when-then format, the distinction between the blocks is lost at runtime and the scenario is reduced to a single sequence of steps.
When executing a program it is perhaps unimportant that certain steps are declarative and others are imperative\footnote{We found it useful to make the distinction with the implementation too. Anaplan is a transactional system and it was useful for us to be able to create the declarative, initial context in a single transaction.}, but when animating a specification we wanted to transform the declarative steps into a single declaration rather than calling a series of functions.
After some time trying to adapt Cucumber to our needs, we found that we were better able to make progress with a custom Kotlin DSL~\cite{subramaniam2021programming}.

In practice, we found the majority of our effort in this area was in creating the DSL and adapting to the specification and the various implementations, rather than in the mechanics of execution.
We briefly describe our approach to these mechanics for completeness, and it is likely that another team employing BDS would benefit from replicating the approach rather than using our tooling directly.

Example of some trivial EACs are given in listings~\ref{lst:create} and ~\ref{lst:add}\footnote{Note that the keyword \texttt{whenever} is used in the scenario as \texttt{when} is a Kotlin keyword.}.

\noindent
\begin{minipage}[t]{.5\textwidth}
\begin{lstlisting}[caption=EAC for creating a list, label=lst:create]
    @Eac("A created list is empty")
    fun create() {
      whenever {
        createAList("list")
      }
      then {
        listContains("list")
      }
    }
\end{lstlisting}
\end{minipage}
\noindent
\begin{minipage}[t]{.5\textwidth}
\begin{lstlisting}[caption=EAC for adding to a list, label=lst:add]
    @Eac("When an entity is added,
      it is contained by the list")
    fun addToAList() {
      given {
        thereIsAList("list", "a")
      }
      whenever {
        addEntityToList("list", "b")
      }
      then {
        listContains("list", "a", "b")
      }
    }
\end{lstlisting}
\end{minipage}

\noindent
A single EAC is run using a custom JUnit runner;
this means that all the usual JUnit tools are available for running from an IDE, a build tool, or a continuous integration environment.
To run an EAC, the runner first identifies what implementations are available on the classpath and whether any constraints have been placed on the EAC.
For example, we might want to skip a particular implementation due to a known bug and use an annotation to indicate this.
Once the list of available implementations has been determined, the runner will attempt to run the EAC against each of them.

The DSL is backed by an API that is implemented by an adapter for each implementation.
The JUnit runner will use this API to convert the DSL into a series of implementation specific commands.
There is not a one-to-one correspondence here as, particularly with the declaration, an adapter may choose to maintain a state and create a single command corresponding to several calls to the API.

The adapter is also allowed to return an `unsupported command' from any API method -- indicating that there is piece of functionality that has not yet been implemented, or perhaps which may never be implemented.
For this reason, during conversion, the runner does not instantiate an implementation and none of the commands are executed.
Only if all commands are supported does the runner actually execute them against the implementation, otherwise it can quickly skip the EAC without having to partially execute the scenario.
This is extremely valuable for BDS where -- unlike BDD -- we are usually working with more than one adapter and development may not immediately follow specification.
When the specification of a behaviour is complete, we can mark that behaviour as unsupported in the implementation's adapter until development can begin.

This capability is even more useful when working with multiple implementations --- as in our Anaplan project.
Consider the example in~\ref{subsec:the-need-for-formal-methods}: we want to be able to specify both the dense and sparse behaviour of calculating $0^0$, but clearly each implementation will only support one outcome.
We can extend our DSL to be able to disambiguate the two methodologies\footnote{To prevent cluttering, we would only specify the arithmetic used in the EAC when it mattered and we would expect identical behaviour otherwise.} and write two EACs as shown in listings~\ref{lst:dense} and~\ref{lst:sparse}.
Our two sparse platforms would return an `unsupported command' for the EAC requiring dense arithmetic and would be skipped, but would be able to run the EAC requiring sparse arithmetic (and vice versa for the two dense platforms).
Our specification would be able to support both arithmetic methodologies and evaluate both EACs.
From a documentation perspective, this is ideal as we can place the EACs side by side, to compare and contrast the two behaviours.

\noindent
\begin{minipage}{.5\textwidth}
\begin{lstlisting}[caption=Dense EAC,label={lst:dense}]
    given {
      thereIsAFormula("f",
        "POWER(0,0)",
        arithmetic = DENSE)
    }
    then {
      formulaEvaluatesTo("f", 1)
    }
\end{lstlisting}
\end{minipage}
\begin{minipage}{.5\textwidth}
\begin{lstlisting}[caption=Sparse EAC,label={lst:sparse}]
    given {
      thereIsAFormula("f",
        "POWER(0,0)",
        arithmetic = SPARSE)
    }
    then {
      formulaEvaluatesTo("f", 0)
    }
\end{lstlisting}
\end{minipage}

\subsection{Adapting to a specification}\label{subsec:adapting-to-a-specification}

A requirement of BDS is the capability to create an adapter from the DSL to the specification, which requires a specification framework where animation is feasible.
We considered a number of formal specification systems with suitable tooling, but -- to reaffirm the arguments presented in section 3 of~\cite{larsen2011formal} -- as engineers in industry we preferred the flexibility and familiarity of VDM~\cite{larsen2010overture} to more alien, proof-oriented methods such as Event-B~\cite{abrial2010modeling}.

VDM's Overture ecosystem provides a choice of IDEs that can animate a specification from the UI, while the published JARs enabled us to perform animation programmatically.
Specifically, the capability to animate any function or operation from the specification was a significant boon as it enabled us to construct focused EACs rather than requiring a full system definition in every instance.
This flexibility, and the maturity of the tooling, enabled us to build the specification adapter iteratively, doing `just enough' to run new EACs.
Having this capability enabled us to be fully agile in our process, and we believe that it made VDM an ideal choice for our project, and for BDS generally.

In \cite{oda2015vdm}, Oda illustrates how animation in VDM can be used to bridge the language gap between that of a project's stakeholders and that of a formal specification.
The EACs used in BDS provide an extra layer of abstraction to provide a single language that can also be used by developers.
Sections 4, 5, and 6 of~\cite{oda2015vdm} demonstrate how animation can be driven interactively through various interfaces, but for our EACs we took an even simpler approach.
Our VDM adapter creates a module that represents the EAC, animates it, and then checks the final returned value to determine success.
The module consists of a single operation that begins with all the declarations made in the {\em given} section of the EAC, followed by the commands corresponding to the {\em whenever} section.
Each check in the {\em then} section is translated to a block that returns \texttt{false} if the check is failed.
If the animation proceeds through all checks successfully then the EAC has passed and \texttt{true} is returned.

\begin{vdmsl}[caption=A translated EAC specification,label=lst:animation,basicstyle=\scriptsize\ttfamily]
AnimateCheck: () ==> bool
AnimateCheck() ==
(
  (
    dcl formulaA: Formula`Formula := mk_Formula`UnsafeFormula([
      DataTypeLiterals`CreateNumberLiteral(0.0),
      DataTypeLiterals`CreateNumberLiteral(0.0),
      AnaplanFormulaFunctions`POWER_SPARSE_2
    ], {1 |-> [], 2 |-> [], 3 |-> [1, 2]});
    (
      dcl f: Formula`Formula := formulaA;
      let
        expected = ValueOption`Create[Number`Number](0)
      in
        if not TestEquality`Equals(FormulaModelling`Evaluate(f, { |-> }), expected)
        then (
          IO`println("* Check 'FormulaEvaluationCheck' failed (tolerant = true)");
          IO`print("    Expected: ");
          IO`println(expected);
          IO`print("    Actual: ");
          IO`println(FormulaModelling`Evaluate(f, { |-> }));
          return false;
        )
        else IO`println("* Check 'FormulaEvaluationCheck' passed");
     );
  );
  return true
)
\end{vdmsl}

\noindent
For example, the EAC in listing~\ref{lst:sparse} generates the VDM operation shown in listing~\ref{lst:animation}.

If an EAC fails, it is trivial to open the workspace in Overture and debug the animation of the specification using its rich tooling.

\section{Integrating into the Agile Process}
An agile project typically starts with the creation of user stories, which are then prioritized and placed into an initial backlog.
This is usually led by a product owner (PO) who collaborates with customers to get a general feel for what a new feature should do and to gauge the importance of each of its aspects.
As work progresses, the team then gets feedback from the customer and new user stories are created to refine and extend the behaviours of the features.

For our project we had a good idea of what the calculation engine should do --- generally the same as the existing engine.
The initialization of the backlog was therefore a little different as the PO was able to quickly identify the features required and rank them according to customer need\footnote{Note that, although our intent was to provide feature parity, like all agile projects we wished to ship when we had a minimum viable product and incrementally broaden the feature set with regular releases thereafter.}.
Specification and development teams worked with the same prioritized backlog, but independently.

There was no need for specification to complete before implementation, if the specification team found bugs or missing behaviours they would simply feed into the backlog of the development teams.
Similarly, the specification was not untouchable, if different choices made sense for the implementation, they could be discussed with the PO, and if approved the required change would be added to the specification team's backlog.
In either case the first step would be to create an EAC that exemplified the expected behaviour, which would be referenced in the ticket created in the backlog and could be immediately committed with an appropriate annotation indicating that it was not supported by relevant implementations.
The fluidity of the specification and the implementations meant that the EACs generated during BDS were considered to be the `source of truth'.
The specification was `complete' when it satisfied all EACs in the system, and an implementation was `complete' when it satisfied the subset of EACs that described its full behaviour\footnote{For this reason it was important to distinguish an action that was not yet supported from one that would never be supported}.

As a project progresses, the PO and team groom the backlog on a regular basis and refine the features at the top of the backlog.
For development work this usually involves establishing the set of AC required for the PO to mark the story as done.
This was a little different for specification work as the ACs were an {\em output} of the BDS process.
Instead, grooming involved identifying the functional elements required for the feature and establishing a broad set of behaviours for those elements.
A trivial example, might be a list functional element -- a data structure which can be used as a dimension of a cube -- with behaviours such as creating a list, adding an item to a list, or moving an item within a list.

The initial set of behaviours would be established from a cursory inspection of the existing system and was created to give an idea of the work involved but, as normal in an agile environment, we would expect change as more detailed analysis and specification were completed.

It was not considered feasible to operate a methodology where sizing was required, as these size estimates are usually guided by the ACs.
Instead, a lean methodology~\cite{shalloway2009lean} was used, with an intent that a team member complete the specification of a single behaviour without interruption.
We would imagine that this would be a choice made by all agile specification teams.

\subsection{Specifying a behaviour}\label{subsec:specifying-a-behaviour}

The general procedure of BDS has been detailed in section~\ref{subsec:what-is-bds?}, while here we elaborate a little more on the process followed in our project.

The first step of BDS is to identify an AC.
For example, the behaviour that allows adding an entity to the end of a list could be described as `when an entity is added, it is contained by the list'.
We would then write an EAC for this -- as previously seen in listing~\ref{lst:add}.
Here, we had the advantage that we could immediately run the EAC against the dense platforms to validate that we had captured the behaviour correctly.

We would then try to run the EAC against the sparse platforms.
If the required functionality had not yet been implemented, we could ensure that the adapter returned the \emph{unsupported} command.
If the EAC ran and succeeded we could conclude that it has been implemented correctly, but if it ran and failed our next step would be to determine whether the difference was believed to be correct or incorrect.
If the difference was acknowledged to be incorrect, we could raise a bug referencing the EAC and feed it into the appropriate team's backlog.
On the other hand, if the development team believed that they had implemented the correct behaviour, we would raise a ticket in a formal difference backlog.
At this point we would mark the EAC as `under review' until the difference was discussed by a working group at a fortnightly meeting, where all the stakeholders would discuss differences and make an informed decision on whether the proposals should be accepted or rejected\footnote{Note that, as this was an agile process, there was always a possibility that a future discovery could invalidate the decision, and members of the team were free to reopen the discussion and present new evidence to the working group before a new decision was made.}.
Once the difference was resolved, it would be necessary to return to the EAC and ensure if reflected the decision made.

The specifier then runs the EAC against the specification to ensure it fails, the EAC is marked as unsupported by the specification, and the author pushes the work to a branch and raises a pull request (PR) for review by another team member.
Once the review of the PR is complete -- and all necessary builds have passed -- the EAC is merged to the mainline, and specification can proceed.

There is not too much to say about writing the specification itself, other than to reinforce that the only changes should be those required to make the EAC pass, and that the presence of EACs does not divest the author of the responsibility to write units tests within the VDM, which should focus on testing the correctness of each individual function, as well as its pre and post conditions.
Having the EACs in place should discourage the specifier from writing end-to-end tests directly in the VDM, and instead ensure the focus is on good coverage of the specification.

\subsection{Continuous validation}\label{subsec:continuous-validation}

An important part of agility is continually integrating the pieces of work from all parts of the team to ensure that changes do not cause problems elsewhere.
The Gradle plugin introduced in~\cite{fraser2018integrating} enables us to build the specification, run all the VDM unit tests in Jenkins, and publish the specification to Artifactory.

Change to the DSL and its associated adapters are managed like any other agile development work -- with frequent commits, PRs, and adoption upstream.
The implementation adapters leverage Java's service loading mechanism to use whichever version of the implementation they find on their classpath, while the VDM adapter was necessarily tied to a specific version of the specification, since part of its function was to package the text-based specification into a Java library that could be `run'.

The EACs are all placed into a single validation repository, and whenever the mainline of this project builds successfully, it also checks for newer versions of the DSL, the adapters and the implementations.
If a new version is found, the system automatically raises a PR to update the version numbers of the dependencies, and if this PR builds successfully, it is considered a team priority to merge it quickly.
On occasion, a version update PR will fail, typically due to either a regression of behaviour in an implementation, or because some previously unsupported action has been implemented, causing an EAC that was previously being skipped to fail.
At this point we follow the same process that we use when a new EAC fails, as described previously.
Again, resolving this difference is considered a team priority.
As with any agile development team, team members will work on a failing mainline or version update build -- rather than regular assigned work -- until it is fixed.

\section{Results}
\section{Results}
\label{sec:results}

This section presents results of applying the MOO support to the Robotti case study described above, using the four scenarios presented in Figure~\ref{fig:res_scenarios} (\emph{sin1}, \emph{turn\_ramp1}, \emph{speed\_ramp1}, \emph{speed\_step1}). In order to ensure comparisons with exhaustive search are valid, and any comparisons of speed were made on the same machine,  the exhaustive search was re-run for the four scenarios. These results are shown in Figure~\ref{fig:res_scenarios2}.

Note that the trends found are the same except for \emph{turn\_ramp1} which found a completely different trend. To try and account for this discrepancy, two manual co-simulations were run with values of 3000 for \emph{m\_robot} and 20000 for \emph{cAlphaF}, and \emph{mu} values of 0.3 and 0.7, respectively. After running these two co-simulations and evaluating their cross-track error using the objective script, the configuration with a mu value of 0.3 returned a considerably lower error value (in line with Figure~\ref{fig:res2_turn_ramp1}). It is therefore possible that Bogomolov et al.~\cite{bogomolov2021tuning} used some different setup that affected these results. In any case, the results from the re-run are used here to ensure comparisons are made on a consistent setup.

\subsection{Optimisation of Robotti using MOO Algorithms}
\label{sec:results:moo}

The MOO functionality was then used to optimise these scenarios using two different algorithms, and compare against exhaustive search. Given its popularity, the NSGA-II (Non-dominated Sorting Genetic Algorithm-II) algorithm was chosen, as well as the SMPSO (Speed-constrained Multi-objective Particle-Swarm Optimisation) since it is the most different from the evolutionary approach of NSGA-II. To ensure fairness, the MOO algorithms were initially limited to the same 125 co-simulations as the exhaustive search. Table~\ref{tab:initial} shows the mean cross track error found by the two algorithms and the exhaustive search.

\begin{figure}[tb]
	\centering
    \subfloat[$\mathsf{sin1}$ Scenario\label{fig:res2_sin1}]
	{\includegraphics[width=0.45\textwidth]{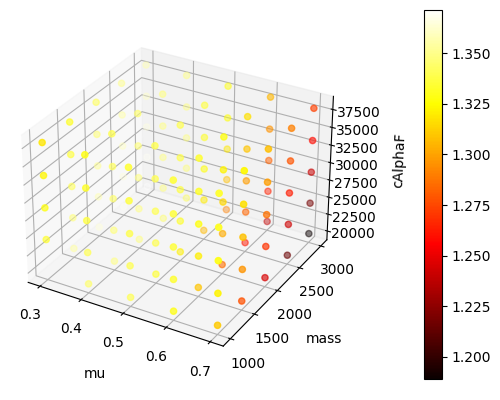}}
    \subfloat[$\mathsf{turn\_ramp1}$ Scenario\label{fig:res2_turn_ramp1}]
	{\includegraphics[width=0.45\textwidth]{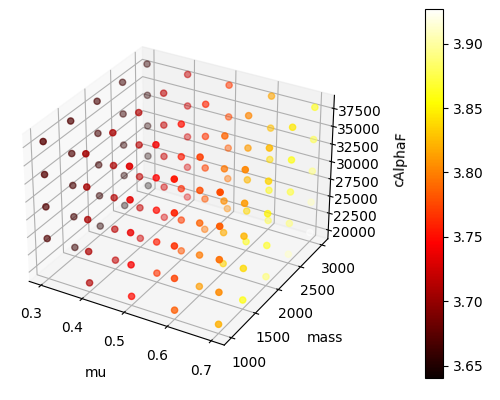}}\\
    \subfloat[$\mathsf{speed\_ramp1}$ Scenario\label{fig:res2_speed_ramp1}]
	{\includegraphics[width=0.45\textwidth]{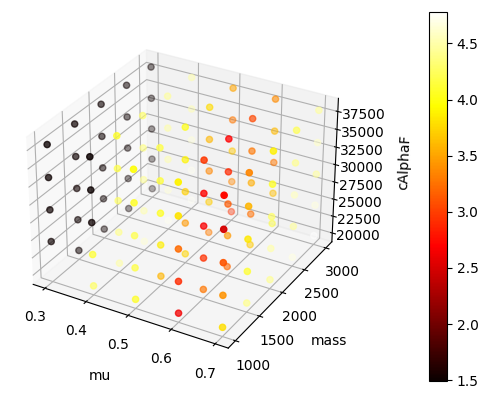}}
    \subfloat[$\mathsf{speed\_step1}$ Scenario\label{fig:res2_speed_step1}]
	{\includegraphics[width=0.45\textwidth]{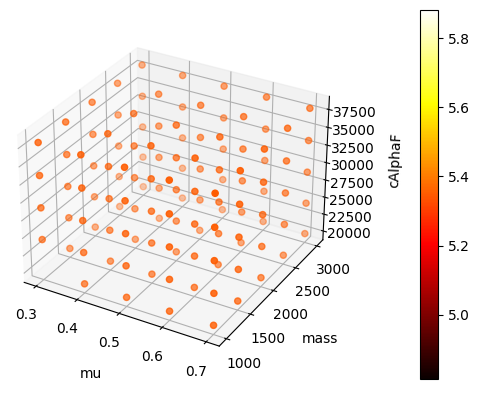}}
	\caption{Results of a re-run exhaustive search of the four scenarios}
\label{fig:res_scenarios2}
\end{figure}

\begin{table}
\centering
\caption{Mean cross track error found by algorithm and scenario}\label{tab:initial}
\begin{tabular}{p{2.5cm}p{2cm}c} \hline
\textbf{Scenario} & \textbf{Algorithm} & \textbf{Mean Cross Track Error} \\ \hline
\multirow{3}{*}{\emph{sin1}}            & NSGA-II       & 1.226 \\
                                        & SMPSO         & 1.225	\\
                                        & Exhaustive    & 1.188 \\ \hline
\multirow{3}{*}{\emph{turn\_ramp1}}     & NSGA-II       & 5.348 \\
                                        & SMPSO         & 5.348	\\
                                        & Exhaustive    & 3.640 \\ \hline
\multirow{3}{*}{\emph{speed\_ramp1}}    & NSGA-II       & 1.508 \\
                                        & SMPSO         & 1.498	\\
                                        & Exhaustive    & 1.491 \\ \hline
\multirow{3}{*}{\emph{speed\_step1}}    & NSGA-II       & 3.653 \\
                                        & SMPSO         & 3.655	\\
                                        & Exhaustive    & 5.348 \\ \hline
\end{tabular}
\end{table}

As can be seen, it appears that the MOO algorithms are unable to find optimal solutions as well as the previous study's exhaustive search, when completing only 125 evaluations. In the \emph{sin1} scenario, both algorithms came somewhat close to the error values found in that study, but in every other scenario, both algorithms calculated a much worse optimal amount of error. Plots of the intermediate solutions are given in Figure~\ref{fig:res_scenarios3}. By inspection, it appears as though the trends are the same as those in Figure~\ref{fig:res_scenarios2}; \emph{sin1} is in the opposite corner to \emph{turn\_ramp1} and \emph{speed\_ramp1}, and \emph{speed\_step1} shows no overall trend.

To investigate whether further co-simulations would find as optimal a solution as the exhaustive search, the optimisations for \emph{sin1} were repeated using NSGA-II and incrementally increasing the maximum number of evaluations. These are shown in Figure~\ref{fig:sin1}. It was not until around 750 evaluations that the NSGA-II consistently produces results as well as the exhaustive search.

This was somewhat surprising, as we expected that using MOO algorithms would find the most optimal configuration much faster in terms of the number of evaluations. Or failing that, in the same number of evaluations, the MOO algorithm would find a more optimal solution. Since such a small set of values on the range of each parameter was tested during the exhaustive study, values between those specified would not be tested. Because of this, we assumed more optimal solutions would be missed, and that the MOO algorithms would find them since they would have the ability to test the full range of values possible.

\begin{figure}[tb]
	\centering
    \subfloat[$\mathsf{sin1}$ Scenario\label{fig:res3_sin1}]
	{\includegraphics[width=0.48\textwidth]{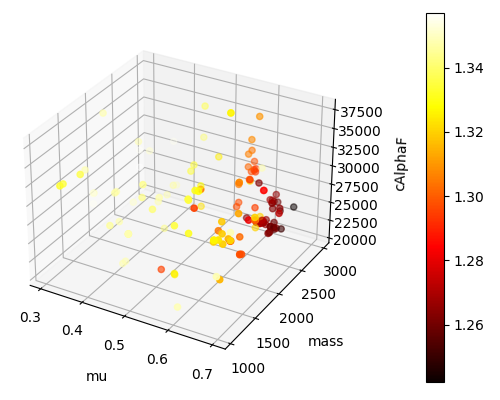}}
    \subfloat[$\mathsf{turn\_ramp1}$ Scenario\label{fig:res3_turn_ramp1}]
	{\includegraphics[width=0.48\textwidth]{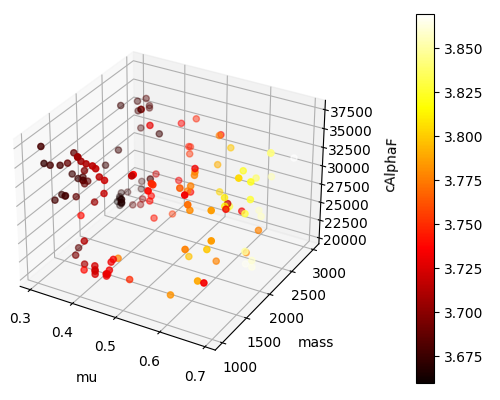}}\\
    \subfloat[$\mathsf{speed\_ramp1}$ Scenario\label{fig:res3_speed_ramp1}]
	{\includegraphics[width=0.48\textwidth]{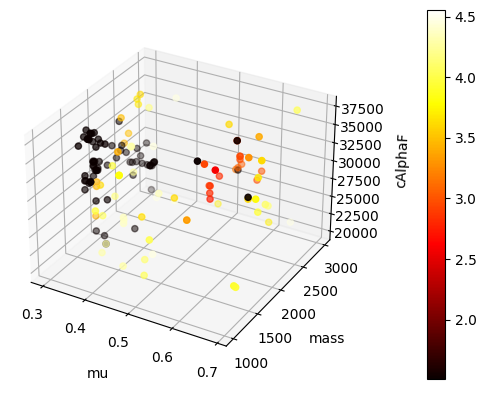}}
    \subfloat[$\mathsf{speed\_step1}$ Scenario\label{fig:res3_speed_step1}]
	{\includegraphics[width=0.48\textwidth]{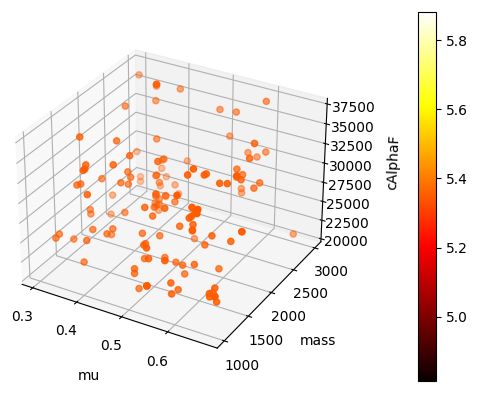}}
	\caption{Results of a NSGA-II results for the four scenarios limited to 125 co-simulations}
\label{fig:res_scenarios3}
\end{figure}

\begin{figure}[tb]
	\centering
    \subfloat[$\mathsf{sin1}$ (125 co-simulations)\label{fig:sin1_125}]
	{\includegraphics[width=0.48\textwidth]{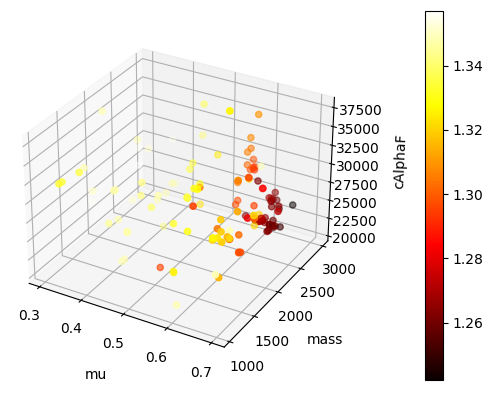}}
    \subfloat[$\mathsf{sin1}$ (250 co-simulations)\label{fig:fig:sin1_250}]
	{\includegraphics[width=0.48\textwidth]{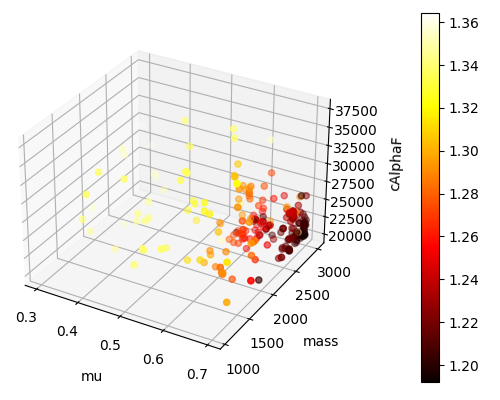}}\\
    \subfloat[$\mathsf{sin1}$ (500 co-simulations)\label{fig:fig:sin1_500}]
	{\includegraphics[width=0.48\textwidth]{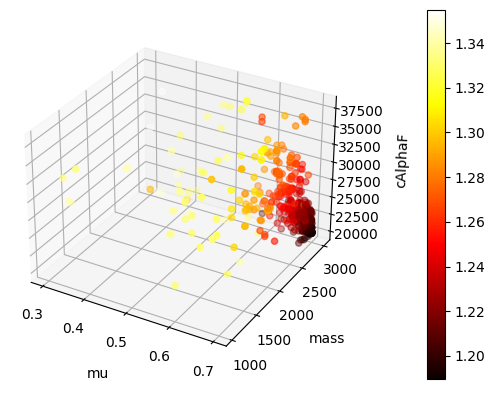}}
    \subfloat[$\mathsf{sin1}$ (750 co-simulations)\label{fig:fig:sin1_750}]
	{\includegraphics[width=0.48\textwidth]{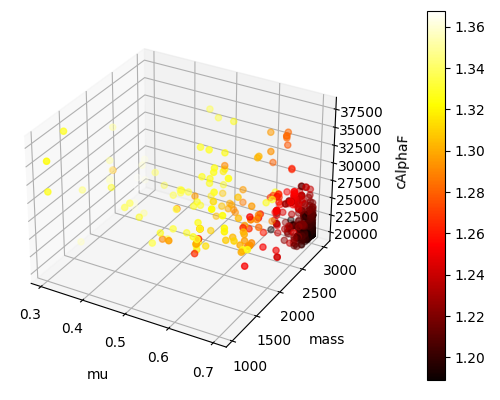}}
	\caption{Results of NSGA-II on \emph{sin1} with increasing numbers of co-simulations}
\label{fig:sin1}
\end{figure}

These findings suggested that, in terms of number of evaluations, the MOO solution was much slower at finding optimal solutions. However, we hypothesise that for models with many more dimensions, MOO algorithms would converge on optimal configurations in fewer evaluations than in an exhaustive search. Due to the nature of the exhaustive search, as the number of dimensions increase, the number of evaluations increases exponentially. This is because the number of evaluations is given by:

\[ | D_1 | \cdot | D_2 | \cdot | D_3 | \ldots \cdot | D_n | \]

\noindent where $N$ and $D_n$ is the set of values that makes up the $n$th dimension. If the cardinality of each set tested is uniform, this would give the exhaustive search a time complexity of $O(c^n)$ in terms of the number of evaluations, where $n$ is the number of dimensions. The time complexity of NSGA-II on the other hand is not affected by the number of dimensions directly, since the time complexity of NSGA-II is given by $O(GMN^2)$, where $G$ is the number of generations, $M$ is the number of objectives and $N$ is the population size~\cite{jensen2003reducing}. While some or all of these three variables would likely need to increase with the number of dimensions to continue providing accurate optimisation configurations, this would not correspond to as much as an exponential increase in evaluations.

Using equation derived above, doubling the number of dimensions of the Robotti design space to six dimensions would increase the number of evaluations performed by the exhaustive search from 125 to 31,250. Even if this increase in dimensions required by NSGA-II to double the number of generations completed, this search would still only comprise of 1,500 evaluations. Because of this, we believe that problems with larger numbers of dimensions, the MOO solution could be much more efficient at DSE when compared with exhaustive search. This however requires further study through development of co-simulation examples that exhibit multi-objective behaviour.

\subsection{Evaluation of Execution Time}
\label{sec:results:time}

Table~\ref{tab:speed} compares the execution speed of the exhaustive DSE, along with the NSGA-II algorithm running on one and six threads. It shows that when running each method for the same number of evaluations, using NSGA-II is slightly slower than using an exhaustive search, likely due to the overhead of using the jMetal library. However, because the MOO solution supports parallel execution, the ability to run multiple simulations at once decreases optimisation time significantly. In this case, utilising six parallel cores decreased optimisation by a factor of three when compared to single thread evaluation. Note that the most recent version of the existing DSE scripts now support parallel execution, so further benchmarks are required to truly compare approaches, using a range of case studies to show the strengths and weaknesses of the two approaches.

\begin{table}
\centering
\caption{Execution time for exhaustive and NSGA-II algorithms}\label{tab:speed}
\begin{tabular}{p{2.5cm}p{3.5cm}cc} \hline
\textbf{Scenario} & \textbf{Algorithm} & \textbf{Mean Time / Evaluation (s)} & \textbf{Mean Total (s)}\\ \hline
\multirow{3}{*}{\emph{sin1}}
& NSGA-II (6 threads)   & 2.2 & 280  \\
& NSGA-II (1 thread)    & 6.1 & 771 \\
& Exhaustive (1 thread) & 5.3 & 664 \\ \hline
\multirow{3}{*}{\emph{turn\_ramp1}}
& NSGA-II (6 threads)   & 3.8 & 480 \\
& NSGA-II (1 thread)    & 10.9 & 1365 \\
& Exhaustive (1 thread) & 9.1 & 1141 \\ \hline
\multirow{3}{*}{\emph{speed\_ramp1}}
& NSGA-II (6 threads)   & 3.9 & 488 \\
& NSGA-II (1 thread)    & 11.8 & 1485 \\
& Exhaustive (1 thread) & 10.4 & 1296 \\ \hline
\multirow{3}{*}{\emph{speed\_step1}}
& NSGA-II (6 threads)   & 4.6 & 584 \\
& NSGA-II (1 thread)    & 12.3 & 1540 \\
& Exhaustive (1 thread) & 10.8 & 1350 \\ \hline
\end{tabular}
\end{table}

\section{Concluding remarks}\label{sec:concludingRemarks}
We started using VDM at Anaplan before beginning the project described in this paper, but the focus had been on investigating new modelling concepts rather than the delivery of working software.
With the switch to a delivery-focus, we knew we needed a process that could align with the workflow of the development teams assigned to the project.
Behaviour driven specification proved to be such a process, enabling us to create a formal specification {\em with} the development teams rather than {\em for} them.

In the initial phases of the project, there was a suspicion that formal methods were intrinsically linked to waterfall development.
This worried developers who believed they would lose the ability to influence the system's requirements.
There was a belief that the specification team would form a barrier between the product owner and the developers, and at early meetings we needed to remind the wider group that the role of the specification team was to record decisions, not to make them.
We believe that the use of BDS helped to overcome this:
not only because it has enabled the specification process to be more flexible, but also because it has made the specifiers feel like a collaborative partner in the implementation effort rather than a siloed team issuing diktats to developers.
The use of executable acceptance criteria has reinforced this;
having a `source of truth' expressed in a language that requires no special understanding enables all members of the project team to engage on an equal footing.

As we approach our first customer-facing release, the feeling across the whole project team is that the integration of formal specification into the agile process has been a success, with specifiers, developers, and product owners all highlighting the value it has provided.

\clearpage
\endgroup

\begingroup
\renewcommand\theHchapter{3-Oda:\thechapter}
\renewcommand\theHsection{3-Oda:\thesection}
\locallabels{3-Oda:}
\setcounter{footnote}{0}
\setcounter{chapter}{0}
\setcounter{lstlisting}{0}
\fontfamily{ptm}\selectfont

\makeatletter
\def\input@path{{3-Oda/}}
\makeatother

\graphicspath{{3-Oda/}}

\title{Refactoring for Exploratory Specification in VDM-SL}

\titlerunning{Refactoring for Exploratory Specification in VDM-SL}

\author{
Tomohiro Oda\inst{1} \and 
Keijiro Araki\inst{2} \and
Shin Sahara\inst{3} \and
Han-Myung Chang\inst{4} \and
Peter Gorm Larsen\inst{5}
}

\authorrunning{Oda, T.; Araki, K.; Sahara, S.; Chang, H.M.; Larsen, P.G.}

\institute{Software Research Associates, Inc.~(\email{tomohiro@sra.co.jp})
\and National Institute of Technology, Kumamoto College~(\email{araki@kyudai.jp})
\and Hosei University~(\email{ss@shinsahara.jp})
\and Nanzan University~(\email{chang@nanzan-u.ac.jp})
\and Aarhus University, DIGIT, Department of Electrical and Computer Engineering,~(\email{pgl@ece.au.dk}) 
}

\maketitle

\begin{abstract}
This paper discusses refactoring as a tool for identifying concepts in the problem domain and defining them with appropriate language elements in appropriate modules through abstraction and generalisation.
Refactoring operations required for exploratory specification are identified, and their implementation on a tool called ViennaTalk is described and discussed.
\end{abstract}

\section{Introduction}
\label{sec:Introduction}

Formal specifications have been reported to increase productivity and reliability by describing system functionality with rigour, enabling early detection and correction of specification-induced problems in software development.
VDM-SL is a formal specification language with an executable subset \cite{Fitzgerald&09}. 
Interpreted executions of the specification, or specification animation, can be used to simulate and try out the behaviour of the system before implementation takes place.

Specification animation is a powerful tool for deepening understanding of the problem domain.
Software development is typically an ill-defined problem where the problem to be solved is not defined in advance.
Although a development project starts with objectives and a tentative goal, the developers do not always have a full understanding of the problem domain.
During the development, the developers learn and understand the problems to be solved, and specify, design, and implement the software as a means to solve them.
Especially in the early stages, it is important to learn about the problem domain by defining and animating the formal specification, and also feedback from stakeholders.
Exploration involves frequent modification to the specification to identify associated concepts and find their definition with appropriate abstraction and concise presentation.

Refactoring is a technique to modify the program source while keeping the behaviour of the program code \cite{Mens&04}.
Refactoring has mainly been used to improve the maintainability of code.
In particular, in agile development, refactoring plays an important role in removing design distortions caused by added features and in preparing for further feature additions.
The tool support for refactoring originated from the programming community of Smalltalk \cite{Roberts&97}.
Smalltalk is by its nature bundled with powerful IDEs featuring class browsers that strongly support prototyping.
It is not a coincidence that Smalltalk also delivered a unit testing framework to the programming community which later derived JUnit \cite{Louridas&05}.

Both refactoring tools and unit testing framework support rapid modification to the program code in prototyping.
Exploratory specification and prototyping share rapid modification to the source as well as learning aspects through trial and error.
Refactoring in agile development support the maintenance aspect of prototyping.
Rapid trials and errors followed by modifications often make the source code less readable and thus make further modifications harder.
On the other hand, the exploratory specification values its learning aspect.
We expect refactoring techniques to help the engineers widen and deepen their understandings of the problem domain.

Section \ref{sec:RefactoringForExploration} will discuss the role of refactoring in the exploratory specification phase and show families of semi-automated refactoring operations for exploratory specification.
In Section \ref{sec:RefactoringSupportInViennaTalk}, refactoring functionalities implemented in ViennaTalk will be described with an example.
Afterwards, Section \ref{sec:RelatedWork} explains existing research on refactoring of VDM-SL.
Finally, Section \ref{sec:DiscussionAndConcludingRemarks} discusses the result of implementation and concludes the paper.

\section{Refactoring for exploration}
\label{sec:RefactoringForExploration}

Refactoring is technically a collection of transformations on abstract syntax trees (ASTs) that preserve the behaviour of the existing code.
Refactoring of program code often benefits the maintainability of the code including readability and extensibility by reorganising the structure of the code.
Refactoring is often applied to code fragments that typically contain magic numbers, frequent expressions that can be abstracted, definitions located in inappropriate modules, misleading identifiers, and so on.
In the context of software maintenance, such code fragments are often called {\it code smell}.

The exploratory specification is the stage of writing a specification with a partial understanding of the application domain.
Software development typically goes parallel with acquiring new concepts in the target domain and exploring how they are associated with other known concepts.
Since a specification engineer who writes a specification does not have an accurate grasp of the entire domain, many parts of the specification remain tentative.
The specification at this stage is not always comprehensive and concise due to limited understanding of the domain and thus contains {\it code smell}.
In the exploratory specification, they are signs of learning opportunities.
While code smell in software maintenance is considered a threat to the maintainability of the code, it is an opportunity for learning in the exploratory specification.
Refactoring on a formal specification does not add new functionality to the system to be developed, but adds more senses to the model.

Our approach to refactoring operations focuses on names.
In programming, refactoring is sometimes carried out to improve performance by modifying to faster code and data structures without changing the functionality.
Although VDM-SL has an executable subset, performance is less emphasised in general.
We value the readability and maintainability brought by appropriate naming and abstraction.
In this research, we identified and implemented a set of semi-automated refactoring operations for naming, renaming, and unnaming a fragment of the specification source.
The goal is to provide a safe way to modify the specification without unintended change of the behaviour and also to reduce the workload of the editing task.

\subsection{Refactoring for naming}

The first group of refactoring operations is to give a name to a fragment of the specification.
In VDM-SL, there are three kinds of scopes of names: (1) global names, e.g. module name, (2) module-wide names defined in types sections, values sections, state sections, functions sections and operations sections, and (3) local names such as local definitions in let expressions/statements, and pattern identifiers for the pattern matching in parameters of functions/operations, local binding in set/sequence/map comprehensions, lambdas, quantifiers, loop constructs and so on.

The {\it Extract} family of refactoring operations generates the language constructs enumerated above from a fragment of the specification.
The {\it Use} family of refactoring operations replaces fragments in the same pattern with another that uses names.

\subsubsection{Extracting module-wide names}

In VDM-SL, type definitions, value definitions, state variables, function definitions, and operation definitions have the module-wide scope.
{\it Extract type} is a refactoring operation to define a new type definition for the specified type expression.
The newly defined type name will be used In the original place of the specified expression.

{\it Extract value} defines a new value definition for the specified expression.
Please note that a refactoring operation should be safe; the resulting specification should not break the well-formedness of the specification.
If the specified expression contains a reference to a state variable or an operation call,  the tool should not apply the operation.

\begin{figure}
\begin{vdmsl}
functions
  priceIncludingTax : real -> real
  priceIncludingTax(displayPrice) == 
    displayPrice + displayPrice * 0.1
\end{vdmsl}
\vspace*{-2em}\begin{center} $\Downarrow$ {\it Extract value on }{\tt 0.1}\end{center}\vspace*{-2em}
\begin{vdmsl}
values
  TAX_RATE = 0.1
functions
  priceIncludingTax : real -> real
  priceIncludingTax(displayPrice) == 
    displayPrice + displayPrice * TAX_RATE
\end{vdmsl}
\vspace*{-1em}
\caption{An example application of the {\it Extract value} refactoring}
\label{fig:spec-extract-value}
\end{figure}

Figure \ref{fig:spec-extract-value} illustrates {\it Extract value} operation that transforms the definition above the arrow into the one below by adding a value definition.
The original definition specifies the computation of price with 10\% sales tax using a magic number {\tt 0.1}.
By extracting {\tt 0.1} as a constant value named {\tt TAX\_RATE}, the definition becomes more expressive to the readers.

{\it Extract state variable} adds a new state variable.
The expression specified by the user will be used in the initialiser of the state definition.
The specified expression will be used as the right-hand side of the type definition.
If the specified expression contains a stateful subexpression, the tool must block the refactoring.
Also, the tool must guarantee that all the record constructor expressions with the state must be given an appropriate element for the newly added field.
A conservative approach is to prohibit the {\it Extract state variable} refactoring when there are one or more record constructors with the state. 

{\it Extract function} and {\it Extract operation} define a new definition of function and operation accordingly.
To define a new function, a let expression is a language construct of VDM-SL that has corresponding functionality with function application: parameters, arguments and a body.
If there is no dependency among local definitions in a let expression, pairs of a local name and its value can be interpreted as pairs of a parameter and an argument as the local bindings to evaluate the body expression.
The tool must check whether the body expression contains a stateful subexpression and also a reference to other local names that are invisible in the top-level scope.
The let statement can similarly be converted into an operation definition.

\begin{figure}
\begin{vdmsl}
functions
  priceIncludingTax : real -> real
  priceIncludingTax(displayPrice) == 
    displayPrice + 
    let rate:real = 0.1, taxable = displayPrice in taxable * rate
\end{vdmsl}
\vspace*{-2em}\begin{center} $\Downarrow$ {\it Extract function}\end{center}\vspace*{-2em}
\begin{vdmsl}
functions
  priceIncludingTax : real -> real
  priceIncludingTax(displayPrice) == 
    displayPrice + tax(0.1, displayPrice);
  tax : real * ? -> ?
  tax(rate, taxable) == taxable * rate
\end{vdmsl}
\vspace*{-1em}
\caption{An example application of {\it Extract function} refactoring}
\label{fig:spec-extract-function}
\end{figure}

Figure \ref{fig:spec-extract-function} shows an example application of {\it Extract function}.
In the original specification, a 10\% sales tax computation is specified using let expression.
{\it Extract function} generates the definition of the function {\tt tax} from the let expression.
Please note that the type name {\tt ?} means so-called {\it any} type.
Any-type is not a legitimate type in the language specification of VDM-SL\cite{Larsen&13b}, but is a feature supported by many VDM-SL interpreters including VDMJ, Overture tool, VDMTools and ViennaTalk.
To turn the specification into a legitimate one, a concrete type should be given for the any-type, i.e. {\tt real} in this particular case.
It is in general difficult to infer a static type of an arbitrary expression though this case can be trivially inferred.
This issue will be discussed in Section \ref{sec:DiscussionAndConcludingRemarks}.

\subsubsection{Extracting local names}

The let expression is a language construct to make local definitions.
Parameters of function or operation also have local scopes.
Defining those local names for expressions add contextual meaning to the expressions and thus are expected to gain readability.
Those names also help abstraction.

{\it Extract let expression} turns an expression {\it exp} into {\tt let} {\it localname} {\tt =} {\it exp} {\tt in} {\it localname}.
Although the resulting form is trivial, this refactoring can be effectively applied along with refactoring operations for scoping described later.
{\it Extract local definition} adds a local definition with the specified subexpression in a let expression.

{\it Extract parameter in function} adds a parameter to the function definition and also adds the specified subexpression as an argument to the function for all applications of the function.
The subexpression ported as the argument must be valid in the function applications.
Also, the subexpression must be side-effect free because the argument is evaluated before the body of the let expression and the change of order of evaluation may produce different results than the original.
{\it Extract parameter in operation} similarly adds a parameter to an operation definition and the extra argument to the callers.

\begin{figure}
\begin{vdmsl}
functions
  tax : real -> real
  tax(taxable) == taxable * 0.1
\end{vdmsl}
\vspace*{-2em}\begin{center} $\Downarrow$ {\it Extract parameter in function}\end{center}\vspace*{-2em}
\begin{vdmsl}
functions
  tax : real * ? -> real
  tax(taxable, rate) == taxable * rate
\end{vdmsl}
\vspace*{-1em}
\caption{An example application of {\it Extract parameter in function} refactoring}
\label{fig:spec-extract-parameter-in-function}
\end{figure}

Figure \ref{fig:spec-extract-parameter-in-function} illustrates an example application of {\it Extract parameter in function}.
The magic number {\tt 0.1} that appears in the original specification is turned into a parameter of the function so that the function gains it generality.
Please note that the {\it any}-type appears also in this refactoring.
Although the literal value {\tt 0.1} is apparently of the {\tt real} type, it is not trivial in general to find an appropriate name of the type among possibly many alias names of the {\tt real} type.

\subsubsection{Using names}

Once a name is given to an expression, the occurrences of the same expression may better use the name.
The {\it Use} family of refactoring operations supports the user to deal with fragments that frequently appear in the specification.
{\it Use type} replaces an occurrence of the right-hand side of a type definition with the type name.
{\it Use value} does the same on a value definition.
{\it Use function} replaces an expression that matches with the definition body of a function with a function application that results in the same expression.
If a function has a precondition, {\it Use function} is not applicable to the function because the resulting function application may possibly violate the precondition.

\subsection{Refactoring for renaming}
In the exploratory stage of specification, the specifier has limited knowledge of the system and the domain, often resulting in misleading names.
Modification to the name, its location and its scope is also expected to be frequently performed to improve the quality of the specification as the knowledge is updated.

\subsubsection{Changing the name}

Renaming is frequently applied among refactoring operations in programming languages \cite{Negara&13}.
Renaming is not mere string replacement because the lexically same name can be semantically different names.
A semi-automated renaming operation that takes account of syntax and semantics of the language is desirable to conduct safe and efficient renaming.
The {\it Rename} family of refactorings supports rename types, quote types, values, functions, state variables, operations, local definitions and parameters.

\subsubsection{Changing the scope}
\begin{figure}
\begin{vdmsl}
functions
  priceIncludingTax : real -> real
  priceIncludingTax(displayPrice) == 
    let rate = 0.1, taxable = displayPrice 
    in displayPrice + taxable * rate
\end{vdmsl}
\vspace*{-2em}\begin{center} {\it Narrow let expression} $\Downarrow$ \hspace{1cm} $\Uparrow$ {\it Widen let expression}\end{center}\vspace*{-2em}
\begin{vdmsl}
functions
  priceIncludingTax : real -> real
  priceIncludingTax(displayPrice) == 
    displayPrice + 
    let rate = 0.1, taxable = displayPrice in taxable * rate
\end{vdmsl}
\vspace*{-1em}
\caption{An example application of {\it Widen/Narrow let expression} refactoring}
\label{fig:spec-widen-narrow-let}
\end{figure}
Every name has its scope and choosing the right scope of a local name is considered generally good for readability and maintainability.
The {\it Widen} family of refactorings widens the body of the {\tt in}-clause in a let expression or a let statement into its enclosing expression or statement.
The {\it Split} family of refactoring separate one local definition from an inner let expression to the outer let expression.
If the let expression is not nested, the {\it Split} refactoring will create another let expression that encloses the original let expression.
The {\it Narrow} family of refactorings narrows the body into its subexpression if and only if the locally defined names appear only in the subexpression.
Figure \ref{fig:spec-widen-narrow-let} shows an example of widening and narrowing a let expression.

\subsubsection{Changing the location}

VDM-SL supports a modular construction of a model.
The choice of the module to define a type, value or function is important but sometimes not trivial.
The {\it Move} family of refactoring operation is to move a definition from one module to another and update the import/export declarations of each module.

\subsection{Refactoring for unnaming}
In the exploration to a better presentation of the model, a function definition or a value definition may have little use or be found confusing.
A let expression may introduce unnecessary syntactical complexity to the presentation.
Removal of a name needs not only just deleting its definition but also replace the name with its flat presentation.
Semi-automated refactoring operation for these tasks is expected to save the user's cognitive workload.

The {\it Inline} family of refactoring operations supports inlining both local and module-wide names.
{\it Inline type}, {\it Inline value} and {\it Inline function application} are inverse of their counterparts in the {\it Use} family, which replace the expression with the definition body.
{\it Inline let expression} and {\it Inline let statement} replace the occurrences of the locally defined names with their definition bodies and remove the local definition in the let expression or statement.
The {\it Remove} family of refactorings deletes the declaration of names when no reference to the names is left in the specification.
A {\it Remove} refactoring can be either automatically performed after an {\it Inline} refactoring or triggered independently.


\section{Refactoring Support in ViennaTalk}
\label{sec:RefactoringSupportInViennaTalk}

ViennaTalk\cite{Oda&17a} is an IDE for the exploratory stage of VDM-SL specification built on top of the Pharo Smalltalk environment.
ViennaTalk has a browser called VDM Browser designed to empower the use of animation in writing a specification, evaluating expressions, unit testing, UI prototyping, web API prototyping, and executable documentation.
While text-based source editors in conventional IDEs provide text-editing operation at the source file tree, VDM Browser holds an interpreting process that contains the source text and the binding information of state variables.
We chose to implement another browser named Refactoring Browser to implement semi-automated refactoring operations. 
The design of the new Refactoring Browser is revised to hold an abstract syntax tree (AST) because refactoring can more naturally be defined on AST rather than the source text.
The source text displayed on the Refactoring Browser is promptly generated from the AST when the user selects the definition to edit.
The source text edited by the user is parsed and merged into the AST, and the source text will be disposed.

It is widely advised that refactoring should be performed with unit testing to ensure that the functionality is not changed.
Unit testing functionality was ported from VDM Browser to Refactoring Browser with a modification to UI.
While VDM Browser has a Unit Testing tab to display the result of unit tests behind the source pane, Refactoring Browser places the unit testing pane aside the source pane so that the result of unit testing is always visible and has more chance to catch the user's eyes.

History functions to record versions of the specification are added to Refactoring Browser.
Semi-automated refactoring may cause unexpected modifications to the specification, and the ability to roll back to previous versions is desirable.
Refactoring Browser records a mirror copy of AST into history at every execution of refactoring operations and manual edits.

\begin{figure}
\begin{center}
\includegraphics[width=1\textwidth]{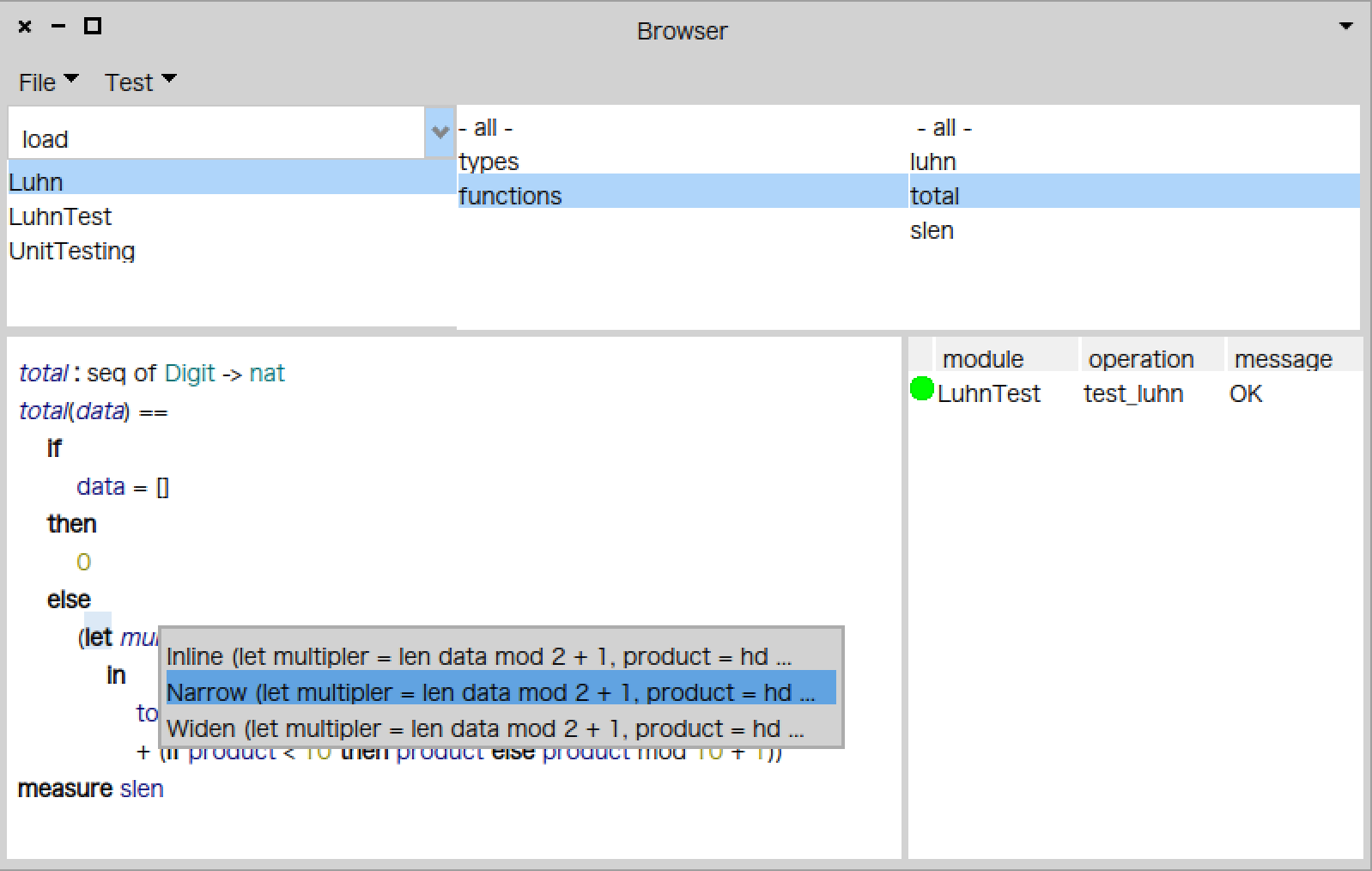}
\end{center}
\caption{Screenshot of Refactoring Browser}
\label{fig:screenshot-refactoring-browser}
\end{figure}

Figure \ref{fig:screenshot-refactoring-browser} is a screenshot of the Refactoring Browser.
The left top dropbox that reads ``load'' is a history selector that lists all versions of the AST with the names of refactoring operations, and let the user choose a version to roll back.
The list widget on the left is called module list that shows all modules defined in the specification, e.g. {\tt Luhn}, {\tt LuhnTest} and {\tt UnitTesting} in the Figure \ref{fig:screenshot-refactoring-browser}.
The up centre list widget shows sections in the selected module, and the right list enumerates top-level definitions in the sections.
The source text of the selected top-level definition is generated from the AST and shown in the lower text pane.
In the screenshot, the function {\tt total} in the functions section of the {\tt Luhn} module is edited.

A refactoring operation is a semi-automated manipulation of AST.
Each refactoring is defined as a subclass of ViennaRefactoring class.
When a user opens a context menu, each subclass of ViennaRefactoring is created and tested executability to the specified AST node.
All refactoring instances that returned true to the executability test will be listed on the context menu.
In Figure \ref{fig:screenshot-refactoring-browser}, the keyword {\tt let} highlighted in the source is selected by the user, and the context menu shows all applicable refactoring operations on the function.

\subsection{Example: LUHN}

\begin{figure}
\begin{vdmsl}
module Luhn
exports all
definitions
types
  Digit = nat inv d == d < 10;

functions
  luhn : seq1 of Digit -> Digit
  luhn(data) == total(data)  * 9 mod 10;
    
  total : seq of Digit -> nat
  total(data) ==
    if data = []
    then 0
    else
       (let 
         multipler = len data mod 2 + 1, 
         product = hd data  * multipler
       in
         total(tl data) + 
           (if product < 10 
           then product 
           else product mod 10 + 1))
    measure slen;
    
    slen : seq of Digit -> nat
    slen(data) == len data;
end Luhn
\end{vdmsl}
\caption{The original source of the {\tt Luhn} module}
\label{fig:luhn-original}
\end{figure}

Luhn algorithm, also known as mod-10,  is a checksum algorithm mainly used to verify a series of numbers and/or letters against accidental errors such as typos.
Figure \ref{fig:luhn-original} shows a part of the specification of the Luhn algorithm published at the Overture website\footnote{\url{https://www.overturetool.org/download/examples/VDMSL/LUHNSL/index.html}} with modification.
The {\tt lune} function computes the check digit for the given sequence of digits.
Although the algorithm is well-known and specification is already mature, the specification can be modified more informative by applying a series of semi-automated refactoring operations and a few manual edits on ViennaTalk's Refactoring Browser.
Two improvements will be illustrated and explained in the rest of this section.

\subsubsection{Improve the definition of {\tt total}}

The first improvement is on the function {\tt total}.
A let expression appears in the definition body of {\tt total}.
The subexpression {\tt total(tl data)} is included in the let expression, but does not refer to any local definition of the let expression.
The {\it Narrow let expression} refactoring can be applied to adjust the scope of the local definitions.
The let expression will be refactored to the below.

\begin{vdmsl}
total(tl data) + 
(let 
  multipler = len data mod 2 + 1, 
  product = hd data  * multipler
in 
  (if product < 10 then product else product mod 10 + 1))
\end{vdmsl}

The expression is now separated into the recursion part and the computation on each element of {\tt data}.
Extracting a parameter  {\tt  datum} from {\tt hd data} in the let expression and extract a function definition from the let expression will simplify the definition of the {\tt total} function.
The result of {\it Extract local definition in let expression} and {\it Split let} in order is shown below.

\begin{vdmsl}
total(tl data) + 
(let datum = hd data in
  let 
    multipler = len data mod 2 + 1, 
    product = datum  * multipler
  in 
    (if product < 10 then product else product mod 10 + 1))
\end{vdmsl}

This time, the Refactoring Browser does not enable {\it Extract function from let expression} because the reference to the local parameter {\tt data} is left in the body of the let expression.
The local definition of {\tt multipler} should be moved to the outer let expression by {\it Split let}.

\begin{vdmsl}
total(tl data) + 
(let 
  datum = hd data,
  multipler = len data mod 2 + 1, 
in
  let product = datum  * multipler in 
    (if product < 10 then product else product mod 10 + 1))
\end{vdmsl}

The Refactoring Browser enables the user to apply {\it Extract function from let expression} to the outer let expression.
The following shows the resulting definition of the {\tt total} function and the newly defined {\tt single} function.

\begin{vdmsl}
total : seq of Digit -> nat
total(data) ==
  if data = []
  then 0
  else
    (total(tl data) + single(hd data, len data mod 2 + 1));
   
single : ? * ? -> ?
single(datum, multipler) ==
  let product = datum  * multipler in 
    (if product < 10 then product else product mod 10 + 1)
\end{vdmsl}

The type of the {\tt single} function is not specified well, and the parameter {\tt multipler} should be constrained by adding a precondition.
The definition of {\tt single} will be manually edited into the below.
\begin{vdmsl}
single : Digit * nat -> Digit
single(datum, multipler) ==
  let product = datum  * multipler in 
    (if product < 10 then product else product mod 10 + 1)
pre multipler in set {1, 2}
\end{vdmsl}

The precondition of {\tt single} is important.
Many explanation of this formula writes that each digit in the 1st, 3rd, 5th, ... place from the right should be doubled.
{\tt multipler}  should be either 1, meaning not doubled, or 2 meaning doubled.
The extraction of the function {\tt single} created a place to formally document the constraint while the constraint can be inferred from {\tt len data mod 2 + 1} in the original specification.

The refactored specification works the same as the original one, and asserts two constraints explicit: (1) the resulting value to be added is in the range of {\tt Digit} type, and (2) the {\tt multipler} should be either {\tt 1} or {\tt 2}.
The refactoring involved five semi-automated operations and one manual edit.
The automated operations do not only save the keyboard typing efforts but also prevent the user from breaking the semantics of the original specification.

\subsubsection{Generalising {\tt 10} as a base}

A literal value {\tt 10} appears in multiple places in the specification.
A constant value can be defined by {\it Extract value} refactoring.
\begin{vdmsl}
values
  base = 10;

types
  Digit = nat inv d == d < base;

functions
  luhn : seq1 of Digit -> Digit
  luhn(data) == total(data)  * 9 mod base;
    
  total : seq of Digit -> nat
  total(data) ==
    if data = [] 
    then 0 
    else 
      total(tl data) + single(hd data, len data mod 2 + 1)
  measure slen;
    
  slen : seq of Digit -> nat
  slen(data) == len data;
    
  single : Digit * nat -> Digit
  single(datum, multipler) ==
    let product = datum  * multipler
    in 
      (if product < base 
      then product 
      else product mod base + 1)
  pre multipler in set {1, 2};
end Luhn
\end{vdmsl}

A literal number {\tt 9} appears in the definition of {\tt luhn}.
The original expression {\tt total(data) * 9 mod 10} is a short form of more explanatory expression {\tt (10 - total(data) mod 10) mod 10}, intending a single-digit value that will be summed up with the last digit of {\tt total(data)} to make a natural number whose last digit is {\tt 0}.
The below is the result of manual editing to {\tt (10 - total(data) mod 10) mod 10} and apply the {\it Use value in all occurrences} operation.

\begin{vdmsl}
luhn : seq1 of Digit -> Digit
luhn(data) == (base - total(data) mod base)  mod base;
\end{vdmsl}

\begin{figure}
\begin{vdmsl}
module Luhn
exports all
definitions
values
  base = 10;

types
  Digit = nat inv d == d < base;

functions
  luhn : seq1 of Digit -> Digit
  luhn(data) == (base - total(data) mod base)  mod base;
    
  total : seq of Digit -> nat
  total(data) ==
    if data = [] 
    then 0 
    else 
      total(tl data) + single(hd data, len data mod 2 + 1)
  measure slen;
    
  slen : seq of Digit -> nat
  slen(data) == len data;
    
  single : Digit * nat -> Digit
  single(datum, multipler) ==
    let product = datum  * multipler
    in 
      (if product < base 
      then product 
      else product mod base + 1)
  pre multipler in set {1, 2};

end Luhn
\end{vdmsl}
\caption{The refactored source of the {\tt Luhn} module}
\label{fig:luhn-refactored}
\end{figure}

The resulting specification is shown in Figure \ref{fig:luhn-refactored}.
The specification gained generality so that it can be easily extended to other variations of the Luhn algorithm to validate a sequence of letters other than just digits.

\section{Related Work}
\label{sec:RelatedWork}

Pedersen and Mathiesen pioneered the application of refactoring to VDM-SL~\cite{Petersen&17a}.
They identified refactoring operations applicable to VDM-SL, including rename, add parameter, convert function to operation, extract definition and remove operations, and built a Proof-of-Concept implementation on the Overture tool.
Refactoring is discussed as means to address {\it model smells} and improve the maintainability of the model while its counterpart in the programming languages is to address {\it code smell} in the field of software maintenance.
This paper pays attention to the same technique, refactoring, for a different purpose.
This paper shed light on refactoring as guidance for the specifier to learn the system and domain, which is more performed at the earlier phase of the specification than the phase to improve the rigour and conciseness of the model.

\section{Discussion and Concluding Remarks}
\label{sec:DiscussionAndConcludingRemarks}

Refactoring, in general, is often performed to improve readability and maintainability in both programming languages and specification languages.
Refactoring in programming is widely practised in the maintenance of program sources to keep them readable and extendable.
The objective of this research is to design a set of semi-automated refactoring operations that support the learning process in the exploratory stage of the specification rather than the maintenance phase of the specification.
The question is whether the automated refactoring techniques widely used in programming is also useful in the exploratory stage of the specification.
The authors used the Refactoring Browser to modify several existing models including the Luhn algorithm and found the semi-automated refactoring operations safe and efficient.
The semi-automated refactoring operations make coarse-grained interaction with the user comparing the conventional text editing operations.

Each refactoring has a different scope and makes a different amount of changes to the source text.
The user's expectation of the changes to be made by a refactoring operation may differ from the actual one.
The automated history mechanism implemented in the Refactoring Browser encourages the use of refactoring operations.

While the use of refactoring tools in the exploratory stage look promising, we observed difficulties to apply the semi-automated refactoring operations to VDM-SL.
One difficulty is assertions, especially invariants and preconditions.
Judging the applicability of a refactoring operation that replaces two expressions often involves the satisfiability of assertions.
Even though an expression matches with the definition body of a function, it is not trivial to confirm that the arguments in the function application always satisfy the precondition of the function and the type invariants of the parameter types.

Automated inference of types on extracted identifiers is also problematic.
For example, the type of constant value in the values section can be omitted while the export signatures of values require explicit typing.
The refactoring operations implemented in this research avoid this issue by using the {\it any}-type ({\tt ?}) which is not supported by the language specification.
To generate a formally valid fragment of specification, the user's manual edit is required.

Multiple refactoring operations are often required to make an apparent improvement as seen in the Luhn example.
Support for planning a series of refactoring operations to achieve an expected achievement is required.
History is a promising tool to show the strategic view of the modification to the specification with a series of fine-grained refactoring operations applied so far.
Further study to design a better interaction model with the combination of refactoring operations and history is needed.

\section*{Acknowledgments}
The authors thank Nick Battle for providing the original specification of Luhn algorithm and supplementary documentation.
We would thank the anonymous reviewers for their valuable comments and suggestions.

 \newcommand{\noop}[1]{}

\clearpage
\endgroup

\begingroup
\renewcommand\theHchapter{4-Kulik:\thechapter}
\renewcommand\theHsection{4-Kulik:\thesection}
\locallabels{4-Kulik:}
\setcounter{footnote}{0}
\setcounter{chapter}{0}
\setcounter{lstlisting}{0}
\fontfamily{ptm}\selectfont

\makeatletter
\def\input@path{{4-Kulik/}}
\makeatother

\graphicspath{{4-Kulik/}}
\lstdefinestyle{VDM}
{
  frame=single,
  basicstyle=\small\ttfamily,
  escapechar=!,
  breaklines=true,
  frameround=false,
  linewidth=\columnwidth,
  morekeywords={atomic,is,inv,values,dcl,forall,in,set,nil,and,let,be,st,set1,pure,nat,pre,post,map,to,of,true,false},
  moredelim={[is][keywordstyle]{@}{@}},
}

\lstdefinestyle{TraceOutput}
{
  basicstyle=\small\ttfamily,
  frame=single,
  captionpos=b
}

\newcolumntype{R}[1]{>{\raggedleft\let\newline\\\arraybackslash\hspace{0pt}}m{#1}}

\title{Extending the Formal Security Analysis of the HUBCAP sandbox}
%
%
\author{Tomas Kulik\inst{1} \and
Prasad Talasila\inst{1} \and
Pietro Greco
 \inst{2}\and
Giuseppe Veneziano
 \inst{2}\and
Angelo Marguglio
 \inst{2}\and
Lorenzo Franco Sutton
 \inst{2}\and
Peter Gorm Larsen\inst{1} \and
Hugo Daniel Macedo\inst{1}}
\authorrunning{Kulik et al.}
%
\institute{DIGIT, Department of Engineering, Aarhus University, Denmark \and
ENGIT, Italy}
%
\maketitle              
\begin{abstract}
    Cloud computing thrives in the modern distributed and model-based design area, where many models and tools are domain specific and seldom interoperable. As shown during the HUBCAP project, clouds thrive because they provides several virtual resources on demand, thus enabling the development of customized workspaces of tools that operate on diverse hardware and software execution environments. Yet cloud resources are public and shared, thus restricting access to private tools and achieving proper isolation of the user. Customized workspaces require the implementation of a sandbox, which ensures the protection of intellectual property rights, flexible licensing/revenue models, and clear sharing policies for the platform assets. The security of a cloud platform like HUBCAP relies on the properties of its sandbox implementation, and 
we applied formal methods techniques to establish theoretical approximations of such properties.  In this paper, we  report on the extension of a previous study to analyze user access and sharing policies of sandboxes provided by the HUBCAP Sandboxing Middleware. 
We provide an account of the extensions/refinements to the previous formal specification
and specify the semantics and properties of the ``private tool'' feature recently added to the middleware.

\keywords{Formal analysis  \and Security \and Sandbox \and Model-based design.}
\end{abstract}
\section{Introduction}


\section{Architecture}\label{sec:arch}
In addition to a social collaboration web portal \cite{Larsen&20c},
the HUBCAP Sandboxing Middleware (HSM) provides repositories of Virtual Machines (VMs), also addressed to as servers within this paper (a combination of a Tool and an OS), from which users can draw to compose their sandboxes, fully accessible via a browser web page to users, who experience MBD tools and models in a sandbox.
A sandbox is implemented as an isolated set of VMs (each one running a CPS tool) that interact with each other sharing a virtual dedicated subnet and a dedicated Network File System (NFS) storage service. No interaction is permitted between the VMs belonging to different sandboxes. The sandbox capability integrated with the web-platform is therefore a sort of private cloud service provider
plus the middleware to manage and mediate the access to those cloud services. In addition, as many cloud service providers offer the capability to select a combination of hardware and operating systems, the HSM allows users to select a combination of OS environments, tools, and models to run an experiment using the HUBCAP sandbox feature.

The sandbox service is outlined in Figure~\ref{fig:sandbox}. The web-platform is enhanced with a broker component (labeled as \emph{Sandboxes Broker} in the figure), which hosts a web application mediating the access of different users (\emph{Client 1} and \emph{Client 2}) to the sandboxes they created (\emph{Sandbox 1} and \emph{Sandbox 2} respectively). All the users will use an Internet browser to access the tools in the sandbox and all the interactions are mediated by the broker.

\begin{figure*}[bt]\centering
\includegraphics[width=\textwidth]{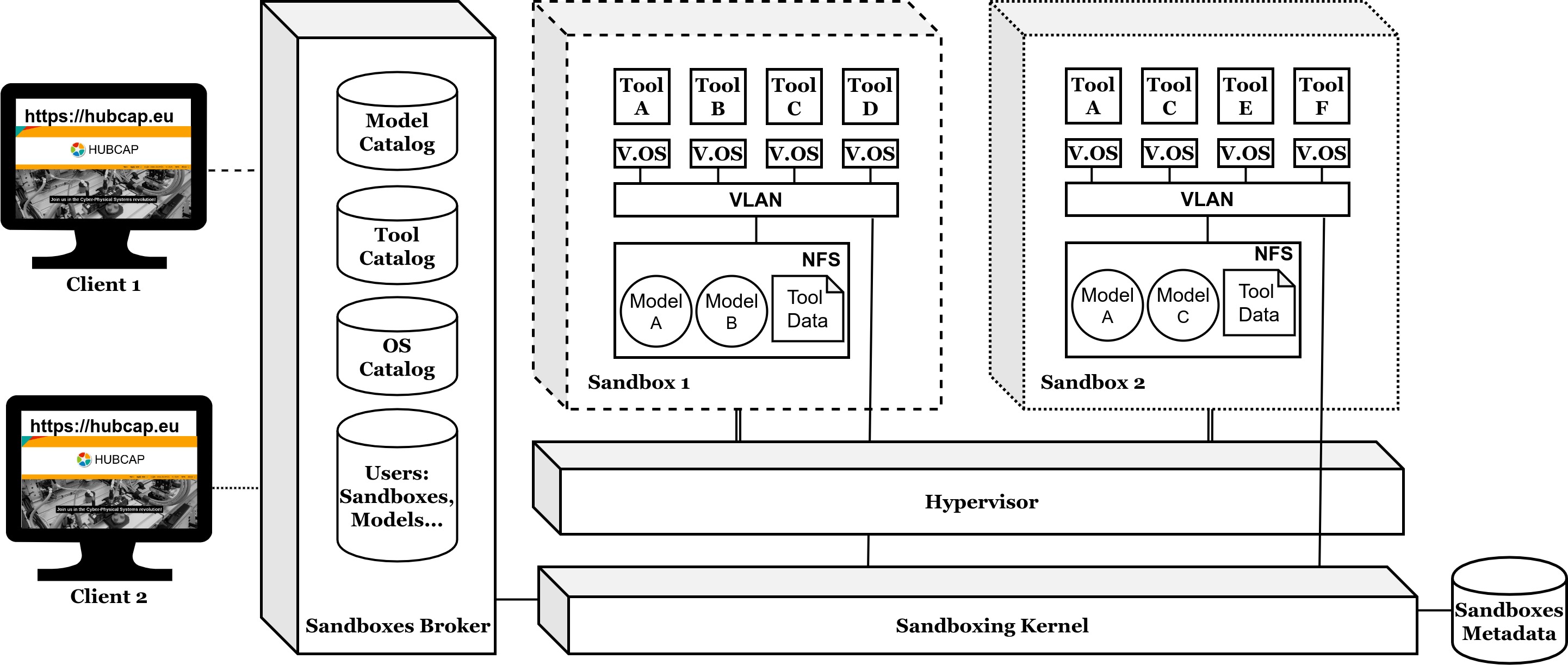}

\caption{\label{fig:sandbox}The HUBCAP Sandboxing Middleware architecture (adapted from  \cite{Larsen&20c})}
\end{figure*}

The \emph{Sandbox Broker} has access to the catalogs of different models, tools, and pre-configured OSs that are available, so an end user can simply pick a valid combination to request a sandbox. In addition to those catalogs, the \emph{Sandbox Broker} keeps all the user information necessary to allow the creation of new sandboxes.

The operation of user requests and the sandboxing logic is provided by the \emph{Sandboxing Kernel}, which is a component that interacts with the system \emph{Hypervisor} to launch the different constituents of a sandbox, namely:
\begin{itemize}
	\item \textit{NFS} - Network File System providing storage in the form of shared folders where model files and tool outputs are placed.
	\item \textit{VLANs} - Virtual networks restricting the communications of the VMs inside a sandbox to the set of VMs composing it and those only.
	\item \textit{VOS} - Virtual machines running the OSs supporting a tool.
	\item \textit{Tools} - In the HUBCAP Sandboxing Middleware, a Tool is a VM obtained by installing CPS software  on a base OS VM offered by the HSM itself.
	\item \textit{Models} - A mathematical/formal description of a component, stored as compressed archives containing a set of folders and files.
\end{itemize}

The operation relies on a database of metadata about the different sandboxes (the \emph{Sandboxes Metadata} component in the figure). This component stores and keeps track of the sandboxes' states (running, suspended, \dots) and user ownership of the resources. It is worth highlighting that the Kernel has direct network connections to the Sandboxes' VLANs. We base our model on the building blocks of the sandboxing platform presented in this section, focusing on the components necessary for analysis of the client access based on roles and profiles, while abstracting away details such as VLAN addresses, specifics of the operating systems, file storage and tool properties.

\section{Access rights and hierarchy}\label{sec:rights}

The HSM has evolved with features such as tool versioning and privacy settings for tools, therefore new versions of existing tools may be uploaded by the users that have initially provided these tools as well as the possibility to set these tools as \textit{private}. The tool providers can always mark their tool(s) as public and hence make them available to all users of the sandbox platform. It is important to note that the HSM could contain more tools than those available within the sandbox, however for the scope of this paper we only consider the tools available for sandboxing.

\begin{table}[h!]
	\begin{center}
		\caption{Overview of HSM user profiles and roles}
		\label{tab:roles}
		\begin{tabular}{c|c c c c} 
			\multirow{2}{*}{\textbf{Feature}} & \textbf{Provider} & \textbf{Provider} & \textbf{Consumer} & \textbf{Consumer}\\
			& \textbf{Owner} & \textbf{Guest} & \textbf{Owner} & \textbf{Guest}\\
			\hline
			Access to remote viewer & X & X & X & X \\
			Upload archive & X &  & X &  \\
			Download archive & X &  & X &  \\
			Invite guests & X & & X & \\
			Destroy sandbox & X & & X & \\
			Select tool & X & & X & \\
			Select model & X & & X & \\
			Select operating system & X & & & \\
			Save tool & X & & & \\
			Upload new model & X & & & \\
			Delete repository item & X (own) & & & \\
			\hline
		\end{tabular}
	\end{center}
\end{table}

Different features of the platform (sandbox and the underlying platform) are governed by a table of access roles and profiles as shown in Table~\ref{tab:roles}. As can be seen in the table, the provider profile represents users that could provide specific tools / models to the HSM. Both the providers and consumers can utilize the operating systems, public tools and models. In addition, providers can also utilize the private tools they own on the HSM. 

In the context of sandbox instantiation and usage, the HSM users have two roles. A user creating a sandbox dons the role of \textit{owner} and any user invited to use that sandbox dons the role of \textit{guest}. The owner, if a provider, selects meaningful subset of available operating systems, tools, models and then creates a sandbox. In addition, owner can also invite other users to become guests in a sandbox. This invitation process is an integral part of sandbox instantiation. Once a tool --- whether public or private --- is instantiated in a sandbox, it's available to all users who have access to that sandbox. Thus both the sandbox owner and guest have access to all the tools available in a sandbox. One user can be the owner of many of sandboxes and can also be a guest in sandboxes owned by the other users, yet one user can only own a single Sandbox at a time.

\section{Formal access model}\label{sec:VDM}
The model of the HUBCAP Sandboxing Middleware (HSM) consists of several subcomponents, specifically (1) Client, (2) Broker, (3) Gateway, (4) Sandbox, (5) Server, (6) Tool and (7) System. Every subcomponent represents a part of the HSM and is modeled as a VDM-SL file with different amount of detail. The model utilizes another VDM-SL file capturing the test traces needed for the analysis, namely the Test VDM-SL file. The model is based upon a flat structure where the a single default module is utilized for tracking of the state of all of the subcomponents. In this study the model is utilized to determine the correctness of the modeled system according to the updated table of roles and profiles. This sections provides a detailed overview of the modeled subcomponents with the goal to specifically pinpoint the updated parts of the system. 

\subsection{Client}
\label{subsec:client}
The client is an entity responsible for making calls to the sandbox. It could be understood as an application that acts upon a behalf of a sandbox user. The HSM is a multi-user system, hence multiple clients can access the different sandboxes and system features. Since the system can handle multiple clients, each client carries a unique identity as \lstinline[style=VDM]{ClientId=nat}. In a current iteration of the HSM, the client can select multiple tools and models from a repository, i.e. a library of preexisting items hosted on the system and launch a new sandbox that distributes these tools across as many newly spawned virtual machines as there are selected tools. In the current HSM analysis model, the models are not associated with a specific tool, what is sufficient as the functionality expressed within the table of roles and profiles does not define interaction between a model and a tool. The selection of tools and models and launching of a new sandbox with these entities is shown in Listing~\ref{lst:selections}. The client can further destroy an existing sandbox, connect and disconnect from a sandbox and upload repository items. 

\begin{vdmsl}[style=VDM, label={lst:selections}, caption={Selection of sandbox entities and launching of a sandbox}] 
SelectToolsFromRepository: ClientId * SelectedTools ==> ()
SelectToolsFromRepository(cId, tIds) ==
  clientst.selectedTools := SelectTools(tIds, cId)
pre cId in set validClients;

SelectModelsFromRepository: ClientId * SelectedModels ==> ()
SelectModelsFromRepository(cId, mIds) ==
  clientst.selectedModels := SelectModels(mIds, cId)
pre cId in set validClients;

LaunchNewSandbox: ClientId ==> ()
LaunchNewSandbox(cId) == 
  StartNewSandbox(cId, clientst.selectedTools, clientst.selectedModels)
pre cId in set validClients;
\end{vdmsl}

In this iteration of the HSM the client can not only upload a tool once to the repository, but can keep the tool up to date by updating the tool with potential new versions. Another option is to upload an archive of files to a specific sandbox and download data generated by a tool running on a specific virtual machine within a sandbox that the client has access to. The uploading of a tool, updating of a tool, uploading of an archive and downloading of an archive is shown in Listing~\ref{lst:repositoryOperations}. It is important to note that the pre-condition \lstinline[style=VDM]{pre cId in set validClients} checks that the client is recognized as valid by the system. After the data download the post-condition \lstinline[style=VDM]{post card clientst.downloadedData = card clientst~.downloadedData + 1} checks that the client has received the downloaded data. While this post-condition would not allow for a repeated download of the same data, it is sufficient within our analysis.

\begin{vdmsl}[style=VDM, label={lst:repositoryOperations}, caption={Upload, update and download of items}] 
UploadTool : ClientId * OSId * Version * Private * OsOnly ==> ()
UploadTool(cId, oId, v, p, oo) == SaveTool(cId, oId, v, p, oo)
pre cId in set validClients and oId in set brokerst.validOSs;

UpdateTool : ClientId * ToolId ==> ()
UpdateTool(cId, tId) == UploadNewToolVersion(cId, tId)
pre cId in set validClients;

UploadArchive : ClientId * SandboxId * token ==> ()
UploadArchive(cId, sbId, arch) == 
  UploadArchiveToSandbox(cId, sbId, arch)
pre cId in set validClients;

DownloadArchive : ClientId * ServerId * SandboxId ==> ()
DownloadArchive(cId, sId, sbId) == 
let d = DownloadArchiveFromServer(cId, sId, sbId) in
  clientst.downloadedData := clientst.downloadedData union {d}
pre cId in set validClients
post card clientst.downloadedData = 
      card clientst~.downloadedData + 1;
\end{vdmsl}

\subsection{Broker}
\label{subsec:broker}

The broker is the part of the HSM that handles the calls from the client and provides replies from the system to the client. The broker is a single entity hence it does not need a specific identity. The broker state is responsible for handling of client/sandbox ownership relation, for example a relation \lstinline[style=VDM]{Guests = map ClientId to set of SandboxId} specifies that the specific clients are guests to specific sandboxes. The broker further specifies other relations such as ownership of tools, assignments of tools to sandboxes and others. One very important function of the broker is to start new sandboxes. This requires starting of new virtual machines under the sandbox. The operation for launching of new sandboxes is shown in Listing~\ref{lst:launch}. This operation also calls \textit{BuildNewServers} operation responsible for spinning out new virtual machines under the system. Another operation responsible for getting the operating systems out of a specific tool used when launching a sandbox is \textit{GetToolOSs}. This is necessary in order to keep track of which operating systems are used under a specific sandbox. These operations are described in more detail in Section~\ref{subsec:system}. The pre-condition on the operation states that	the sandbox can only contain valid tools and valid models, i.e. items that are recognized by the sandboxing system. The post-condition states that new sandbox is added upon completion of the operation, the calling client is added as an owner of the newly spawned sandbox.

\begin{vdmsl}[style=VDM, label={lst:launch}, caption={Launching of a new sandbox}] 
StartNewSandbox : ClientId * set of ToolId * set of ModelId  ==> SandboxId
StartNewSandbox(cId, tId, mId) == 
let sId = GenerateNewSandboxId()
in
  (brokerst.sandboxOSs := brokerst.sandboxOSs munion {sId |-> GetToolOSs(GetToolsByToolIds(tId, brokerst.validTools))};
      brokerst.sandboxModels := brokerst.sandboxModels munion {sId |-> mId};
      (brokerst.sandboxTools := brokerst.sandboxTools munion {sId |-> tId};
      systemSandboxes := systemSandboxes munion {sId |-> mk_Sandbox(sId, BuildNewServers(tId,mk_token(nil)), {})});
    if cId in set dom brokerst.owners then
      brokerst.owners(cId) := brokerst.owners(cId) union {sId}
    else
      brokerst.owners := brokerst.owners munion {cId |-> {sId}};  
    return sId)
pre tId <> {} and tId subset dom brokerst.validTools
    and (mId <> {} => mId subset brokerst.validModels)
post card dom systemSandboxes = card dom systemSandboxes~ + 1
    and cId in set dom brokerst.owners
    and RESULT in set dom systemSandboxes;
\end{vdmsl}

The broker is an integral part of the system often responsible for handling operation calls according to the table of roles and profiles, encoded as pre and post conditions.

\subsection{Gateway}

The gateway is a connectivity component. It provides connections between clients and virtual machines within the sandbox. The gateway holds a state of connections, while it is the broker that assigns these connections. The gateway is defined as shown in Listing~\ref{lst:gateway} with the connections handling as shown in Listing~\ref{lst:gatewayConnections}. All of the sandbox connections and disconnections are recorded within the gateway state. The operation called during connections and disconnections is the \textit{UpdateConnections} that is defined within system and is described in detail in Section~\ref{subsec:system}. Within the Listing~\ref{lst:gatewayConnections} the precondition checks that the client actually has rights as defined within the table of roles and profiles and the client is either an owner of the sandbox it is connecting to, or a guest that was invited towards this sandbox.
\begin{vdmsl}[style=VDM, label={lst:gateway}, caption={Gateway state}] 
GateWaySt ::
	connectedClients : ConnectedClients
	connectedServers : ConnectedServers;
ConnectedClients = set of ClientId;
ConnectedServers = set of ServerId;
\end{vdmsl}

\begin{vdmsl}[style=VDM, label={lst:gatewayConnections}, caption={Sandbox connection initiation}] 
BrokerInitiateSandboxAccess: ClientId * SandboxId ==> bool
BrokerInitiateSandboxAccess(cId, sId) == 
  let sandboxes = GetSystemSandboxes(),
      servers = dunion {dom s.sandboxServers 
                       | s in set rng sandboxes 
                       & s.sandboxId = sId} 
  in 
	  (for all x in set servers do
	     UpdateConnections(cId, x, sId, true);
	   return true)-- used to simulate error handling
pre not ClientIsNull(cId,brokerst.providers,brokerst.consumers, 
                     brokerst.owners, brokerst.guests)
and ((cId in set dom brokerst.owners and sId in set brokerst.owners(cId))
or (cId in set dom brokerst.guests and sId in set brokerst.guests(cId)));
\end{vdmsl}

\subsection{Sandbox}
\label{subsec:sandbox}
The sandbox is considered to be an isolated entity, a container that contains several virtual machines hosting tools and models that user can interact with remotely. Each sandbox isolates its data from other sandboxes and access to it is governed by the table of roles and profiles. The sandbox itself can receive data remotely via a client once it is launched, this data is only present within a selected sandbox and cannot be shared to other sandboxes. An important notion within the sandboxing system is the ownership model, i.e. the user could be an owner of a sandbox and be able to invite other as guests and also destroy the sandbox. The sandbox is defined as shown in Listing~\ref{lst:sandbox}.

\begin{vdmsl}[style=VDM, label={lst:sandbox}, caption={Sandbox definition}] 
SandboxId = nat;
SandboxServers = map ServerId to Server;
UploadedData = set of token;
Sandbox:: 
 sandboxId : SandboxId
 sandboxServers : SandboxServers
 uploadedData : UploadedData
\end{vdmsl}

Each of the sandboxes has a unique identity, defines virtual machines that belong to this sandbox and has a set of data that has been uploaded from a client (starting as an empty set). Since each sandbox must have an unique identity it is necessary to generate new identity each time a new sandbox is launched. This is handled by functions and operations shown in Listing~\ref{lst:sandboxId}. This shows that the identity of sandboxes is incremental.

\begin{vdmsl}[style=VDM, label={lst:sandboxId}, caption={Sandbox identity}] 
Max: SystemSandboxes -> nat
Max(ss) ==
  if ss = {|->} then 0
  else let max in set dom ss be st 
           forall d in set dom ss & d <= max
       in
         max;

GenerateNewSandboxId: () ==> nat
GenerateNewSandboxId() == return Max(systemSandboxes) + 1;
\end{vdmsl}

The current iteration of the model allows for sandboxes consisting of multiple virtual machines, increasing the complexity from all previous attempts to model the system.

\subsection{Server}
\label{subsec:server}

The server is considered to be a virtual machine within the system used to host a specific tool. Each server within the system has a unique identity where the identity is generated similar to the sandbox identities, i.e. they are incremental. Within the system each server needs to be a part of a sandbox in order for a connection to be established via a client. The server definition is shown in Listing~\ref{lst:server}. Each server defines what tool it carries and potentially there could be a piece of abstract data that has been generated by the interaction of a tool and a model. This data can then be obtained by a client and when the server is started is starts empty as \lstinline[style=VDM]{BuildNewServers(tId,mk_token(nil))}. Within the model this is used for analysis of clients ability to download data from a server according to the table of roles and profiles.

\begin{vdmsl}[style=VDM, label={lst:server}, caption={Server definition}] 
ServerId = nat;
Data = token;
Server::
	serverId : ServerId
	toolId : ToolId
	data : Data
\end{vdmsl}

\subsection{Tool}
\label{subsec:tool}

The tool is the newest addition to the modeling effort. The tool is considered to be an application that is used to interact with specific models and is deployed within a specific virtual machine under a sandbox. In some cases the tool does not contain an application but only an operating system. The tool can further be set as private, meaning that only the user that provided it to the repository of tools could select it and deploy it within a sandbox. The tool further defines a version as in this iteration of the system it is possible to update existing tools rather than keep all of the different tool versions within the platform. The tool is defined as shown in Listing~\ref{lst:tool}. The operating system of the tool is specified by the identity of the operating system.

\begin{vdmsl}[style=VDM, label={lst:tool}, caption={Tool definition}] 
Version = nat;
Private = bool;
OsOnly = bool;
Tool::
	osId : OSId
	version : Version
	private : Private
	osOnly : OsOnly
\end{vdmsl}

\subsection{System}
\label{subsec:system}
Finally the system could be understood as an abstraction encompassing the core system functionality. For example the system defines operations for generation of new identities, starting of new virtual machines and assigning connections to the gateway. The operation utilized to create new servers is shown in Listing~\ref{lst:newserv}. Here a new virtual machine is created with its properties when required due to a new sandbox being launched. This operation creates a new server for every tool selected by the client.

\begin{vdmsl}[style=VDM, label={lst:newserv}, caption={New server (virtual machine) creation}] 
BuildNewServers: set of ToolId * Data ==> map ServerId to Server
BuildNewServers(tId, d) ==
(dcl servers : map ServerId to Server := {|->}; 
 for all t in set tId do 
	let newServerId = GenerateNewServerId()
	in(
		systemServers:= systemServers union {newServerId};
		servers(newServerId) := mk_Server(newServerId, t, d);
);
return servers;);
\end{vdmsl}  

The system furthermore reads out the operating system identities out of the specific tools, information that is used to setup relation between sandboxes and operating systems. This function is shown in Listing~\ref{lst:getos}.

\begin{vdmsl}[style=VDM, label={lst:getos}, caption={Get operating system identities}] 
GetToolOSs: set of Tool -> set of OSId
GetToolOSs(tools) == {t.osId | t in set tools};
\end{vdmsl}  

Finally an important operation is an operation responsible for changes in connections, updating the state of the gateway. This operation is shown in Listing~\ref{lst:updatecon}. Both connections and disconnections are handled by this operation.

\begin{vdmsl}[style=VDM, label={lst:updatecon}, caption={Update the gateway connections}] 
UpdateConnections: ClientId * ServerId * SandboxId * bool ==> ()
UpdateConnections(cId, sId, sbId, connect)== 
  if connect 
  then (if cId in set dom gatewayConnections 
       then atomic
            (gatewayConnections(cId):= gatewayConnections(cId) union {sId};
             gatewayConnectionsSandbox(cId) := gatewayConnectionsSandbox(cId) union {sbId};
             brokerst.activeSandboxes := brokerst.activeSandboxes union {sbId})
             else atomic
             (gatewayConnections := gatewayConnections munion {cId|-> {sId}};
             gatewayConnectionsSandbox := gatewayConnectionsSandbox munion {cId |-> {sbId}};
             brokerst.activeSandboxes := brokerst.activeSandboxes union {sbId}))
  else atomic
       (gatewayConnections(cId) := gatewayConnections(cId) \ {sId};
        gatewayConnectionsSandbox(cId) := gatewayConnectionsSandbox(cId) \ {sbId};
        brokerst.activeSandboxes := brokerst.activeSandboxes \ {sbId})
pre (connect=false) => cId in set dom gatewayConnections;
\end{vdmsl}  

The file System.vdmsl also holds the initialization of the model building up initial states of different entities and holds an invariant \lstinline[style=VDM]{dom ss.gatewayConnections = dom ss.gatewayConnectionsSandbox}, specifying that all of the sandbox connections and server connections are established using the same clients.

\section{Formal analysis}\label{sec:analyse}
This section specifies the different traces that have been utilized within the analysis of the system functionality. Several new features have been analyzed such as launching of sandboxes containing multiple servers, updating of an existing tool and download of data from a specific virtual machine. The traces have been analyzed using the combinatorial testing feature of the VDM as it allowed for expansion of tests but utilization of simple parameters. The goal of the analysis is to determine the consistency of the table of roles and profiles as well as compare the results with interaction with the actual sandboxing system.

The scenarios considered within this paper have been selected to analyze the new features of the sandboxing system as well as updates to the model. 





The first trace shown in this paper has been used to analyze the private tool feature. This trace is shown in Listing~\ref{lst:privateTool}. In this trace a client uploads a tool that is set to private. The trace uses the combinatorial testing to an extent where it tries to select the tool from a repository on behalf of two clients. The first client is the tool owner, hence the result of the trace is a success, while the second client is not the owner of the private tool and the result of this trace is inconclusive as it violates the outermost pre-condition on the \lstinline[style=VDM]{SelectToolsFromRepository} operation of the client.

\begin{vdmsl}[style=VDM, label={lst:privateTool}, caption={Trace for checking of the private tool functionality}]
CheckPrivateToolAccess:
SetupClients(1);
SetupClients(2);
SetupProviders(1);
SetupProviders(2);
SetupOSs(1);
UploadTool(1, 1, 1, true, false);
let clientId in set {1,2}
in( 
 SelectToolsFromRepository(clientId, {1}););
\end{vdmsl}

Another interesting action shown within this paper is the ability of the client to upload an archive (i.e. a compressed archive of files) to a sandbox. The trace used to analyze this situation is shown in Listing~\ref{lst:archiveUpload}. First the client selects tools and models, then launches a new sandbox and finally uploads an archive to the sandbox. Within the test the upload of an archive happens without any issue, what is reflected within the implemented platform. 

\begin{vdmsl}[style=VDM, label={lst:archiveUpload}, caption={Trace of upload of archive to a sandbox}] 
UploadArchiveTrace:
SetupClients(1);
SetupProviders(1);
SetupTools(1,1,1,false,false);
SetupModels(1);
let clientId in set {1}
in(
	SelectToolsFromRepository(clientId, {1});
	SelectModelsFromRepository(clientId, {1});
	LaunchNewSandbox(clientId);
	UploadArchive(clientId, 1, mk_token(1)););
\end{vdmsl}

The analysis had utilized several more traces, in order to provide high feature coverage, these have been omitted from this paper due to space considerations. These are however publicly accessible via~\cite{SystemModel}.
	

\section{Related work}\label{sec:related}

The Access Control (AC) defines the limits of users actions and transactions of an information system \cite{hu2014sp,hu2015attribute,radack2006minimum}. The broad guidelines of access control are applicable to single node centralized information systems like the current implementation of HSM. The provider / consumer user profiles and the sandbox owner / guest roles must have well-defined access control policies to the HSM functionality. The formal model and the sandbox implementation consider the standard-compliant access control policies. The latest NIST SP 800-210 standard provides recommendations on access control guidelines for cloud computing services \cite{hu2020general}. The formal model and HUBCAP implementation of the future sandbox environments will incorporate some of the recommendations given in the NIST standard.

The Infrastructure as a Service (IaaS), Platform as a Service (PaaS) and Software as a Service (SaaS) are three popular cloud computing service models \cite{hwang2013distributed}. The HSM provides PaaS to its users. The HSM platform manager provides some functionalities to support providers in protecting the intellectual property (IP) of their tools. For example, the HSM prevents a provider from (re)saving a tool owned by another provider. 
As has been previously observed, software licensing and IP protection are a crucial stumbling block in the migration of desktop software to native cloud computing software \cite{fox2009above}. Consequently, as in other PaaS cloud environments, the HSM users currently face the same licensing issues.

\section{Conclusion}\label{sec:conclude}


In this paper we have investigated the efforts needed to support \gls{vdm} in \gls{vscode} using the standardised protocols \gls{lsp} and \gls{dap} and our \gls{lsp} extension \gls{slsp}.
Also, we evaluated to which extent the \gls{vdm} language features can be covered by the protocols.

The implementation efforts were estimated using a LoC measure.
In total, the \gls{vscode} extension consists of 1880 LoC, where only 210 are related to the support of the standardised protocols.
This is possible because \gls{vscode} provides generic modules for supporting the protocols.
Similar generic support is found in other \glspl{ide}, we therefore believe that the implementation efforts for these \glspl{ide} are comparable to our results.
The protocol support for VDMJ totals 7855 LoC which we find appropriate as the server allows the language support to be easily reused in other \glspl{ide} as they only have to adhere to the protocols to use the server, which requires little effort compared to creating native support for \gls{vdm}.

In the language feature coverage evaluation we find that only a subset of the \gls{vdm} language features are covered by standard protocols.
To support features related to specification languages, such as \gls{pog} and \gls{ct} we propose the protocol extension \gls{slsp}. 

Going forward we believe that it would be beneficial to also support other specification language features related to validation, verification and translation, using a similar architecture to decouple the language features from the development tool using a language-agnostic protocol.
This would make it easier for multiple development tools to support specification languages, reduce the long term efforts needed to maintain language support, and increase the uptake of specification languages by allowing users to use their preferred development tool.

\subsubsection*{Acknowledgements.}
The work presented here is partially supported by the HUBCAP Innovation Action funded by the European Commission's Horizon 2020 Programme under Grant Agreement 872698. We would also like to express our thanks to the anonymous reviewers.

 \newcommand{\noop}[1]{}

\clearpage
\endgroup

\begingroup
\renewcommand\theHchapter{6-Bottjer:\thechapter}
\renewcommand\theHsection{6-Bottjer:\thesection}
\locallabels{6-Bottjer:}
\setcounter{footnote}{0}
\setcounter{chapter}{0}
\setcounter{lstlisting}{0}

\fontfamily{ptm}\selectfont

\makeatletter
\def\input@path{{6-Bottjer/}}
\makeatother

\graphicspath{{6-Bottjer/}}
%
\title{Modelling an Injection Moulding Machine using the Vienna Development Method} 
%
%
\author{
Till Böttjer \and Michael Sandberg  \and
Peter Gorm Larsen \and Hugo Daniel Macedo}
\authorrunning{T. Böttjer et al.}
%
\institute{DIGIT, Aarhus University, Department of Electrical and Computer Engineering, Åbogade 34, 8200 Aarhus N, Denmark\\
\email{\{till.boettjer,ms,pgl,hdm\}@ece.au.dk}\\
}
\maketitle              
\begin{abstract}
The  injection  moulding  machine has changed the plastics manufacturing industry. Its design and control evolved from a mechatronics artefact orchestrating solenoids to a smart cyber-physical system, where the control relies on process modelling and computation. This evolution took place in an industrial context, thus designs, implementations and models of the machine and its controller are usually not available to the academic and public domains.
In this paper, we report on the first model of a controller for an injection moulding machine resulting from the application of the Vienna Development Method.  The work is done in the context of a larger research project on Digital Twins in the  manufacturing domain, and our model, being executable, supplements the academic literature with an open design of the machine process, quality and safety control. Our preliminary findings confirm that the model fits its purpose, it adequately orchestrates a continuous time model of the process developed in MATLAB, and it may benefit both the community developing injection moulding machine models, the community 
interested in formal modelling, and researchers in the area of Digital Twins in manufacturing. We expect this model to be extended in the future and used in the training of future engineers.

\keywords{model-based design  \and injection moulding machine \and Digital Twins \and co-simulation.}
\end{abstract}

\section{Related Work}
\label{sec:related}

In this section, we present a brief review of related work. We follow a top-down approach, where we list the works about Digital Twins followed by models of IMM. Finally, we delve into 
modelling examples in VDM and cover relevant works.

The authors of \cite{Bibow&20} use a Domain Specific Language for specifying events, connecting to an injection moulding machine, and automate the otherwise time-consuming setup of new production jobs. Our work uses standard VDM to specify the behaviour of a controller for an IMM and validates it with a continuous-time model developed in MATLAB. Although our work towards a Digital Twin setup is in progress, we are still far from Hardware In the Loop (HIL) development. 

Autonomous vehicles are also an area where Digital Twins bring added value. The work in \cite{LumerKlabbers&21} presents a framework resourcing to some 
common technologies that we use, as the standard VDM supporting IDEs \cite{Rask&21} are part of the INTO-CPS Application \cite{Macedo&19b} ecosystem. Our work has more emphasis on the specification of the controller and is less oriented towards HIL, as both a desktop model of a commercial IMM or the usage of one is less feasible than the kind of research platform we are comparing ourselves with.

In the following, we provide a small account of scientific works on models of injection moulding and IMMs. The works are listed chronologically.
The only executable model we found is a  MATLAB\footnote{\url{https://se.mathworks.com/help/physmod/hydro/ug/injection-molding-actuation-system.html}}
 example based on a machine design from 1989. Contrary to ours, the model relies on purely hydraulics actuation designs. New machines are now full Cyber-Physical Systems with several computation units inside and the actuation is now done using electric pumps and drives. 

The model in \cite{Woll&97} is used as a key reference in our work. In their work, the authors lay a precise mathematical formulation of the behaviour of an IMM, yet the authors do not model the controller logic, nor repeated iterations. In our work, we have developed a MATLAB implementation and use it to derive test cases/scenarios to our VDM model.

A step towards a co-simulation model of parts of an injection moulding machine is taken in \cite{Hostert&05}. The dynamic model of the clamping unit combines a multi-body simulation of the mechanical structure (ADAMS/ANSYS), a hydraulic model (DSHplus), and a control model (MATLAB/Simulink). The model allows testing of interactions between sub-components and analysis with varying design parameters, but overlooks the injection unit and interactions of machine and process parameters. Our work focuses on the injection unit and the influence of machine parameters on the process flow. 

The authors of \cite{Steinegger&13} present standardised (according to IEC 61499) function blocks for typical faults in manufacturing systems that decouple fault handling and control code and reduce the complexity of the control architecture. The methodology is evaluated on an injection moulding machine.


The usage of VDM modelling to develop incremental models of CPS systems span inceptional cases such as \cite{Macedo&08} or \cite{MacedoNL2019}, to extensive guidelines in the form of complete books \cite{Fitzgerald&13a}.

VDM has been enabled with recent tooling support in Visual Studio Code \cite{Rask&20,Rask&21} and together with the INTO-CPS application and the Functional Mock-up Interface-based co-simulation, it may become a valuable tool to build Digital Twins \cite{Fitzgerald&19}.

\section{Injection Moulding Process}
\label{sec:immp}

The typical aspect of a horizontal IMM used in a manufacturing setting is shown in Fig.~\ref{fig:my_label}. The machine consists of a workbench enclosed by panels and a gate used by operators to access the clamping unit, which alongside the injection unit sits at the top of the workbench. The two units are the main components involved in the production of plastic parts that are ejected from the mould inside the clamping unit to a side bucket below the gate. Modern machines may use a conveyor belt situated under the clamping unit to collect and route the piece to a quality control check and automatically scrap non-compliant pieces. 
\begin{figure}
    \centering
    \includegraphics[width=\textwidth]{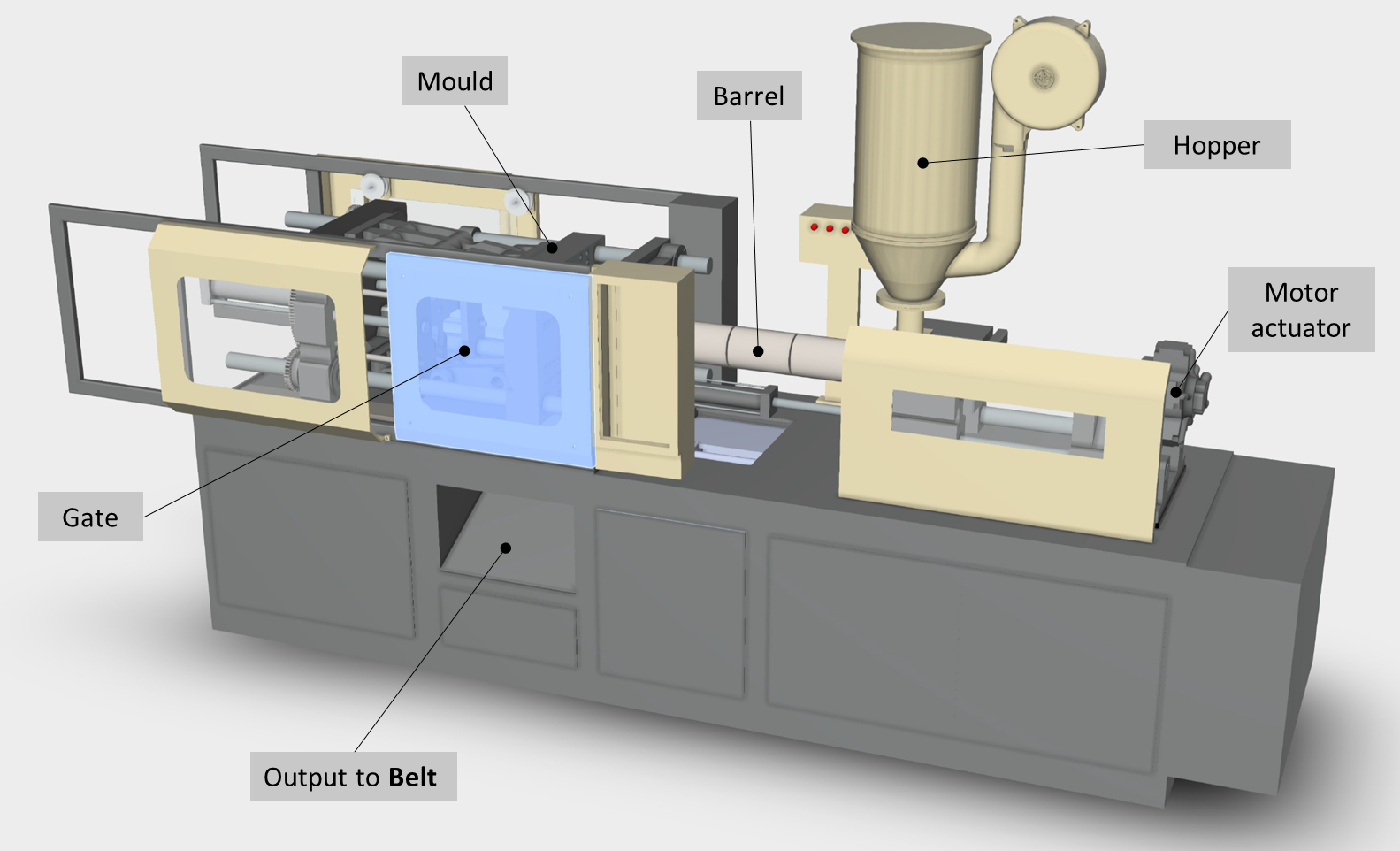}
    \caption{3D representation of an injection moulding machine with the gate highlighted in blue. Below the gate, the part is ejected from the machine and passed onto the belt.}
    \label{fig:my_label}
\end{figure} 
The machine controller and other components are packaged under the workbench with sensors and actuators connected to different parts of the machine. For example, given that the gate near the clamping unit must be locked during operation to avoid damage and safety hazards to the operators, it is common to have a sensor that gauges the state of the gate (open/closed) and feed it to the controller among other interlocking mechanisms. This is one of the first properties (a safety property) that any IMM controller must satisfy:

\begin{prop}
    The machine must be idle when the gate is open.
\end{prop}

\begin{figure}
    \centering
    \includegraphics[width=0.9\textwidth]{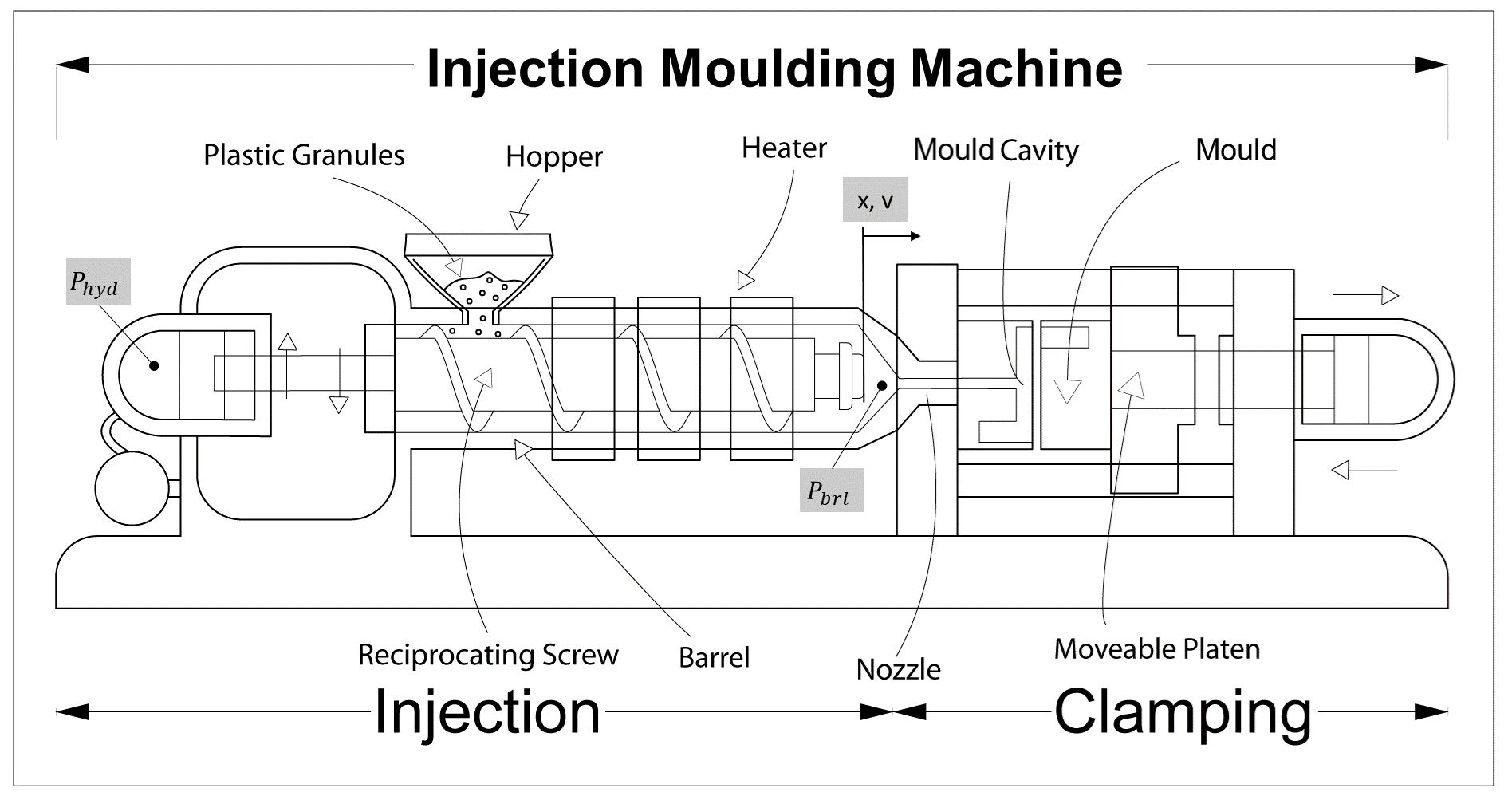}
    \caption{Schematics of the components of an IMM} 
    \label{fig:IMM}
\end{figure}

In Fig.~\ref{fig:IMM} we consider the main components of a single screw, hybrid IMM composed of a clamping unit and an injection unit. The units are further split into actuators.
The clamping unit consists of an actuator and a control system to close the mould that provides the clamping pressure and ejects the part. 
The injection unit consists of various components - the barrel, the screw, the hopper, heat bands, the nozzle - and an electrical and hydraulic actuator, for plastification and forcing the polymer into the mould cavities respectively. 

\subsection{The Injection Moulding Process}

The IM process is cyclic and consists of four states which are associated with the activation of the different machine tool components and amenable to an automaton representation as shown in Figure \ref{fig:automaton}. At the beginning of the process, the screw tip is assumed to be at a position of $x=0$. At this point, the barrel is filled with molten polymer material, and the mould is closed in the clamping unit. At the start of the process, the machine is in the Filling state and is controlled based on the screw velocity. Hydraulic pressure is applied to the screw to achieve the set screw velocity. After the screw reaches a certain position, $x_3$, or the expected time, ($tfill$), the filling phase is complete. Subsequently, the machine transitions to the Packing phase and switches from velocity to pressure control. For the Packing, Cooling/Classification, and Ejection phases, the state transition is time controlled. After the Ejection phase, the machine becomes Idle and a new cycle may start. In the following paragraphs, we describe the different phases of the IM process and some of the properties that one can check or assure in the development of a VDM model.

\begin{figure}
    \centering
    \includegraphics[width=0.95\textwidth]{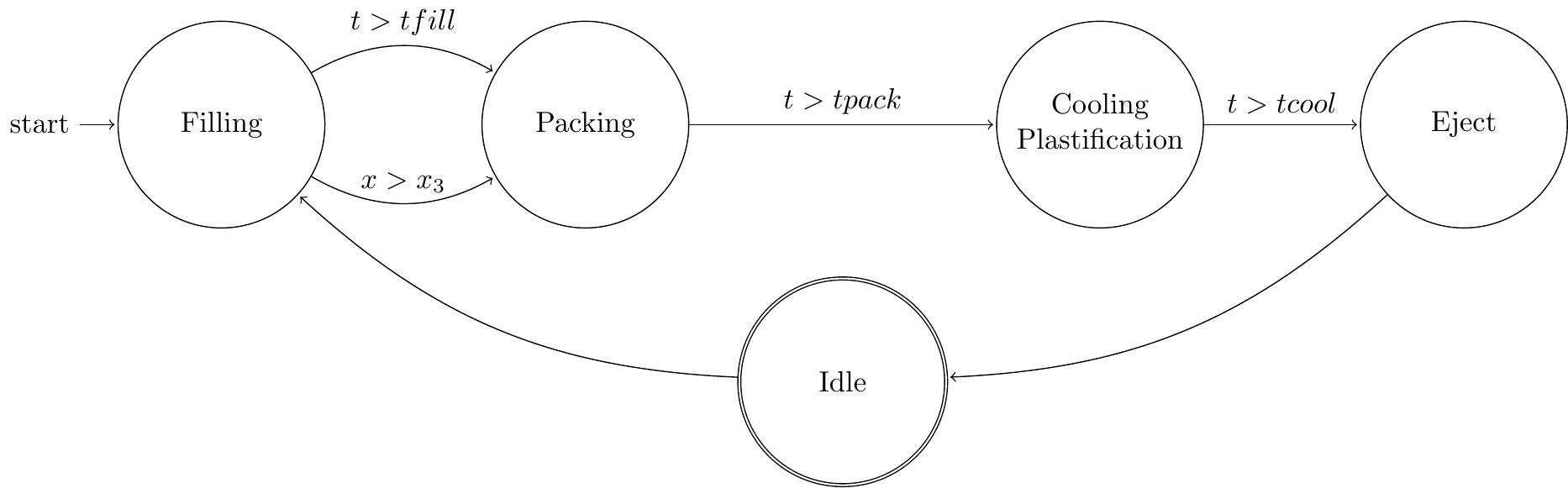}
    \caption{States and phase transitions of an IM process depicted as an automaton}
    \label{fig:automaton}
\end{figure}

\paragraph{Injection/Filling Phase:} The material is filled into the mould cavity (form) under high pressure and speed. During injection, the material starts to solidify. In this phase, the screw is position/velocity controlled, and deviations from the final position reached by the screw are used as an input to the quality control.  
For example, an undershoot of the screw position means a lack of material in the mould cavity, thus a defect in the part being produced.  This is the second property (the first on quality control) we illustrate about our IMM controller model:

\begin{prop}
    The screw must reach EndPositionFill. For the scenario in this paper, EndPositionFill equals 20.3 mm.
\end{prop}

\paragraph{Packing Phase:} Based on a pre-defined position, the machine control switches to the packing phase. The material inside the cavity is still flowable along the cross section and it is possible to influence the final part weight e.g. to compensate for shrinkage. During packing, the remaining material is further compressed into the cavity at constant pressure. At the end of the packing phase, the gate freezes off such that no more material can flow into or out of the mould cavity.
Excessive and insufficient packing pressure leads to various defects of the moulded part such as mould sticking and empty holes, respectively.  
Therefore, we monitor the values achieved and check if they are inside a target interval. This is the third property (the second on quality control) we illustrate about our IMM controller model:

\begin{prop}
    The hydraulic pressure peak must occur at PeakTimeIntervall and be PeakPressureBand. For the scenario this paper, these are set to $0.9 < t < 1.1$ s and $9 \pm 1$ MPa.
\end{prop}

\paragraph{Cooling Phase \& Plastification Phase:} 
After the packing and holding phases, cooling is applied through the coolant channels of the mould to lower the temperature and completely harden the material. In parallel, a new charge of material for the next shot is heated, mixed (homogenised) and transported by screw backward rotation. The transition to the ejection phase is specified by a defined cooling time.

\paragraph{Ejection phase:} The mould opens, the solidified part is ejected, and the mould is closed. The final part is transferred via a belt to a storage box. A quality controller automatically rejects parts that do not meet defined standards based on monitoring relevant process variables. 

This concludes our brief presentation of an IMM and the IM process. For more details about this application domain, we refer the readers to the work of \cite{KULKARNI2017,Woll&97}. In terms of the process, we will provide more details on the continuous behaviour of the machine in Section \ref{sec:validation}, where we introduce the continuous time model of screw motion and injection process.

\section{Model Validation}
\label{sec:validation}

In this section, we describe the steps that were followed to perform a first check of the suitability of the model. We cross-check the output behaviour of the model execution while running it against scenarios simulating the events/inputs to the system sensors at each time instant. To generate the scenarios, we have implemented a continuous model of the expected behaviour of an IMM in MATLAB and exported the events to a CSV file that is in turn read by the VDM model.

\subsection{Continuous Time Model}

We follow the work by \cite{Woll&97}, where the authors develop a first-principles model of an IMM linking the controllable machine parameters (e.g., the motion of the screw) to measurable process variables (e.g. injection pressure). Accordingly, we only introduce the equations containing the parameters used to generate the scenarios for validating our VDM model. For a comprehensive overview of the first-principles model, we refer the reader to \cite{Woll&97}. 

The model describes the injection process based on a set of selected equations for the Filling, Packing and Cooling phases. The set of equations consists of Newtons Second Law and thermodynamic relationships such as the equations of continuity, momentum and energy. In each phase the equations are solved for hydraulic pressure, nozzle pressure, ram velocity, polymer flow, and cavity pressure. 

During the Filling phase, Newton's Second Law is used for describing the velocity of the screw using first order dynamic equations. As shown in Fig. \ref{fig:ramvel}, the screw velocity is changed at positions $x_1, x_2, x_3$ to the associated velocity setpoints $v_1, v_2, v_3$. 
The screw velocity $v$ is described by
\begin{equation}\label{6-Bottjer:eq:1order_fill}
    \tau_1 \diff{v}{t} + v = K_1 v_f
\end{equation}
where $v_f$ describes the velocity trajectory, and $\tau_1$, $K_1$ are constants.
Using Newton's Second Law the hydraulic pressure is found (compare Fig. \ref{fig:hp})
\begin{equation}\label{6-Bottjer:eq:Newton}
    \diff{v}{t} = \frac{\sum F}{M} = \frac{P_{hyd}A_{hyd} - P_{brl}A_{brl} - C_f sign(v) - B_f v}{M}
\end{equation}
where hydraulic pressure is applied to the screw, and barrel pressure, $P_{brl}A_{brl}$, Coulomb friction, $C_f sign(v)$, and viscous damping, $B_f v$, resist the motion of the screw. $M$ is the combined mass of the actuators.

During the Packing phase, the process switches to pressure-based control to reach holding pressure, $P_{hold}$. Eq. \ref{eq:1order_fill} is replaced by
\begin{equation}
    \tau_2 \diff{P_{hyd}}{t} + P_{hyd} = K_2 P_{hold}
\end{equation}
describing the transition of hydraulic pressure to the holding pressure, $P_{hold}$. Eq. \ref{eq:Newton} is solved for describing the deceleration of the screw. Fig. \ref{fig:ct_out} shows the pressure response and screw velocity as outputted from the continuous model. 

\begin{figure}
\centering
\subfloat[Combined nozzle pressure response (MPa) for Packing, Filling, and Cooling Phase\label{fig:hp}.]{\includegraphics[width=5.75cm]{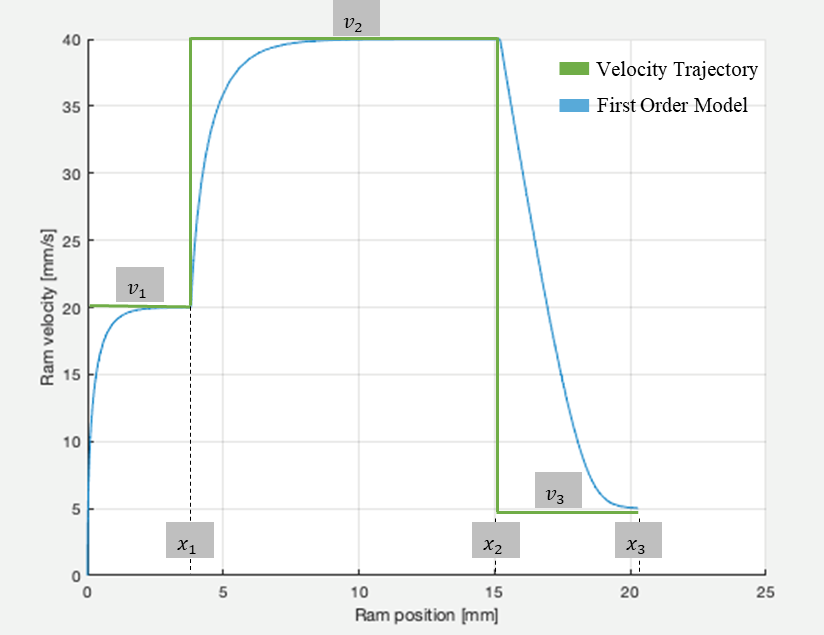}}
\hfill
\subfloat[Ram velocity (mm/s) with the three velocity set points $v_1, v_2, v_3$ for the transition positions $x_1, x_2, x_3$\label{fig:ramvel}.]{\includegraphics[width=5.75cm]{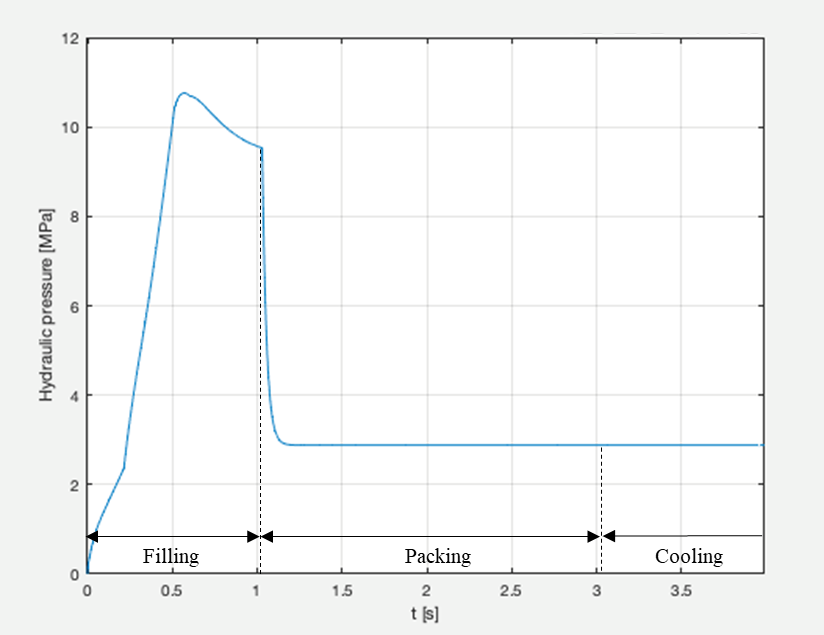}}
\hfill
\caption{Output generated by the continuous time model. These plots were obtained from our MATLAB implementation of \cite{Woll&97}'s model}
\label{fig:ct_out}
\end{figure}

\subsection{Extracting Scenarios from the Continuous Time Model}
We validated the model using three scenarios generated using the physical model of an injection moulding machine developed in MATLAB.
Fig. \ref{fig:profiles} shows the hydraulic pressure during actuation of the injection ram for three different scenarios. Scenario A) uses the original parameters introduced in \cite{Woll&97}, while Scenario B) and Scenario C) simulate an injection process where the position setpoint, $x_2$, is not reached or overshoot, respectively. 

Slight variations in the process conditions during the injection cycle can lead to bad quality parts. Variation in hydraulic pressure can cause quality issues with the polymer part. Hydraulic pressure above the pre-set value lead excessive polymer flow into the mould. This causes a lid surrounding the injection part at the parting line of the mould halves. Lower pressure than the pre-set value leads to underfilling of the mould cavities and causes incomplete parts. We varied the hydraulic pressure, a controllable variable actuating the injection screw during the Filling phase, for validating the main functionality of our VDM model that are:

\begin{enumerate}
    \item switching of the injection phases at defined conditions
    \item setting of the velocity set points during filling 
    \item returning of the screw in the cooling phase, i.e. plasticising new polymer for the next injection cycle
    \item the auto scrap control responsible for detecting quality issues from process data and automatically reject corresponding elements
\end{enumerate}

\begin{figure}
    \centering
    \includegraphics[width=\textwidth]{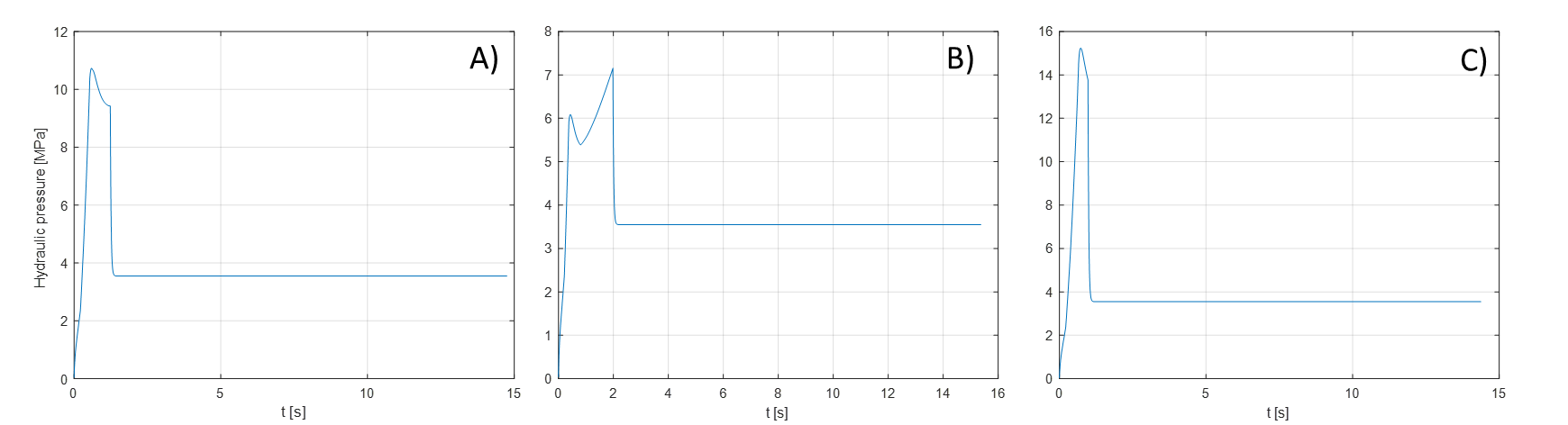}
    \caption{Three inputs to the VDM model that were generated using a deterministic injection moulding machine model. The hydraulic pressure is shown for A) optimal operating conditions (x2 = 15.2 mm, the same behaviour as in Fig. \ref{fig:ct_out}), B) the second position set point is not reached (x2 = 9 mm), and C) second position is overshoot (x2 = 20 mm).}
    \label{fig:profiles}
\end{figure}

\subsubsection*{Acknowledgements.}
The work presented here is partially supported by the the Poul Due Jensen Foundation for funding the project Digital Twins for Cyber-Physical Systems (DiT4CPS). This work is also supported by the Manufacturing Academy
of Denmark (MADE) in the MADE FAST project (see \url{http://www.made.dk/}). We would also like to thank Daniella Tola for feedback on the first draft of this paper.

\bibliographystyle{splncs04}

 \newcommand{\noop}[1]{}

\clearpage
\endgroup

\begingroup
\renewcommand\theHchapter{1-Frasheri:\thechapter}
\renewcommand\theHsection{1-Frasheri:\thesection}
\locallabels{1-Frasheri:}
\setcounter{footnote}{0}
\setcounter{chapter}{0}
\setcounter{lstlisting}{0}

\fontfamily{lmr}\selectfont

\makeatletter
\def\input@path{{1-Frasheri/}}
\makeatother

\graphicspath{{1-Frasheri/}}
\title{RMQFMU: Bridging the Real World with Co-simulation\\
  \large For Practitioners\thanks{We are grateful to the Poul Due Jensen Foundation, in supporting the establishment of a new Centre
for Digital Twin Technology at Aarhus University,  and Innovation Foundation Denmark for funding the AgroRobottiFleet and UPSIM projects.}}

\author{Mirgita, Henrik, Casper, Lukas}
\titlerunning{RMQFMU: For Practitioners}

\author{Mirgita Frasheri \inst{1}\orcidID{0000-0001-7852-4582} \and
Henrik Ejersbo\inst{1}\orcidID{0000-0003-4753-3800} \and
Casper Thule\inst{1}\orcidID{0000-0001-6606-9236}\and Lukas Esterle\inst{1}\orcidID{0000-0002-0248-1552}}
\authorrunning{M. Frasheri et al.}
%
\institute{Centre for Digital Twins, DIGIT, Aarhus University \\ 
Finlandsgade 22, 
8200 
Aarhus N, 
Denmark \\
\email{\{mirgita.frasheri,hejersbo,casper.thule,lukas.esterle\}@ece.au.dk}\\
\url{https://digit.au.dk/centre-for-digital-twins/} }

\maketitle
\begin{abstract}
    In this paper we present an experience report for the RMQ\-FMU, a plug and play tool, that enables feeding data to/from an FMI2-based co-simulation environment based on the AMQP protocol. 
    Bridging the co-simulation to an external environment allows on one side to feed historical data to the co-simulation, serving different purposes, such as visualisation and/or data analysis. 
    On the other side, such a tool facilitates the realisation of the digital twin concept by coupling co-simulation and hardware/robots close to real-time.
    In the paper we present limitations of the initial version of the RMQFMU with respect to the capability of bridging co-simulation with the real world. To provide the desired functionality of the tool, we present in a step-by-step fashion how these limitations, and subsequent limitations, are alleviated. We perform various experiments in order to give reason to the modifications carried out.
    Finally, we report on two case-studies where we have adopted the RMQFMU, and provide guidelines meant to aid practitioners in its use.

\end{abstract}

\section{Introduction}\label{sec:intropap}

Cyber-physical systems (CPSs) refer to systems that combine computational and physical processes, and play an important role in the development of intelligent systems. 
Using real world operation data could possibly facilitate smart decision-making~\cite{banerjee2012ensuring}.
Harnessing this potential is by no means trivial, with challenges including safety~\cite{baheti2011cyber}, reliability, and security~\cite{ecsel2016multi}, among others, that need to be tackled.
Additionally, CPS development is a multi-disciplinary process, where different components will be modelled and validated by different tools used in the different involved disciplines. 
The evaluation of such complex systems can be performed through co-simulation~\cite{Gomes&18}. 
This is made possible by developing the components according to some standard, e.g. the Functional Mock-up Interface (FMI) adopted in our work, with components referred to as functional mock-up units (FMUs).
However, this is not sufficient, as CPSs together with their environment(s) are subject to continuous change, and evolve through time, possibly diverging significantly from the initial co-simulated results~\cite{fitzgerald2019multi}.
Digital twins (DTs), defined as digital replicas of the physical components, also known as physical twins (PTs), can be used to follow the behaviour of the PTs and CPSs during operation, potentially through learning~\cite{Fitzgerald&14f}, by adapting the models, as well as performing different tasks such as monitoring or predictions.
The DT and the aforementioned (or more) operations can be developed via the co-simulation of a modelled system that corresponds to the deployed CPS. 
In order to connect the deployed CPS to the DT, the implementation of data brokering between them becomes a necessity.
Such data brokering can be useful in different scenarios; in this paper we identify four such:
\begin{enumerate}
    \item For data analysis and visualisation of system behaviour, where the user is interested in feeding back to the co-simulation data recorded from the operation of the hardware, i.e. historical data, to perform some analysis on the data, potentially supporting the visualisation of said data.

    \item For co-simulation, where the user is interested in coupling systems that are simulated in different environments, {in those cases for which it is easier to use data-brokering than to implement the FMI interface for a non FMI-compliant component, e.g. connecting to a Gazebo simulation}~\footnote{In other cases, the generation of an FMU might be preferable. Tools like UNIFMU~\cite{legaard2020rapid} for quick FMU prototyping, or PVSio-web based tool-kits~\cite{palmieri2018flexible}, are designed to help users prototype FMUs for their co-simulations.}.

    \item For the realisation of the digital shadow (DS), where the digital models of the physical twin are contained within the co-simulation environment.
The user is interested in getting the live data from the PT into the co-simulation environment in order to estimate the difference between actual and simulated data along with visualising the operation of the physical twin.
This is similar to the first usage, however the digital shadow would operate with live data.

    \item For the realisation of the digital twin, where the user is interested in enabling the communication link from the DS to the PT, thus fully realising the implementation of the DT concept.
Indeed, components within the co-simulation environment, based on the live data, could send live feedback to the PT.
This can be useful for monitoring the behaviour of the PT, or during CPS development, to lower the cost of testing with Hardware in the Loop.

\end{enumerate}

In this publication we extend an existing Data-Broker called RabbitMQ FMU (RMQFMU~\footnote{\url{https://github.com/INTO-CPS-Association/fmu-rabbitmq.git}})~\cite{thule2020formally}, initially suitable for the first scenario, to enable data brokering both to and from an FMI2-enabled co-simulation, applicable for all the aforementioned scenarios. 
We present an experience report on applying the RMQFMU to two actual cases undergoing development to become digital twin systems: a tabletop robot arm and an autonomous agriculture robot. 
As part of this experience report we present guidelines on how to configure the RMQFMU parameters by presenting various experiments that show their effects and relation to one another.
Through the application of RMQFMU to the aforementioned cases and experiments several needs were discovered such as:
\begin{enumerate}
    \item Data Platform - Get all available data instead of minimally-needed 
    data to enable decision making, possibly allowing to jump ahead to 
    future data. 
    \item Performance - RMQFMU shall be as fast as possible.
    \item Data Delay - RMQFMU shall output the newest data if available.
\end{enumerate}

To address need (1) and (2), the existing RMQFMU, henceforth referred to as 
RMQFMU\textsubscript{0}, was realised with multi-threading instead of 
single-threading. This new multi-threaded version is referred to as 
RMQFMU\textsubscript{1}. Experimentation related to detailing the 
configuration of RMQFMU\textsubscript{1} and its related effects led to the 
discovery of need (3).
In order to mitigate (3), yet another version of RMQFMU was realised 
referred to as RMQFMU\textsubscript{2}, which represents the latest version 
of the RMQFMU. 

The rest of this paper is organised as follows.
The next section provides an overview of the FMI standard, the first version of the RMQFMU (RMQFMU\textsubscript{0}), as well as some of the technologies employed in the realisation of the case-studies.
Afterwards, we informally derive the requirements for RMQFMU, based on the usage scenarios.
Section~\ref{sec:rabbitmq_fmu} presents RMQFMU\textsubscript{1,2}, how it mitigates some inadequacies of RMQFMU\textsubscript{0}, and how it compares to RMQFMU\textsubscript{0}. 
Thereafter, in Section~\ref{sec:experiments} the individual case-studies are described, and the results gained with RMQFMU\textsubscript{1,2} are presented.
Moreover, a detailed comparison between RMQFMU\textsubscript{1} and RMQFMU\textsubscript{2} is made. 
Section~\ref{sec:guidelines} provides a set of guidelines meant to aid the use of RMQFMU\textsubscript{2}, whereas Section~\ref{sec:conclusion} concludes the paper.

\section{Background}\label{sec:background}

In this Section we provide a brief summary on the relevant concepts for this work, such as co-simulation, FMI standard, master algorithms, and the tools we use for their realisation. 
Thereafter, we present the first released version of the RMQFMU (RMQFMU\textsubscript{0}), which we have extended as presented in this paper. 
\subsection{Concepts and Tools}
The realisation of CPSs and constituent systems is a cross-disciplinary process, where the different components are modelled using different formalisms and modelling tools~\cite{gomes2018co}. 
In order to evaluate the behaviour of such systems as a whole, co-simulation techniques are used, which require the integration of the separate models into what are called multi-models~\cite{7496424}.
For the latter to be possible, the tools used to produce the individual models need to adhere to some standard. 
One such standard is the Functional Mock-up Interface 2.0 for Co-simulation (FMI)~\cite{FMIStandard2.0.1} that defines the C-interfaces to be exposed by each model, as well as interaction constraints, packaging, and a static description format. 
An individual component that implements the FMI standard is called Functional Mock-up Unit (FMU).
A co-simulation is executed by an orchestration engine that employs a given master algorithm. The master algorithm defines the progression of a co-simulation in terms of getting outputs via \texttt{getXXX} function calls, setting inputs via \texttt{setXXX} function calls, and stepping the individual FMUs in time via the \texttt{doStep} function calls.
Common master algorithms are the Gauss-Seidel and Jacobi (the curious reader is referred to~\cite{gomes2018co} for an overview of these algorithms and a survey on co-simulation).
In this work, we use the open-source INTO-CPS tool-chain~\cite{Fitzgerald&15}, for the design and execution of co-simulation multi-models, with Maestro~\cite{Thule&19} as orchestration engine employing an FMI-based Jacobi master algorithm. 
\subsection{Overview of the RMQFMU\textsubscript{0}}
RMQFMU\textsubscript{0} was implemented to enable getting external/historical data into the co-simulation environment, for either replaying such data in simulation, or for performing different kinds of analysis, e.g. checking its difference from expected data (the reader is referred to the Water-Tank Case-Study for an example\footnote{Available at \url{https://github.com/INTO-CPS-Association/example-single_watertank_rabbitmq}, visited May 6, 2021}). 
Essentially, RMQFMU\textsubscript{0} subscribes to messages with a specified routing key and outputs messages as regular FMU outputs at specific points in time based on the timestamp of the messages.
At every call of the \texttt{doStep} function, the RMQFMU\textsubscript{0} attempts to consume a message from the server.
A retrieved message is placed in an incoming queue, from which it is thereafter processed, according to the quality constraints (\textit{maxage} and \textit{lookahead}), described shortly.
In case there is no data available, the RMQFMU\textsubscript{0} will wait for a configurable timeout before exiting.
Note that, the \texttt{doStep} is tightly bound with the consuming operation, i.e. they are both contained within the main thread of execution. 
If on the other hand there is data, its validity with respect to time is checked.
The \textit{maxage}, main parameter of the RMQFMU\textsubscript{0}, allows to configure the age of data within which it is considered valid at a given time-step.
Additionally, the \textit{lookahead} parameter specifies how many messages will be processed at a step of RMQFMU and thereby at the given point in time\footnote{In this version, the functionality based on the \textit{lookahead} is not fully implemented.}. 
The behaviour of the RMQFMU\textsubscript{0} has been formally verified in previous work~\cite{thule2020formally}.

\section{The RMQFMU}\label{sec:rabbitmq_fmu}
The design of the RMQFMU is based on a set of informally derived requirements, covering those functionalities needed for the adoption of the RMQFMU in the scenarios described in
Section~\ref{sec:intropap}.
These requirements are given as follows:

\begin{enumerate}
    \item The RMQFMU is able to get the data published by an external system to the RabbitMQ server. 
    This is relevant for all scenarios.
    \item The RMQFMU is able to publish data to the RabbitMQ server, thus closing the communication link. 
    This is relevant for the second and fourth scenario, i.e. for co-simulation and DT realisation.
    

    \item The FMU steps as fast as possible. Indeed the role of the RMQFMU is that of a data broker, as such it is not part of the system being simulated, rather a facilitating entity. 
    Therefore, the delays it causes should be minimal and impact the overall co-simulation environment as little as possible.
    \item Provide quality constraints, e.g. with respect to age of data (\textit{maxage}).
    \item Provide performance constraints, e.g. number of messages processed per step (\textit{lookahead}).
    
\end{enumerate}
 RMQFMU\textsubscript{0} already fulfils requirements 1 and 4.
 Requirement 5 is partially implemented in this version as well, in that while the \textit{lookahead} determines how many messages to retrieve from the incoming queue at every time-step, there will not be more than one message at a time in this queue, given that a consume call is performed at the \texttt{doStep} call until a valid message is retrieved. 
In order to tackle requirement 2, we enable the configuration of the inputs of the RMQFMU\textsubscript{1,2} as needed, i.e., a user can define as many inputs as desired, of one the following types: integer, double, boolean, and string. 
These inputs will be sent to the RabbitMQ server, on change; in other words, the values of the inputs will be forwarded if they have changed as compared to the previous step taken by the RMQFMU\textsubscript{1,2}. 
This check is performed within the \texttt{doStep}. 

In order to tackle requirement 3, we investigate the potential benefit of a threaded configuration of the RMQFMU\textsubscript{1,2}. 
Indeed, the FMU allows for a build-time option to enable a multi-threaded implementation, with a separate thread to interface and consume data from the RabbitMQ server, parsing incoming data and placing it in the FMU incoming queue. 
The main \texttt{doStep} function of the FMU still executes in the main thread context of the calling simulation orchestration application, and it reads and processes data from the FMU incoming queue and produces outputs. If the multi-threaded option is disabled (default), the \texttt{doStep} function will consume data directly from the RabbitMQ server in the context of the calling main thread. 

In general the potential performance benefits of the threaded implementation will depend on the amount of data being consumed, parsed, and provided in the internal incoming FMU queue when the \texttt{doStep} function is called. 
Hence, the benefits will depend largely on the co-simulation environment. 
If the co-simulation is fast, in vague terms if the delay between each call of the FMU \texttt{doStep} function is short compared to the rate of the incoming data, then the FMU will be mostly blocking for I/O. 
In this situation, the separate consumer thread may not provide much benefit as the \texttt{doStep} function will still need to wait/block for incoming data.
On the other hand, if the co-simulation is slower compared to the rate of data, then the separate thread may be able to consume, parse, and provide data to the internal incoming FMU queue, in parallel to the main co-simulation thread running its orchestration engine and executing other potential FMUs or monitors. 
In addition, any performance benefits of a threaded implementation also depends largely on the execution platform having a multi-core processor and the OS/thread environment to be able to take advantage.

The threaded implementation allows for the full realisation of the \textit{lookahead} functionality, thus tackling requirement 5. 
The consumer thread is continuously retrieving data -- when available -- and placing these messages in the incoming queue, from where said data is processed in chunks of \textit{lookahead} size. 

\section{Experiments}~\label{sec:experiments}
The RMQFMU\textsubscript{1,2} has been evaluated with a series of experiments, across different combinations of parameters such as the \textit{maxage} and \textit{lookahead}, and in two different case-studies. 
The purpose of these setups is to provide an understanding on the performance of the RMQFMU\textsubscript{1,2}, its parameter tuning, as well as provide some insight into the effect of external factors.
The first case-study presents a scenario based on data from a deployed industrial robot, thus adequately representing an industrial case. 
As a result, we perform the performance evaluation in this case-study. 
Additionally, we add mockup components to the co-simulation structure, to mimic a more realistic co-simulation with different components for different purposes, such as data analysis, prediction among others. 
We also consider the impact of external factors, such as the frequency of sending data to the RMQFMU\textsubscript{1,2}. 
Whereas, the second case-study presents a setup with data from the Gazebo simulation of the Robotti agricultural robot~\cite{foldager2018design}, rather simple both in terms of the size of the messages being sent and the internal structure of the co-simulation.
The parameters of interest are the \textit{lookahead}, \textit{maxage}, and the frequency with which we send data into the RMQFMU\textsubscript{1,2}.

The experimentation is divided in two phases. 
In the first, we evaluate the impact of the multi-threaded implementation on the performance of the RMQFMU.
Specifically, we compare RMQFMU\textsubscript{1} to RMQFMU\textsubscript{0}, and argue for the suitability of adopting the multi-threaded configuration as the primary implementation for the RMQFMU.
Additionally we consider the effects of the \textit{maxage} and \textit{lookahead} for RMQFMU\textsubscript{1}.
In the second phase, we evaluate the alterations to RMQFMU\textsubscript{1}, i.e. RMQFMU\textsubscript{2}, in order to tackle the side effects observed in phase 1, providing both performance and behaviour results.

\subsection{Case-study 1: Universal robot 5e (UR)}
This case study concerns an industrial robot called UR5e\footnote{Available at \url{https://www.universal-robots.com/products/ur5-robot/}, visited April 7, 2021} (Figure~\ref{fig:ur5e}). 
The robot has a reach of 85 mm, payload of 5kg and 6 rotating joints. 
It is chosen because the UR5e is in production, thus represents a realistic component in a digital twin setting, and thereby a realistic amount of data. 
The data set is generated from a series of movements used for calibrating the digital twin to a related physical twin. 
Therefore, all joints are exercised. 
The test data from the UR robot used is present in csv format (35123 lines including the header) and consists of messages (excluding the header) with 107 float values (one being time) and 10 integer values. 
The messages have a sample rate (frequency $f_{data}$) of $2$ms and will be replayed/produced into the RabbitMQ broker with this rate. 
The test data contains a substantial number of time gaps, i.e.\ places where two consecutive messages in the input data are spaced with more than $2$ms. 
Data is replayed at a constant frequency of $2$ms and as such some messages are produced faster than their timestamps indicate. These effects are discussed in the test results. 
Tests are also performed on a \textit{cleaned} test data with a constant message spacing of $2$ms between all pairs of messages - i.e.\ no additional time gaps.

To investigate the behaviour of the FMU, we perform tests 
with different variations and relations between the simulation step size and the delay imposed by the simulation environment. {The tests are run on a 2.6GHz Intel Core i7 (6 cores), with 32GB RAM.}
We perform the tests with and without the threading option to investigate its effect, $t_{\mathit{on}}$ and $t_{\mathit{off}}$ respectively. 
We first consider simulation step duration and environment delays that simulates a setup where data is expected to be consumed as fast as its produced - i.e.\ every $2$ms and the overhead of the simulation delay is minimal. 
In this case we consider both the simulation step and the simulation delay to match the input data rate of $2$ms. 
Then we consider situations with larger simulation step duration and larger simulation delays - to mimic co-simulation environments with additional FMUs or monitors. 
The full list of test data used is provided in Table~\ref{tab:exp1}.
In short, the results will provide an insight into how these different parameters influence the performance of the RMQFMU.
Note, that for these tests we use a fixed \textit{lookahead} of $1$ and a fixed \textit{maxage} of $300$ms. The max age is chosen to cover for time gaps in the input data up to this range.

\begin{minipage}{\textwidth}
  \begin{minipage}[t]{0.4\textwidth}
    \centering
    \raisebox{-15ex-\height}{\includegraphics[scale=0.15]{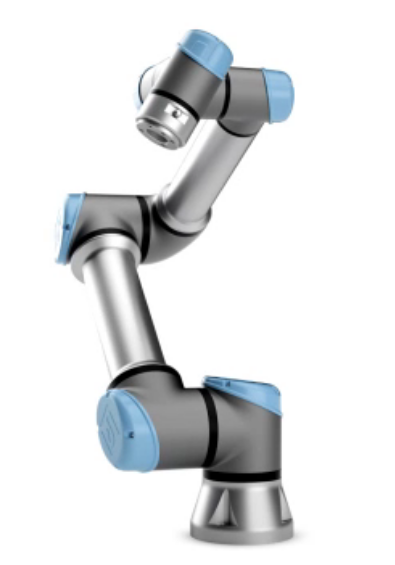}}
    \captionof{figure}{UR5e robot
    }
    \label{fig:ur5e}
  \end{minipage}
  \hspace{0.5cm}
  \begin{minipage}[t]{0.4\textwidth}
    \centering
     \begin{table}[H]
\centering  
\caption{Tests for Case-study $1$~\protect\footnotemark}\label{tab:exp1} 
\begin{tabular}{cccccccc}

\multirow{2}{*}{} & \multicolumn{3}{c}{}                                                                                                                    \\
\toprule                                                                        
            & {Sim step} & {Sim delay} & {Thread} \\
\midrule
Case 1      & 2ms  & 2ms & $t_{\mathit{off}}$ \\
\midrule
Case 2      & 2ms  & 2ms & $t_{\mathit{on}}$  \\
\midrule
Case 3$^*$  & 100ms & 100ms & $t_{\mathit{off}}$ \\
\midrule                                                                                                 
Case 4$^*$  & 100ms & 100ms & $t_{\mathit{on}}$  \\
\midrule
Case 5      & 100ms & 113ms & $t_{\mathit{on}}$  \\
\midrule                                                                                                 
Case 6      & 100ms & 120ms & $t_{\mathit{on}}$  \\
\bottomrule
\end{tabular}
\end{table}

    \end{minipage}
  \end{minipage}
 \footnotetext{Due to space constraints, we present in the paper results for some selected cases marked with $*$. The results as a whole can be found in the technical report~\cite{frasheri2021rmqfmu}.}

\subsection{Case-study 2: Gazebo Simulation of Robotti}
The purpose of this case-study is to provide a basic example of a co-simulation receiving {data recorded~\footnote{We record the data to ease the testing process across different runs.} from }the Gazebo simulation of the Robotti agricultural robot~\cite{foldager2018design}, through the RMQFMU, in order to gain insight into how the configurable parameters of the RMQFMU affect its behaviour.
{Tests are run on a 2GHz Quad-Core Intel Core i5, with 16GB RAM.}
In this case the co-simulation environment consists of the RMQFMU and a monitor FMU. 
The data of interest in this scenario, i.e. the data sent through the RMQFMU consists of: the $x$ and $y$ positions of the robot, and the $x$ and $y$ position of the nearest obstacle. 
Additionally, a sequence number is attached to each message to keep track of the outputted messages during the processing of the results.
The monitor takes as input such data, and computes the distance between the robot and obstacle for every time-step.
In case such distance is below a predefined safety threshold, the monitor will issue an emergency stop to the Gazebo simulation of the robot.

\begin{table}[htb]
\centering
\caption{Overview of experiments for Case-study (CS) 2}\label{tab:exp2}
\begin{tabular}{c | c c c c || c | c c c c || c | c c c c}
\toprule
CS 2 & \multicolumn{4}{c ||}{Parameters}                    & CS 2 & \multicolumn{4}{c ||}{Parameters}   & CS 2 & \multicolumn{4}{c}{Parameters}\\
             & la  & ma   & $t_s$ & $f_{data}$ &                   & la   & ma    & $t_s$ & $f_{data}$   &    & la   & ma    & $t_s$ & $f_{data}$ \\
\midrule
Case 1       & 1   & 0.2s & 0.1   & 100ms &  Case 5                 & 50  & 0.2s & 0.1   & 2ms       &  Case 9$^*$             & 2   & 0.4s & 0.1   & 200ms       \\
\midrule
Case 2       & 50  & 0.2s & 0.1   & 100ms & Case 6                  & 200 & 2s   & 0.1   & 2ms     &  Case 10$^*$            & 5   & 0.4s & 0.1   & 200ms\\
\midrule
Case 3       & 1   & 2s   & 0.1   & 100ms &  Case 7                 & 50  & 0.2s & 0.1   & 2ms   &  Case 11$^*$      & 2   & 2s   & 0.1   & 200ms\\
\midrule
Case 4$^*$   & 50  & 2s   & 0.1   & 100ms &  Case 8                 & 200 & 2s   & 0.1   & 2ms   & Case 12$^*$      & 5   & 2s   & 0.1   & 200ms\\

\bottomrule
\end{tabular}
\end{table}

For this case-study we are interested in the following parameters (see Table~\ref{tab:exp2}), the \textit{maxage} $ma$ of the messages, and the \textit{lookahead} $la$, and how they affect the behaviour of the RMQFMU, i.e. in terms of the sequence of messages outputted at every time-step. 
Note that for all the cases the threaded option is used ($t_{\mathit{on}}$).
Parameters such as the time-step size $t_s$, the frequency of sending data $f_{data}$ to the co-simulation are fixed for cases $1-4$, such that they align. 
We argue such decision with the simplicity of the case-study, which could allow a user in reality to align these values. 
The time-step of the simulation is fixed to $0.1$s for adequate granularity.
Whereas for cases $5-12$ we consider $f_{data}$ that doesn't match the size of the simulation step, and observe the effects of the \textit{maxage} and \textit{lookahead}.

\subsection{Phase 1 Results}

\begin{figure}[t]
  \centerline{\includegraphics[scale=1]{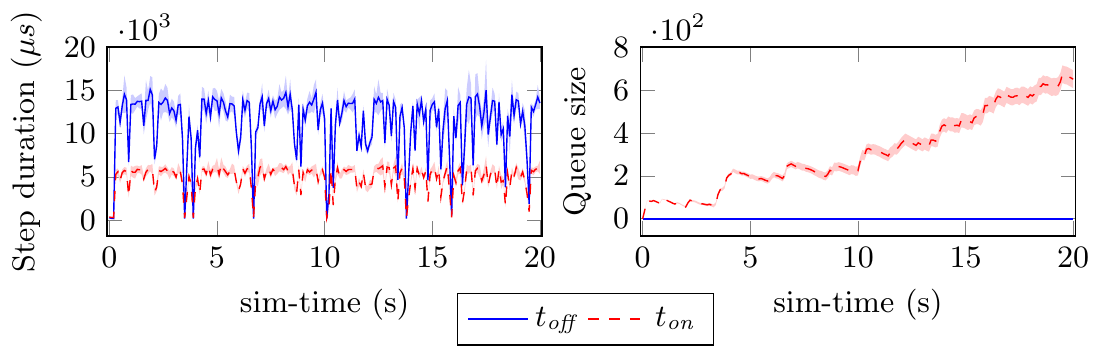}}
  \caption{Cases 3 and 4: step=100ms, delay=100ms, 
  thread=$t_{\mathit{off}}$/$t_{\mathit{on}}$ \vspace{-2mm}}
  \label{fig:log_f2_s100_z100_orig}
  \vspace{-2mm}
\end{figure}
\subsubsection{Case-study 1}

In this paper we will present only the results from cases $3$ and $4$ (Table~\ref{tab:exp1}) due to space constraints, and refer the reader to the technical report~\cite{frasheri2021rmqfmu} that contains the results over all cases. The results not included in this paper are coherent with the ones included.
In the included scenarios, the simulation step and the simulation delay are set each to $100$ms, in order to mimic a co-simulation environment that imposes a step size of $100$ms and has a simulation delay also of $100$ms, caused e.g.\ by other FMUs or monitors. 
The results of these two tests are shown in Figure~\ref{fig:log_f2_s100_z100_orig}.
The number of messages to process in each step is $50$, given the input frequency of $2$ms. 
Additionally, given the simulation delay equal to $100$ms, for the threaded configuration all $50$ messages will be available in the RMQFMU\textsubscript{1} queue when the step function is called. The artificial simulation delay of $100$ms covers enough time for the separate consumer thread to have consumed approximately $50$ messages. While for the non-threaded configuration, the step function itself needs to consume approximately $50$ messages off the socket interface inside a single step. 
From Figure~\ref{fig:log_f2_s100_z100_orig} left-hand side graph it can be observed that for this test, the threaded configuration (red) shows an improvement in average step duration of approximately $10$ms compared to the unthreaded configuration (blue).
This difference corresponds to the overhead of consuming a single message from the socket interface via the rabbitmq client library, parsing the message, and finally adding it to the internal RMQFMU\textsubscript{1} queue. 
As approximately $50$ messages need to be consumed in a single step, this accounts to an overhead of approximately $200\mu$s per message ($10$ms $/$ $50$ = $200\mu$s). 
The right-hand side graph in Figure~\ref{fig:log_f2_s100_z100_orig} shows the internal FMU queue size at the exit of each simulation time step. The increase of queue size in the threaded configuration (red) is mostly an effect of gaps occurring in the input data. 
Large gaps in the input will cause the RMQFMU\textsubscript{1} to stay at its current output for a predefined \textit{maxage} time period ideally covering the input gap. 
This effect can also be observed in the left-hand side graph, when the step duration occasionally lowers to around $0$ when the RMQFMU\textsubscript{1} stays at its current output to cover a gap. 
For this period, data will still be consumed by the separate RMQFMU\textsubscript{1} consumer thread and added to the internal queue. 
The RMQFMU\textsubscript{1} implementation must therefore include guards to respect internal queue size limitations. 
The internal RMQFMU\textsubscript{1} queue size in the unthreaded configuration (blue) is always $0$ at simulation step exit, since that configuration consumes only a single message off the socket interface per step. 
In this configuration, a queue size build up will occur in the socket layer rather than the FMU layer.

\subsubsection{Case-study 2}

Similarly to Case-study 1, we will present only the results from cases $9-12$ (Table~\ref{tab:exp2}) due to space constraints, and refer the reader to the technical report~\cite{frasheri2021rmqfmu} that contains the results over all cases. 
Note that the results not included remain coherent with the ones presented in the current paper.
The results for cases $9-12$ are displayed in Figure~\ref{fig:fig4}.
For a \textit{maxage} equal to $400$ms, we can see that there is not much an effect of the value set for the \textit{lookahead}.
This is due to the fact that with lower \textit{maxage}, the data becomes invalid sooner than for higher \textit{maxage}, thus triggering the RMQFMU\textsubscript{1} to fetch newer data from the incoming queue.
Nonetheless, for a $ma=2$s, it is possible to note a difference for the two \textit{lookahead} values, where for $la=5$, there is a bigger jump in the sequences of data for specific time-steps, when a new value is needed.
Additionally, sequence numbers remain fixed for consecutive number of time-steps, whereas for $la=2$ more in-between value are outputted, as expected. In the latter however, there is more delay in the output values.
Beside the \textit{maxage} value, the frequency of sending data also impacts the effect of the \textit{lookahead}, as it directly affects the number of messages found in the incoming queue.

\begin{figure}
    \vspace{-3mm}
    \centering
    \includegraphics[scale=1]{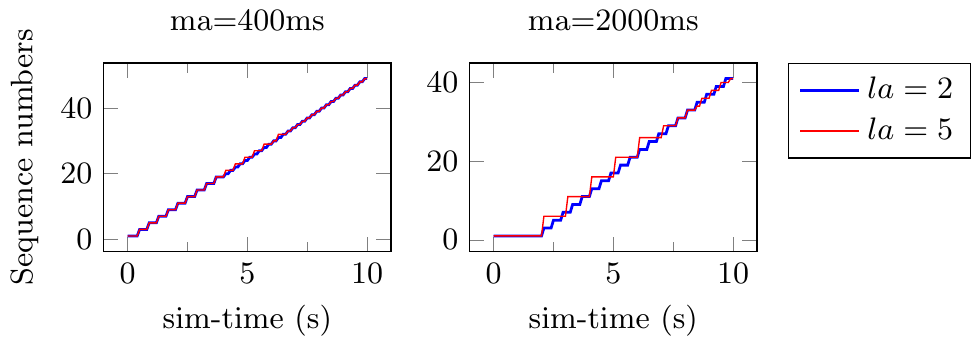}
    \caption{Messages outputted by the RMQFMU\textsubscript{1}, for varying \textit{ma}, \textit{la}, and $f_{data}=200$ms.\vspace{-2mm}
    }
    \label{fig:fig4}

\end{figure}
The number of skipped messages is not exactly the same from run to run, as observed in Figure~\ref{fig:fig2}, with a \textit{lookahead} equal to $50$ and \textit{maxage} fixed to $2000$ms, over 5 independent runs.
One factor that impacts the state of the queue is the speed of the execution of a simulation step. 
In Figure~\ref{fig:fig2}, one can observe the wall-clock time for two of the simulation runs. 
The first run is executed faster than real-time, at least until time step $2.0$, whereas the second is executed slower than real-time. 
This depends on the underlying execution environment, as well as the load on said environment.
As such, a specific \textit{lookahead} will not necessarily output the same sequence numbers at each time step over different independent runs. 
This effect is also visible in Figure~\ref{fig:fig4}, where near the end, for both \textit{lookahead} values, the sequence numbers that are outputted are similar.
This means that, there were no newer messages in the queue at those particular times for \textit{lookahead}$=5$, as the simulation was running faster than real-time.

\begin{figure}[tbh!]
    \centering
    \includegraphics[scale=1]{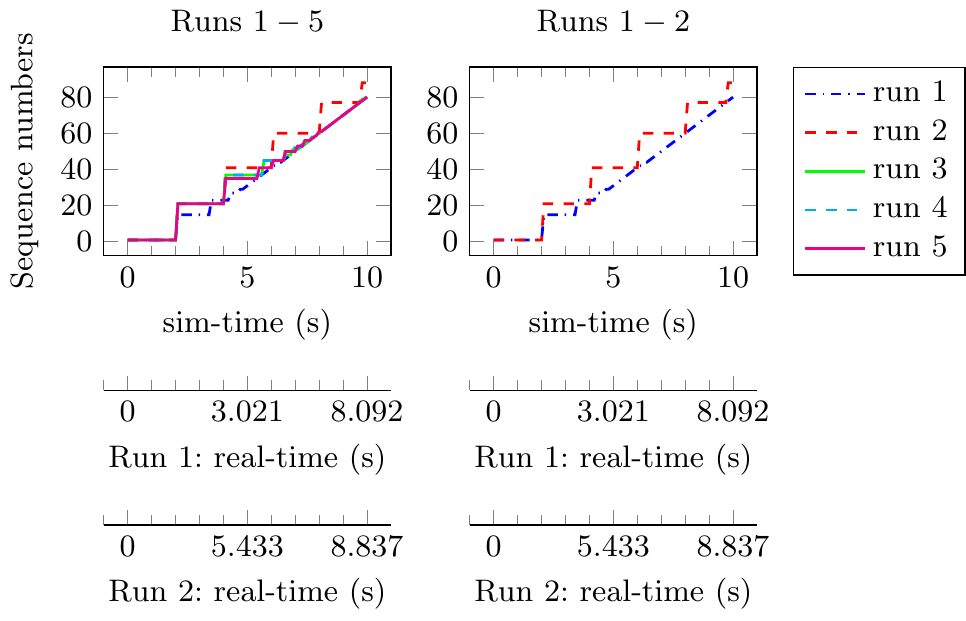}
    \caption{Messages outputted by RMQFMU\textsubscript{1}, over 5 runs, for $ma = 2$s and $la = 50$.}
    \label{fig:fig2}
\end{figure}

\subsection{Phase 2 Results}
As discussed in the previous results, the RMQFMU\textsubscript{0,1} has a rather conservative interpretation of the \textit{maxage} parameter. 
The FMU will \textit{always} stay at its current output, if the timestamp of the output is   within \textit{maxage} range from the current simulation step timestamp. 
This means that, the timestamp of the output plus the \textit{maxage} is greater than or equal to the current simulation timestamp. 
However, this interpretation is rather strict, in the sense that it doesn't take into account if newer data is present in the incoming FMU queue. 
Thus, we want to relax this condition such that the FMU will only stay at its current output, if the output is within the \textit{maxage} and in addition there is no newer data available. 
As such, the FMU should always move ahead to the latest data valid within the current step, and the \textit{maxage} is only considered when there is no new input.

We have implemented this change of \textit{maxage} semantics in the RMQFMU\textsubscript{2} version of the FMU. 
The change itself had some side-effects impacting the performance of the FMU step function. 
In short, the change means that the FMU has to consider if data is present before considering the \textit{maxage}. 
However, situations may occur where data is present in the incoming queue, but it may have time-stamps in the future of the current simulation time. This can occur e.g.\ when replaying with a fixed interval historical data including gaps. 
If this situation occurs, the RMQFMU\textsubscript{0,1} versions of the FMU would move the future data from the incoming queue   into an internal processing queue, that would gradually keep increasing in size and thereby impacting the step performance. 
To remedy this issue, RMQFMU\textsubscript{2} has an additional change to the internal queue structure.
In the following paragraphs we present results of experiments with our two case-studies using RMQFMU\textsubscript{2}.

\subsubsection{Case-study 1}
The effect of the RMQFMU\textsubscript{2} changes to the case-study 1 can be observed in Figure~\ref{fig:fig5}. The test configuration for this test is identical to Case 4 of Table~\ref{tab:exp1}. The results of this case with RMQFMU\textsubscript{1} are present in Figure~\ref{fig:log_f2_s100_z100_orig}.
To demonstrate the effect of moving to the latest input data available, we have specified a \textit{lookahead} size of $100$ in this test. 
This covers more messages than being produced on average within a simulation time step of $100$ms and a data frequency of $2$ms, which is approximately $100 / 2 = 50$ messages. 
Practically, the RMQFMU\textsubscript{2} consumes as many messages as possible within a time step.

It is possible to observe from Figure~\ref{fig:fig5}, 
graph on the left, 
that in contrast to Figure~\ref{fig:log_f2_s100_z100_orig}, the step duration now has decreased from around $6000$us to around $1000$us. 
This is due to the fact that we now use the larger (100 = move as close to step time as possible) \textit{lookahead} size as opposed to the original size 1, additionally to the improvement in the internal queue implementation. Also, the larger \textit{lookahead} size has the effect, that the queue size (right-hand sideL graph) stays low for a longer simulation time compared to earlier test results.
However, the queue size still increases due to both data being produced more frequently than consumed and to gaps in the input data. Thus, as mentioned before, the FMU needs to guard any internal queue size limitations. 
\begin{figure}[tbh!]
    \centering
    \includegraphics[scale=1]{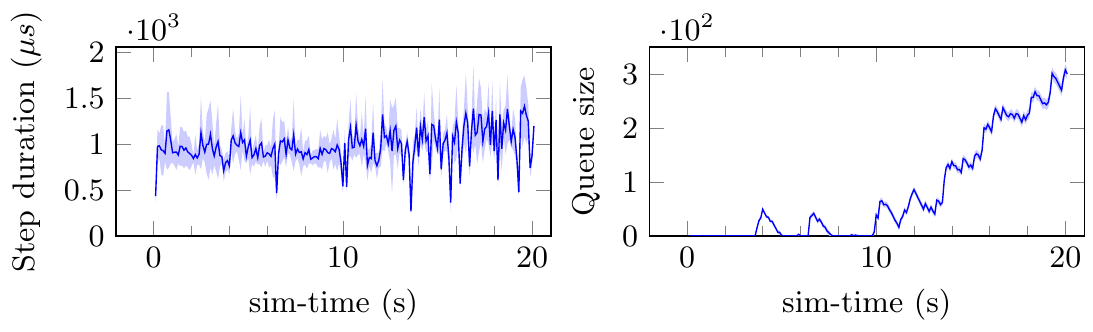}
    \caption{Step duration and sequence numbers outputted by RMQFMU\textsubscript{2} at every step, for step-size of $100$ms, and delay of $100$ms.}
    \label{fig:fig5}
\end{figure}
\vspace{-4mm}
\subsubsection{Case-study 2}
The new version of the RMQFMU\textsubscript{2} removes the initial delay in the outputted sequence numbers noticeable for large values of the \textit{maxage}, e.g. $2000$ms (Figure~\ref{fig:fig6}, for Cases $1-4$), as compared to RMQFMU\textsubscript{1} (Figure~\ref{fig:fig4})~\footnote{Note, the initial delay is present in all cases for RMQFMU\textsubscript{1}.}. 
The utility of the \textit{maxage} parameter becomes evident, once gaps are introduced in the data, in these experiments equal to $500$ms, Figure~\ref{fig:fig6b}, left graph, and $1000$ms, right graph.
It can be observed how when there is no input, the sequence value that is outputted is the last value that is valid from a time perspective. 
Naturally, for larger gaps, the sequence numbers outputted are more delayed, e.g. for $t=5$s, the RMQFMU\textsubscript{2} with $500$ms gaps is at number $10$, whereas for $1000$ms is at $5$.
Note that, with RMQFMU\textsubscript{1}, given the initial delay, we would expect a lower sequence number outputted at $t=5$s.

\begin{figure}[bth!]
    \centering
    \includegraphics[scale=1]{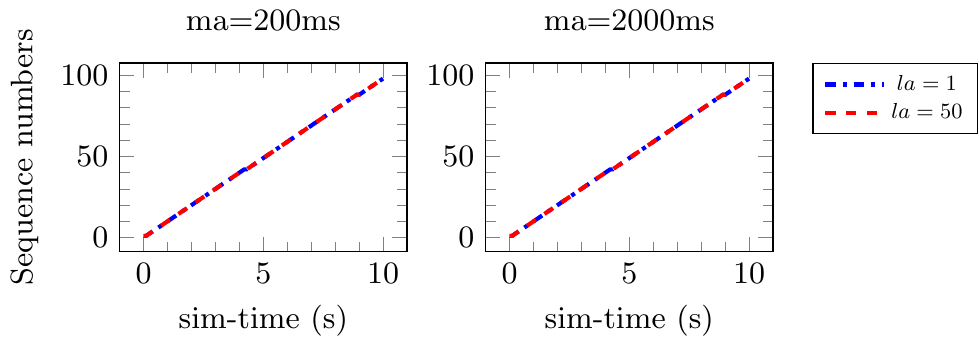}
    \caption{Messages outputted by the RMQFMU\textsubscript{2} at every step, for $f_{data}=100$ms.}
    \label{fig:fig6}
\end{figure}
\begin{figure}[tbh!]
    \centering
    
    \includegraphics[scale=1]{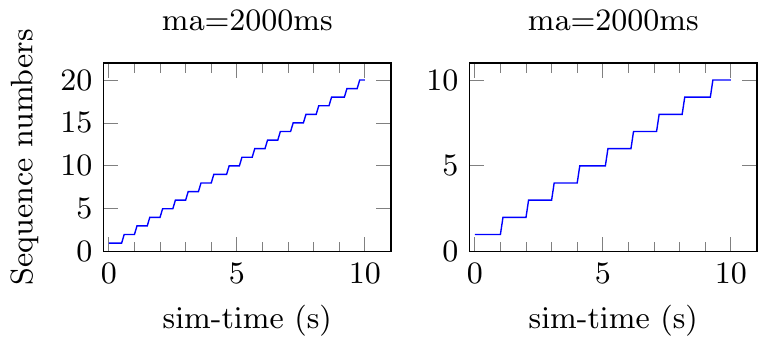}
    \caption{Messages outputted by the RMQFMU\textsubscript{2} at every step, for $f_{data}=100$ms and $la=1$, with gaps of $500$ms (left), and $1000$ms (right).}
    \label{fig:fig6b}
\end{figure}

%
In order to show the utility of the \textit{lookahead}, we ran the experiments with simulation step-size equal to $100$ms, for values of \textit{maxage} in \{200ms, 2000ms\}, with frequency of sending data equal to $2$ms (Figure~\ref{fig:fig7}). 
For smaller values of the \textit{maxage}, the delay of the RMQFMU\textsubscript{2} for \textit{lookahead} equal to $1$, while consistently present, is rather low.
However, for larger values of the \textit{maxage}, the delay for low \textit{lookahead} values is non-negligible.
It is possible to observe that for the duration of the \textit{maxage} the sequence numbers gradually increase by $1$, as expected for \textit{lookahead} equal to $1$.
After the initial \textit{maxage} duration has passed, there is a jump in the sequence numbers, that continue to follow the results for \textit{lookahead} equal to $50$, albeit with a more or less constant delay until the end.
A similar effect was observed in Figure~\ref{fig:fig4}, with the difference that in the current version, the RMQFMU\textsubscript{2} checks if there are newer messages in the queue.
This is due to the fact that after $2$s, older in-between values become invalid, and thus are not outputted, with RMQFMU\textsubscript{2} going back to the queue for more recent messages. 

\begin{figure}[tbh!]
    \centering
  { \includegraphics[scale=1]{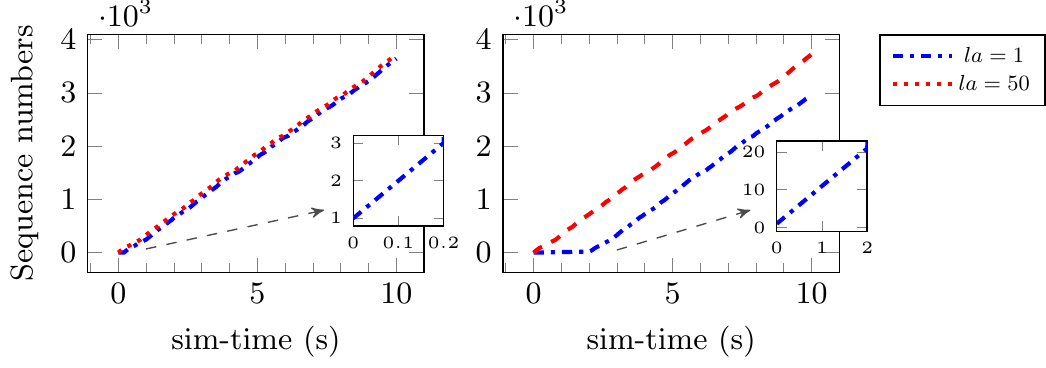}}
   
    \caption{RMQFMU\textsubscript{2} output at every step, for $f_{data}=2$ms, over $2$s.}
    \label{fig:fig7}
\end{figure}


\section{RMQFMU\textsubscript{2} Configuration Guidelines}\label{sec:guidelines}

The parameters of the RMQFMU\textsubscript{2} should be configured based on the use-case in order to result in desired behaviour. 
If possible, the simulation step-size should match the frequency of sending data $f_{data}$.
However, depending on the time granularity needed by different FMUs, achieving such alignment might not be possible. 
The \textit{maxage} value has to be big enough to make up for any time gap between data, but also low enough to ensure consistency in domain of use. 
E.g. for some applications, it might the case that data should not be older than $200$ms to ensure correct operation.
The \textit{lookahead} can be set to counter-balance the effect of the \textit{maxage}, as it allows to jump to the latest data that is available in the queue (within the limits of the \textit{lookahead}).
Note however, larger values of the \textit{lookahead} will result in bigger time jumps between the messages outputted.
If more intermittent values are required, then a lower value should be adopted. 
The frequency of sending data influences the meaningful range of values for the \textit{lookahead}, e.g. sending data every $2$ms, will result in $500$ messages per second.
For a simulation step-size of $100$ms, a reasonable value of the \textit{lookahead} would be circa $50$.
Finally, the speed of the simulation also affects what is present in the queue at any time, as a result impacting RMQFMU\textsubscript{2} outputs.



\section{Concluding Remarks}\label{sec:conclusion}
In this paper we have described an extended RMQFMU data-broker that enables coupling an FMI-based co-simulation environment to a non-FMI external component, via the AMQP protocol.
The RMQFMU supports communication in both directions, and can be used in different contexts ranging from replaying historical data into a co-simulation, to enabling the realisation of the DT concept by connecting the DT to its physical counterpart. 
We evaluate this component in terms of its performance, i.e. the real-time duration of a simulation step.
Our results show the benefit of implementing a threaded solution, that effectively decouples the \texttt{doStep} logic from the consumption from the rabbitMQ server. 
Moreover, we explore different values of its configurable parameters, and provide usage guidelines for practitioners. 

There are {five potential} directions for future work. 
First, we are interested in enabling the RMQFMU to take bigger step-sizes if necessary.
In some monitoring applications, where only the latest data-points are relevant, such jump could be of use to mitigation components that deal with out of synchronisation situations.
Second, we plan to formally verify the presented RMQFMU, thus extending previous work that tackled its initial version~\cite{thule2020formally}. 
Third, we intend to profile the overhead of RMQFMU as data-broker compared to an FMU that directly queries a database, for example in the context of large amounts of data.
Fourth, we will enable the master-algorithm to learn the optimal step-sizes automatically and deal with recurring variation at run-time. 
{Fifth, we are interested in the implementation of a more general solution, that can be configured to use different libraries and protocols, e.g. ZeroMQ.}


\clearpage
\endgroup

\begingroup
\renewcommand\theHchapter{7-Rose:\thechapter}
\renewcommand\theHsection{7-Rose:\thesection}
\locallabels{7-Rose:}
\setcounter{footnote}{0}
\setcounter{chapter}{0}
\setcounter{lstlisting}{0}
\fontfamily{ptm}\selectfont

\setlength\textfloatsep{8.0pt plus 2.0pt minus 2.0pt}
\setlength\intextsep{4.0pt plus 2.0pt minus 2.0pt}
\setlength\floatsep{4.0pt plus 2.0pt minus 2.0pt}
\setlength\abovecaptionskip{4.0pt plus 2.0pt minus 2.0pt}
\setlength\belowcaptionskip{0pt}

\lstset{basicstyle=\scriptsize,tabsize=2,frame=trBL,frameround=fttt}

\makeatletter
\def\input@path{{7-Rose/}}
\makeatother

\graphicspath{{7-Rose/}}
\title{Genetic Algorithms for Design Space Exploration of Cyber-Physical Systems: an Implementation in INTO-CPS}
\titlerunning{Genetic Algorithms for Design Space Exploration of CPS}

\author{Maximilian Rose \and John Fitzgerald}
\institute{School of Computing, Newcastle University, United Kingdom\\
\email{\{M.Rose, John.Fitzgerald\}@newcastle.ac.uk}
}
			
\maketitle

\begin{abstract}
We describe the initial implementation and evaluation of a facility to support Design Space Exploration (DSE) over multi-paradigm models of cyber-physical systems, using Genetic Algorithms (GAs). Existing features of the INTO-CPS toolchain have been updated to allow concurrent exploration of DSE parameters and a new GA-based approach to DSE. Based on a comparison with the previously reported DSE of the multi-model of an agricultural robot, the developed solution appears to reduce the time required to perform a DSE by providing user-configurable parameters for the number of concurrent DSE simulations. It also provides a modular, user-extendable, GA-based DSE system that can be used with the INTO-CPS toolchain.
\end{abstract}

\begin{keywords}
digital twin, FMI, co-simulation, Design Space Exploration, Genetic Algorithms
\end{keywords}

\section{Introduction}
\label{sec:intro}

Modelling and Simulation are important and powerful tools when developing systems of any kind, especially within the field of engineering, and increasingly so in the field of software development. Simulating models of the systems we design allows us to gain a solid understanding of how they might behave once deployed in the real world. This is of particular importance for Cyber-Physical Systems (CPSs)~\cite{Lee10}, where software must interact with hardware and it is difficult to test exhaustively and in a non-destructive way.

One of the key benefits of using simulation in this way is the ability to rapidly test different design parameters of a system without having to first make the financial and time commitments of creating or changing a physical prototype. This places simulation as an invaluable tool for performing Design Space Exploration (DSE). The `design space' of a system can be defined as the set of all configurations of its `design parameters', and therefore DSE is the act of exploring and assessing the range of configurations present in this design space, with the goal of discovering ones in which the system meets certain performance criteria most optimally.

DSE is one feature of INTO-CPS~\cite{Larsen&17a}, a tool chain for model-based design of CPS, built around a co-simulation engine called Maestro~\cite{Thule&192}. INTO-CPS currently offers exhaustive search (i.e. all combinations of a given set of design parameters), and genetic search (solving through natural selection of the fittest designs). Depending on the complexity of the model in question, the number of simulation runs needed to perform these searches accurately enough can make this process increasingly time consuming.

In industry, it is often paramount that tasks take the minimum amount of time needed to be performed properly, to keep costs down and to ensure that crucial deadlines are met. There are a wide variety of Multi-Objective Optimisation (MOO) that attempt to efficiently search for optimal solutions in complex design spaces. Many of these are implemented in existing, well-maintained libraries (see Section~\ref{sec:back:moo}). This paper describes work in extending the DSE capabilities of INTO-CPS with features that attempt to widen the range of DSE options for users of INTO-CPS.

By incorporating state-of-the-art MOO library to INTO-CPS, the user can now choose from the most appropriate and up-to-date optimisation algorithm for their CPS model, allowing them to discover optimal parameter configurations in the most efficient way possible. Furthermore, integrating an open-source, third-party library would allow the application to be easily updated to stay up to date with the most cutting-edge algorithms, eliminating the need to manually implement newly released ones from the ground up.

This paper also applies the MOO support to the Robotti optimisation case study, which was reported as a previous workshop~\cite{bogomolov2021tuning}. This case study demonstrates that the MOO integration works, but does not demonstrate the full benefits that MOO could have on co-simulations with larger design spaces and more competing objectives that require trading-off.

This remainder of the paper is structured as follows. Section~\ref{sec:back} provides some background detail on INTO-CPS and MOO. Section~\ref{sec:what} describes integration of a MOO library into the INTO-CPS framework. Section~\ref{sec:study} provides a refresher on the Robotti optimisation case study. Section~\ref{sec:results} compare the results from the previous Robotti DSE with those form the new MOO feature. Finally, Section~\ref{sec:conc} presents from conclusions and future work.

\section{Updating INTO-CPS to Support GA-based DSE}
\label{sec:update}
\label{sec:update-threading}
\label{sec:update-misc}

Prior to the work reported here, the main means of performing DSE in INTO-CPS was by using outdated Python~2 scripts and an exhaustive search method. Indeed, the authors of~\cite{Bogomolov&21} have noted the overhead caused by the reliance on Python~2. In addition, Python~2 support ceased in 2020. The goal of the work reported here was to update the DSE scripts to use modern Python~3, to improve usability, and to provide INTO-CPS users with more options based on GAs when performing DSE.

The code that is provided with INTO-CPS for DSE consists of 8 scripts that interact with each other as shown in Fig.~\ref{fig:DSEscripts}. Algorithm Selector is used to select the desired DSE algorithm to use. The selected algorithm script then calls helper functions from the Common script to run and evaluate a specific configuration using the COE. To produce a human readable output in both CSV and HTML the Output HTML script is run.

\begin{figure}
\centering
\includegraphics[width=0.75\columnwidth]{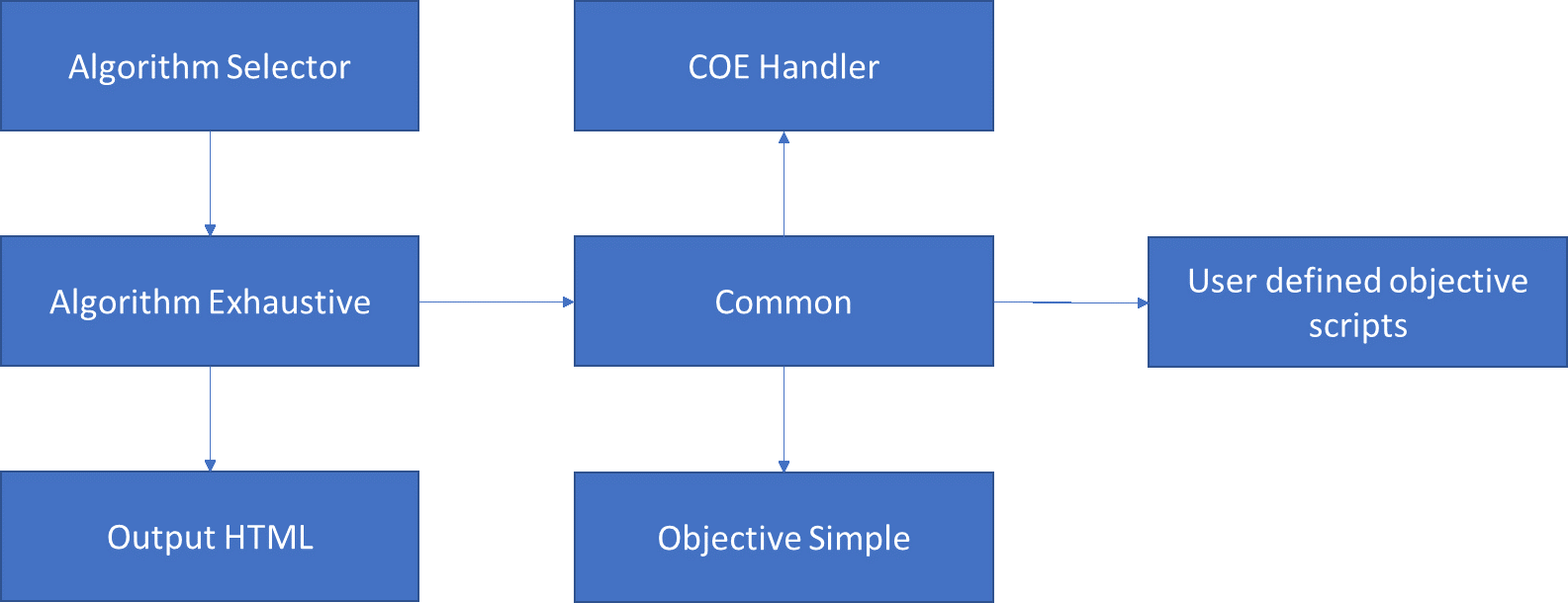}
\caption{\label{fig:DSEscripts} Outline of DSE Scripts in INTO-CPS.}
\end{figure}

\paragraph{Conversion to Python 3.} It was decided that, to maintain compatibility, the overall structure of the scripts would remain the same with only minor changes to improve flexibility. Any new options that would be added would also default to the original behaviour of the scripts. The largest change in respect to Python 2 and 3 was the requirement for brackets on print statements and also the http library used to connect to the COE had also changed its name. As such the upgrade process was not difficult and is fully compatible with anything that may be reliant on the way the original Python 2 scripts executed.

\paragraph{Multi-Threading} The theoretical speed improvements associated with multi-threading are exponential with increasing thread counts to a threshold, specifically the time taken for a single simulation. Although the original scripts and COE were not designed with multi-threading in mind, they are mostly thread safe, and those parts that were not (mainly to do with writing simulation results to a file ready for ranking) required some updates to properly implement the multi-threading.

From Fig.~\ref{fig:threadedtestperf} indicates that the threading performs as expected with a test controller that simply echoes its input and with the case study described in Section~\ref{sec:case-oldVnew}.

\begin{figure}[h!]
\centering
	\includegraphics[width=.4\paperwidth]{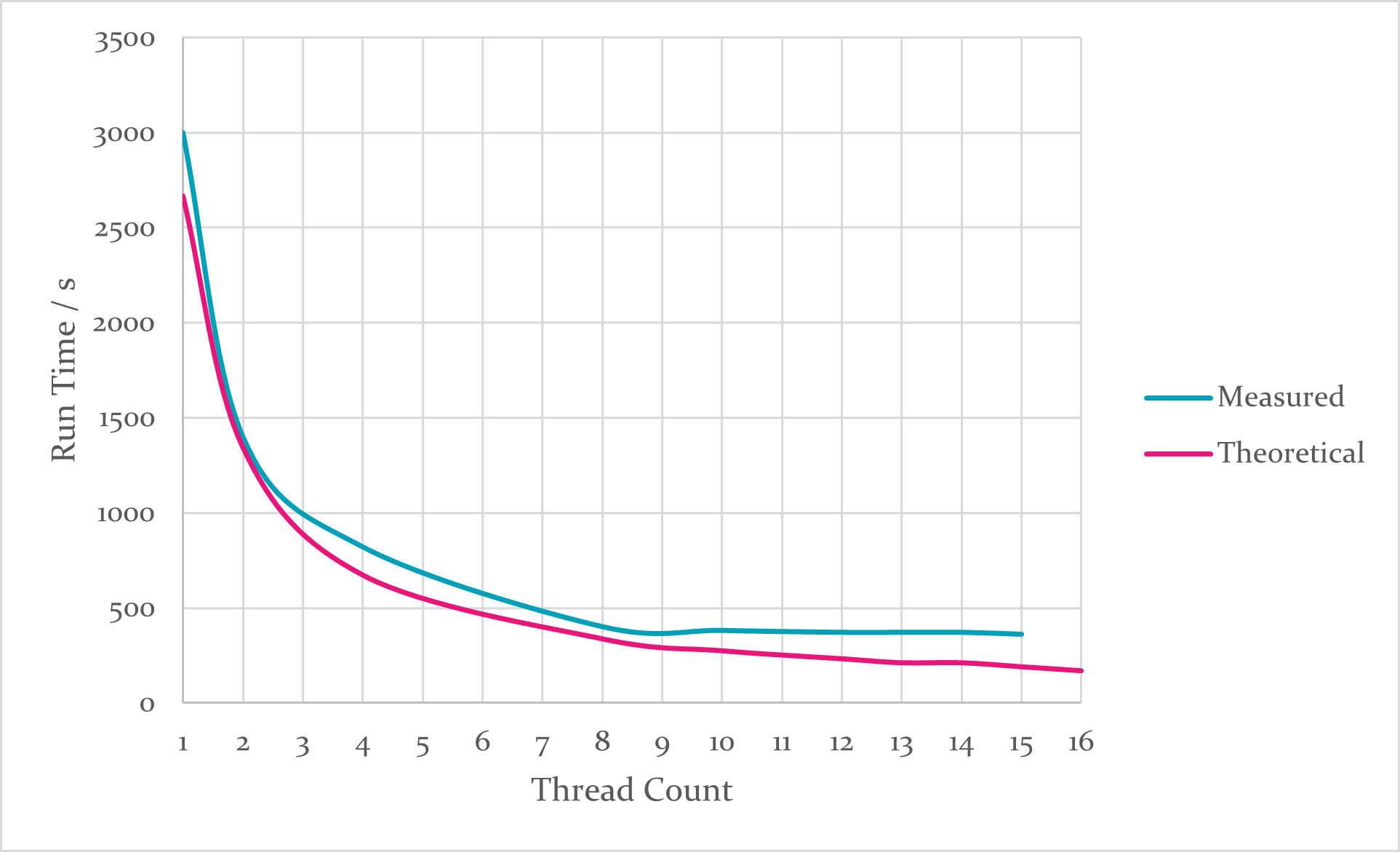}
	\caption{Actual Multi-Threaded Performance}
\label{fig:threadedtestperf}
\end{figure}

More options and flexility were also added to the upgraded DSE scripts:
\begin{description}
\item[Splitting Results Generation.]  Previously there was no option to prevent the generation of the HTML \& CSV results files. This has been changed so that the user may specify how they would like their results to be output.
\item[Arguments.] As it is possible to run the DSE scripts easily via a command line, it is desirable that they produce useful errors. With the old Python 2 scripts the error outputs were simply from Python telling the user that a command was not found. With the updated scripts this has been changed so that the script will notify the user not only that a parameter is missing but also which parameter is missing. A \texttt{--help} command has also been added so that the user can see all the options available to them.
\item[Progress Bar.] A progress bar was added to the console output of the scripts to give the user some sense that the scripts are running and things are happening.

\end{description}

\section{Implementation of Genetic Algorithms in INTO-CPS DSE}
\label{sec:implement}
Several operations have been implemented in order to allow the straightforward construction of GAs for DSE. We here describe their functionality in terms of the GA elements described above in Section~\ref{sec:background}. These have been selected mainly on the grounds of their extensive use in the literature. We do not yet have sufficient experience with DSE in INTO-CPS to have developed heuristics on which combinations of operators  might be used in different design spaces.

\subsection{Selection Operators}
\label{sec:selection}

To simulate the selection of mating pairs in a meaningful way a fitness function must be created and then applied to each organism in the generation. In the case of DSE with INTO-CPS, the fitness function is expressed in terms of the results obtained from each co-simulation.

\subsubsection{Random}
\label{sec:selection-random}

This is the simplest operation and does not actually require the ranking form the previous generation. This algorithm is included for completeness; it may be useful to start the algorithm with in situations in which it may not be practical to create a ranking for organisms in generation 0 in some scenarios.

\subsubsection{Roulette Wheel}
\label{sec:selection-roulette}

Initially proposed by Holland~\cite{Holland92}, this operation can be thought of in terms of a pie chart with the more fit individuals taking up a larger portion of the pie. From this the probability of a specific organism being chosen is controlled by Eqn.~\ref{eqn:roulette} where $p_i$ is the probability of the $i^{th}$ organism being selected relative to its fitness, $f_i$.

Eqn.~\ref{eqn:roulette} does assume that a higher fitness value is always better. This may not be the case for the diverse applications of INTO-CPS. Thus a reversed version of this was also implemented that offers a higher probability of selection for organisms with a lower fitness value.

\begin{equation}
	p_i = \frac{f_i}{\sum^{n}_{j=0}{f_j}}
\label{7-Rose:eqn:roulette}
\end{equation}

\subsubsection{Ranked}
\label{sec:selection-ranked}

This is a variation of Roulette Wheel that, instead of using the raw fitness value for ranking, will instead normalise the values first~\cite{Obikto98}. This solves an issue that when using the raw fitness value it is possible for a single organism to dominate the process by having an extremely high fitness.

The ranking is a total ordering from $1$ to $n$ where $n$ is the organism with the worst raw fitness value. Organisms having the same fitness are ranked in the order they are processed. As can be seen in Fig.~\ref{fig:ranking-tab}, `a' has an extremely high fitness value compared to the other organisms and thus dominates the selection being selected $74.2\%$ of the time. When normalised, `a' is still chosen $33.6\%$ of the time.

\begin{figure}[h!]
	\begin{center}
		\begin{tabular}{c c c c c}
			\multirow{2}{*}{Organism} & \multirow{2}{*}{Fitness} & \multirow{2}{*}{Normalised Fitness} & \multicolumn{2}{c}{Times Selected} \\
			& & & Roulette Wheel & Ranked \\
			\hline
			a & $300$ & $1$ & $742$ ($74.2\%$) & $336$ ($33.6\%$)  \\
			b & $50$ & $2$ & $118$ ($11.8\%$) & $257$ ($25.7\%$)  \\
			c & $30$ & $3$ & $52$ ($5.2\%$) & $196$ ($19.6\%$)  \\
			d & $35$ & $4$ & $78$ ($7.8\%$) & $127$ ($12.7\%$)  \\
			e & $5$ & $5$ & $10$ ($1\%$) & $84$ ($8.4\%$) \\
		\end{tabular}
	\end{center}
	\caption{Difference in selection bias between using raw fitness values and normalised values.}
	\label{fig:ranking-tab}
\end{figure}

\subsubsection{Tournament}
\label{sed:selection-tournament}

This operator is based on the bracketing systems that are seen in sports where $x$ individuals are chosen to compete between each other from all $n$ individuals in the population.

The size, $x$, of each tournament should be a small number else the fittest individual in the population will dominate, leading to premature convergence. This value will be left to be user defined in the implementation and taken as an argument but with a default value of $2$.

To simplify implementation, this operator will accept only the normalised rankings. This is because using the normalised ranking, an organism with a ranking of $1$ will always be more fit than an organism with ranking $2$, ranking $2$ is always more fit than $3$, etc. Thus, we do not need to account for if the DSE is looking to minimise or maximise the fitness in the implementation of the operator as it has already been dealt with by the normalisation operation.

\subsection{Mating Operators}
\label{sec:mating}

A traditional GA would select from the gene pool created by the previous selection operators to create mating pairs. However, due to the implementation in this paper, this step is combined into the selection operation.

\subsection{Crossover Operators}
\label{sec:crossover}

In our DSE setting, we think of the parameter values that characterise a given design as a sequence, analogous to chromosomes. Each parameter value in the sequence is a key to a specific design characteristic. Crossover operators emulate the recombination stage in natural reproduction. During this stage in natural systems two homologous chromosomes come together touching chromatids. Where these chromatids touch, they break and re-join exchanging DNA between the two chromatids~\cite{Pearson15}. In computational crossing the process does not need to simulate chromosomes moving around and touching each other. Instead we can select contiguous keys (which in our setting are design parameter values) from each parent to add into the child. Note that this process does not change the values of the inherited parameters: each inherited value comes from a parent.

\subsubsection{$n$-Point}
\label{sec:crossover-npoint}

One of the simplest crossing functions that mimics biology is n-point crossover. The simple case, 1-point crossover, is the traditional operation for a GA to implement~\cite{Sivanandam&08-2}. To give the user more options when using the GA library, it was decided to implement the more general n-point version.

This operator selects $n$ keys from the parent, takes contiguous keys from the parent and adds them to the child. This process is illustrated in Fig.~\ref{fig:npointcrossing}. In the diagram each coloured rectangle represents a key in the parent. In this example 3-point crossing is selected so three points to cross are selected, shown in yellow, blue and purple. The child is created with contiguous keys from A until the first crossing point is reached, when it is reached, we switch to taking keys from B, this process repeats until there are no keys left.

\begin{figure}[h!]
\centering
	\includegraphics[width=.3\paperwidth]{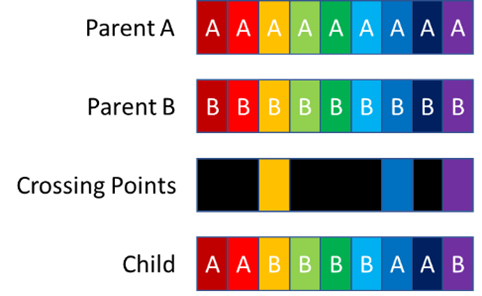}
	\caption{Example of $n$-point crossover.}
\label{fig:npointcrossing}
\end{figure}

\subsubsection{Uniform}
\label{sec:crossover-uniform}

This crossover strategy differs from n-point in that it uses a bitmask to decide which genes are inherited by the child. Fig.~\ref{fig:uniformcrossing} illustrates this strategy. Again, the coloured rectangles represent genes. The crossing mask would be a randomly generated bit string with the value 0 (white rectangles) meaning take a gene value from A and value 1 (black rectangles) meaning take a gene from B.

\begin{figure}[h!]
\centering
	\includegraphics[width=.3\paperwidth]{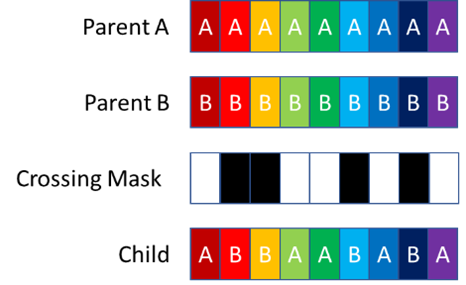}
	\caption{Example of uniform crossover.}
\label{fig:uniformcrossing}
\end{figure}

\subsubsection{BLX}
\label{sec:crossover-blx}

Blend Crossover \cite{Eshelman&93,Takahashi&01} is a more advanced operator that also incorporates elements of mutation into the crossover phase. This incorporation of changing gene values might argue for this being a mutation operation rather than a crossover operation. However, due to the need for two separate individuals, we will treat it as a crossover operation~(the same argument can also be made for the SBX operator discussed below).

This operation takes the two parent gene values and uses them as a range from which the child's value can be selected~(See Fig.~\ref{fig:blxcrossing}). The extents of the ranges past the parent values are controlled by $\alpha$. The child value, $g$, is selected from this range as shown by Eqn.~\ref{eqn:blx}. Common values of $\alpha$ for this equation are $0.5$ \cite{Eshelman&93} and $0.366$ \cite{Takahashi&01}, $0.5$ is chosen as the default value.

\begin{figure}[h!]
\centering
	\includegraphics[width=.3\paperwidth]{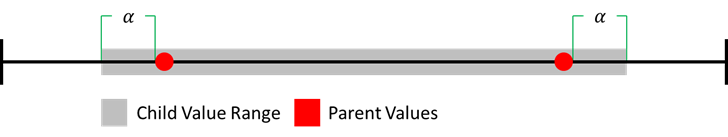}
	\caption{BLX crossover.}
\label{fig:blxcrossing}
\end{figure}

\begin{equation}
	g \in{(\min{(p_{1g}, p_{2g})} - \alpha |p_{1g} - p_{2g}|, \max{(p_{1g}, p_{2g})} + \alpha |p_{1g} + p_{2g}|)}
\label{7-Rose:eqn:blx}
\end{equation}

\subsubsection{SBX}
\label{sec:crossover-sbx}

This is similar to BLX crossover however biases the child gene value to be closer to the parent gene values as depicted in Fig.~\ref{fig:sbxcorssing}. This operation is designed for real value genes proposed in~\cite{Deb&00}. The distribution is achieved using Eqns.~\ref{eqn:sbx1} and~\ref{eqn:sbx2}. The value $\beta$ is used to control the spread of the resulting gene with $\eta$ called the distribution index, with larger values meaning the resulting offspring values should be closer to the parents.

\begin{figure}[h!]
\centering
	\includegraphics[width=.3\paperwidth]{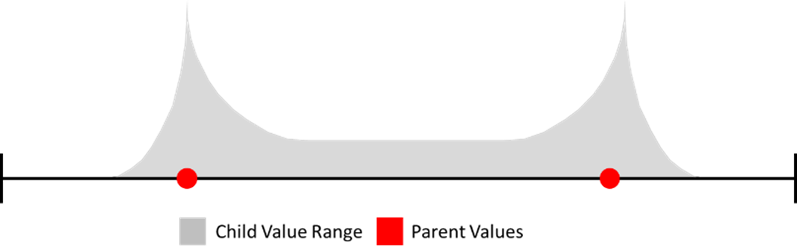}
	\caption{SBX crossover.}
\label{fig:sbxcorssing}
\end{figure}

\begin{equation}
	\beta =
	\left\{
		\begin{array}{@{}lr@{}}
			\mu \leq 0.5, & (2\mu) \\
			\mu > 0.5, & {\frac{1}{2(1-\mu)}}
		\end{array}
	\right\} ^\frac{1}{\eta+1}
\label{7-Rose:eqn:sbx1}
\end{equation}

\begin{equation}
	\beta = 0.5 \times
	\left\{
		\begin{array}{@{}l}
			(1 + \beta)p_{1g} + (1 - \beta)p_{2g} \\
			(1 - \beta)p_{1g} + (1 + \beta)p_{2g}
		\end{array}
	\right.
\label{7-Rose:eqn:sbx2}
\end{equation}

\subsection{Mutation Operators}
\label{sec:mutation}

Aside from the BLX and SBX operations that some may consider mutation operations, the only realistic option is to select values from a normal distribution in the case of a real value algorithm, as a bit flipping operation would effectively result in a random search. Selection from a normal distribution can be done in two ways: either absolute selection as in Eqn.~\ref{eqn:mutationabs} or a relative distribution as in Eqn.~\ref{eqn:mutationrel}.

It is recommended that small values of $\sigma$ are normally used, as larger values will effectively turn the algorithm into a random search. Larger values of $\sigma$ are do not accurately represent the reality of evolution since large changes do not happen suddenly. Using a larger $\sigma$ value can be used to prime the search space in the initial generation to help ensure that the full search space is explored.

As the fitness of organisms converges, it is possible to decrease the value of $\sigma$ slowly to yield a process like simulated annealing.

\begin{equation}
	g' = g + \Phi(0, \sigma)
\label{7-Rose:eqn:mutationabs}
\end{equation}

\begin{equation}
	g' = g \times (1 + \Phi(0, \sigma))
\label{7-Rose:eqn:mutationrel}
\end{equation}

\subsection{Elitism}
\label{sec:elitism}

This operator ensures that a certain percentage of the best individuals from the previous generation will be carried over into the next generation. This is to ensure that the best individual will always be produced by the algorithm no matter what stage it was generated at. Organisms can be chosen at random for removal as they have not yet been ranked. Whilst this does introduce the possibility of removing the new best organism it is probable that it will be recreated in the generation after and the probability of it being removed a second time is low especially with large generation sizes and a low number of organisms being carried over from the previous generation.

\subsection{Diversity Control}
\label{sec:divcontrol}
Diversity control is the mechanism that attempts to prevent premature convergence by applying penalties to organisms based on how similar they are to the rest of the population. This paper implements the first operator from the GOSET manual~\cite{USNA&07}. The diversity operator is based on calculating a mean distance between all organisms, using that to calculate a threshold and then using that threshold to count the number of neighbours an organism has. Although its $O(n^2)$ complexity makes it possibly the most expensive part of the entire GA implementation, it is highly accurate.


\begin{figure}[h!]
\centering
	\includegraphics[width=.5\paperwidth]{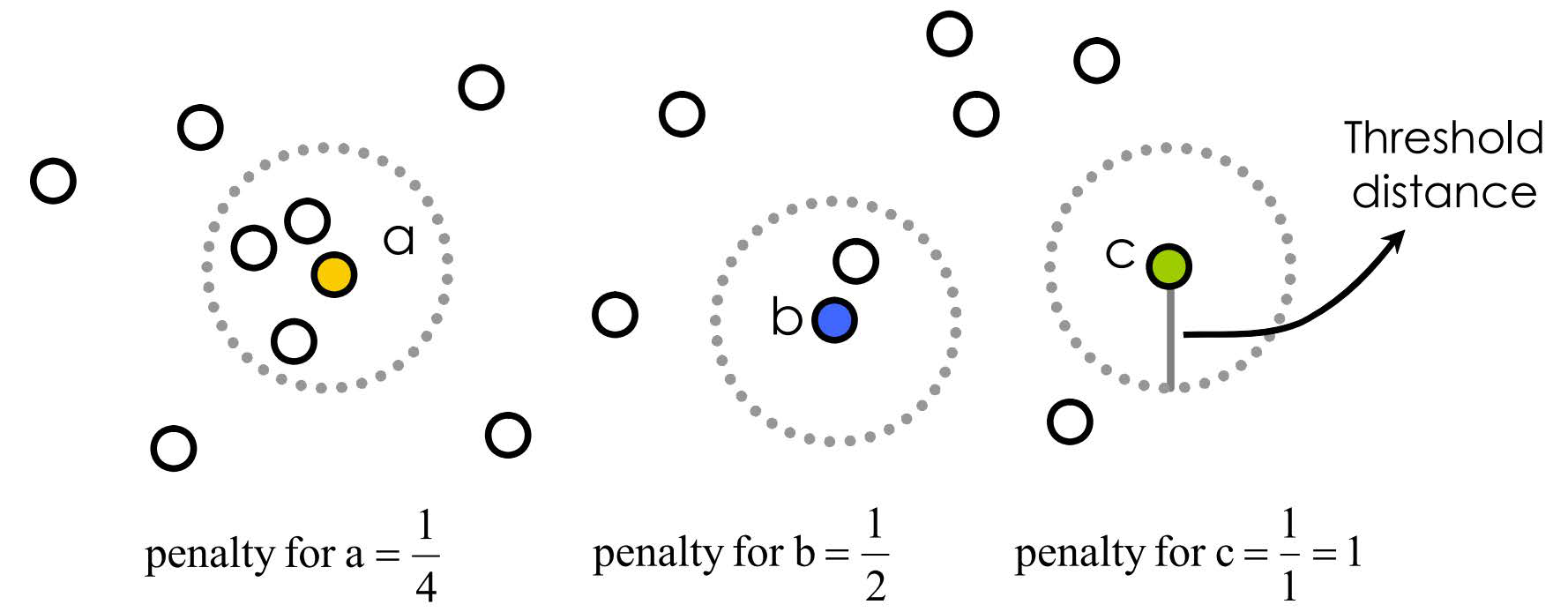}
	\caption{Diversity control.}
\label{fig:divcon}
\end{figure}

\subsection{Controller}
\label{sec:controller}

There are several possible solutions to controlling the iterative process. We can stop after a set number of generations, or after a fitness threshold is reached by $n$ organisms, or after there has been little change is fitness for $n$ generations (convergence stopping), or after the algorithm has run for a certain amount of time. The final option does not appear to be realistic in the context of DSE for INTO-CPS, but the other options have been implemented:

\begin{description}
\item[$n$ Generation.]
\label{sec:controller-ngen}
This is the simplest of the controllers to implement is stopping after n generations. This approach whilst guaranteed to stop at some point is not guaranteed to produce the fittest individuals.

\item[Fitness Threshold.]
\label{sec:controller-firthres}
This requires a certain percentage of organisms to pass a threshold before stopping. This controller is guaranteed to produce organisms of at least some minimum quality, however, this alone does not guarantee the algorithm to stop. To guarantee that the algorithm will eventually stop we have implemented a limit to the number of generations that is configurable by the user.

\item[Convergence.]
\label{sec:controller-convergence}
To find if the search has GA has converged, we need two things: firstly, the threshold at which we consider things to have converged, and secondly the number of previous generations to look at to check for convergence. As convergence is when the average fitness of $n$ sequential generations is relatively similar we keep a list of the previous $n$ generations in the algorithm and then check the standard deviation of that list to check for convergence.
\end{description} 

\section{Evaluation on the Robotti Case}
\label{sec:case}

\subsection{Existing DSE Algorithms against Upgraded Versions}
\label{sec:case-oldVnew}

Fig.~\ref{fig:oldvnew-results} shows the results of speed testing done using Robotti data, which suggest that the upgrades described in Section~\ref{sec:update} have had a positive effect on runtime. Each scenario was run $5$ times to find the average runtime. As no functionality was changed the results of the ranking, HTML and CSV output are also consistent with the original version, apart from a change in the order of the columns in the outputs, which we suggest may be caused by a change in how sort operates between Python~2 and~3, although we cannot find evidence to support this.

\begin{figure}[h!]
	\begin{center}
		\begin{tabular}{c c c c c c c}
			Scenario & Original & \multicolumn{2}{c}{Python 3} & \multicolumn{3}{c}{Python 3 Threaded} \\
			 & Average Time / s & Average Time / s & Diff / \% & Threads & Average Time / s & Diff / \% \\
			\hline
			Sin 2 & $2977.611$ & $2988.454$ & $0$ & $10$ & $496.001$ & $-83$  \\
			Speed Ramp 1 & $2054.087$ & $1756.910$ & $-14$ & $10$ & $329.819$ & $-84$  \\
			Speed Step 1 & $1960.502$ & $1360.295$ & $-31$ & $5$ & $356.744$ & $-82$  \\
			Turn Ramp 1 & $2389.692$ & $1604.571$ & $-33$ & $8$ & $359.412$ & $-85$  \\
		\end{tabular}
	\end{center}
	\caption{Average Run Time Comparisons}
	\label{fig:oldvnew-results}
\end{figure}

\subsection{GA-based DSE of Robotti}
\label{sec:case-robottiGA}

Now we consider the use of GA-based DSE on Robotti. The GA setup was initially the same as for the test functions used to validate the implementations of the GA operations. However, as can be seen in \ref{fig:initialresults} this did not produce the expected results. The issue here is that the diversity control operation was too aggressive, preventing convergence by essentially turning the algorithm into an exhaustive search. This is suggested by the mean line in Fig.~\ref{fig:initialresultsC}. At around $35$ generations the mean fitness begins to rise and maximum fitness remains relatively constant, suggesting that the search is not converging. This highlights the need for parameter tuning to ensure GAs run effectively. To resolve this, the diversity control was disabled as it was found it was not needed for this simulation.

\begin{figure} 
    \centering
    \subfloat[Gene values across all generations.\label{fig:initialresultsA}]
    {\includegraphics[width=0.32\textwidth]{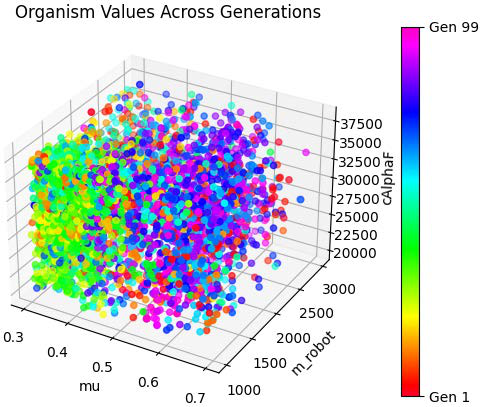}}
    \subfloat[Fitness against gene values.\label{fig:initialresultsB}]
    {\includegraphics[width=0.32\textwidth]{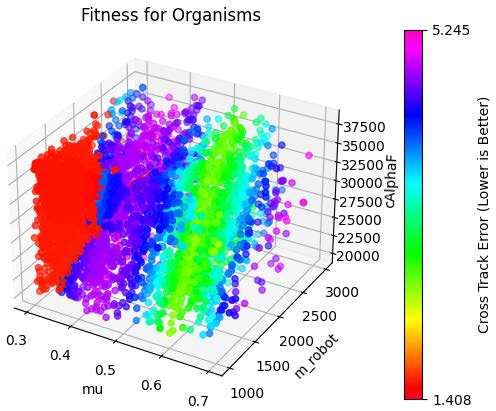}}
    \subfloat[Fitness trends.\label{fig:initialresultsC}]
    {\includegraphics[width=0.32\textwidth]{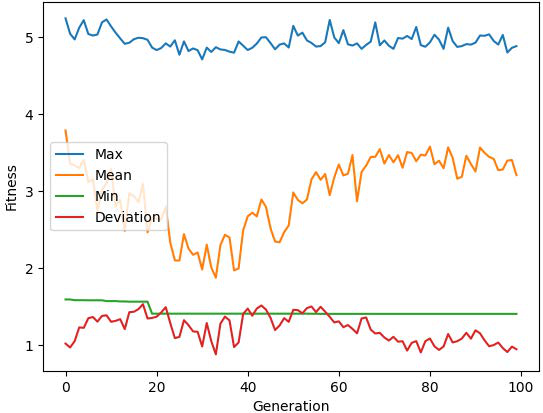}}
	\caption{Initial Speed Ramp results.}
	\label{fig:initialresults}
\end{figure}

Fig.~\ref{fig:robottiGA-exhaustiveVga} summarises the results of DSE in each of the Robotti scenarios compared against the best found by the exhaustive search. As can be seen in Fig.~\ref{fig:robottiGA-exhaustiveVga} the GA-based DSE performs as expected, finding the best parameter combinations within acceptable margins. As we can see, although in the order of $\times10^{-6}$, the genetic algorithm-based search is able to find slightly better parameter combinations for both the Sin and Turn Ramp scenarios. This is a consequence of the ability to select from a continuous range of values, so the GA found a combination that could not have been evaluated in the exhaustive DSE. The results also highlight that DSE scenarios must be selected with care, as if DSE was only performed on Speed Step then we would come to the false conclusion that the chosen parameters do not affect the running of the robot.

\begin{figure}[h!]
	\begin{center}
		\begin{tabular}{c c c c c c c c c}
			\multirow{2}{*}{Scenario} & \multicolumn{3}{c}{Exhaustive Best Combination} & \multicolumn{3}{c}{GA Best Combination} & \multirow{2}{*}{Cross Track Error Difference} \\
			 & cAlphaF & $\mu$ & Mass & cAlphaF & $\mu$ & Mass &   \\
			\hline
			Sin 2 & $20000$	& $0.7$ & $3000$ & $20000$ & $0.70$ & $3000$ & $-1.46\times10^{-6}$ \\
			Speed Ramp 1 & $38000$ & $0.4$ & $1000$ & $37019$ & $0.45$ & $1038$ & $8.40\times10^{-2}$ \\
			Speed Step 1 & $20000$ & $0.3$ & $1000$ & $34725$ & $0.45$ & $1027$ & $0.00$ \\
			Turn Ramp 1 & $20000$ & $0.3$ & $3000$ & $20000$ & $0.30$ & $3000$ & $-9.21\times10^{-7}$ \\
		\end{tabular}
	\end{center}
	\caption{Exhaustive Best Combination vs GA Best Combination. Results are rounded appropriately. Cross track error difference calculated as Exhaustive$-$GA.}
	\label{fig:robottiGA-exhaustiveVga}
\end{figure}


\begin{figure}
    \centering
    \subfloat[Gene values across all generations.\label{fig:sinresultsA}]
    {\includegraphics[width=0.32\textwidth]{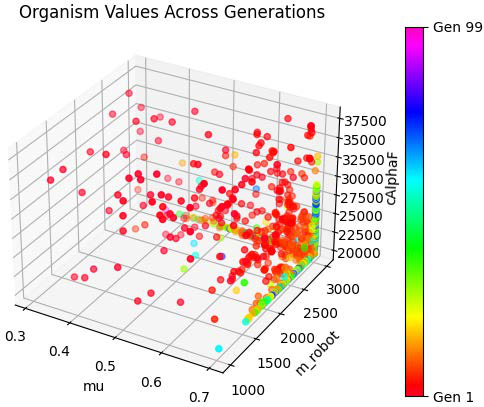}}
    \subfloat[Fitness against gene values.\label{fig:sinresultsB}]
    {\includegraphics[width=0.32\textwidth]{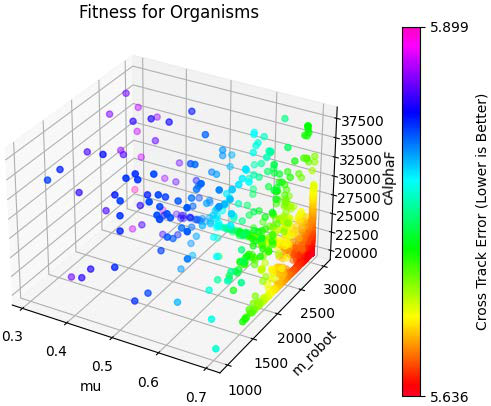}}
    \subfloat[Fitness trends.\label{fig:sinresultsC}]
    {\includegraphics[width=0.32\textwidth]{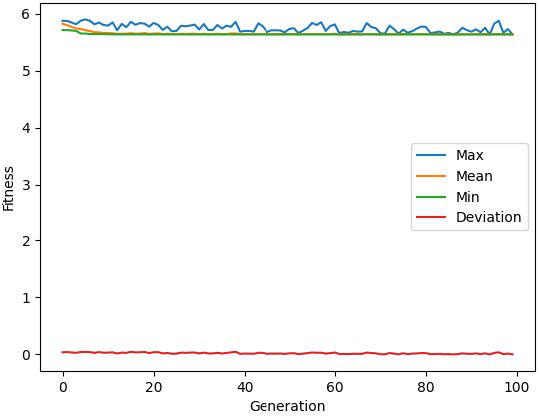}}
	\caption{Sin Results.}
	\label{fig:sinresults}
\end{figure}
\begin{figure}
    \centering
    \subfloat[Gene values across all generations.\label{fig:speedrampA}]
    {\includegraphics[width=0.32\textwidth]{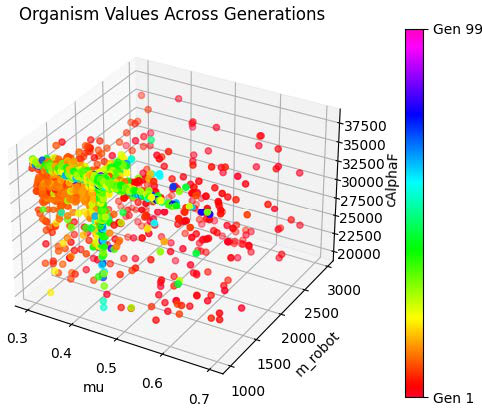}}
    \subfloat[Fitness against gene values.\label{fig:speedrampB}]
    {\includegraphics[width=0.32\textwidth]{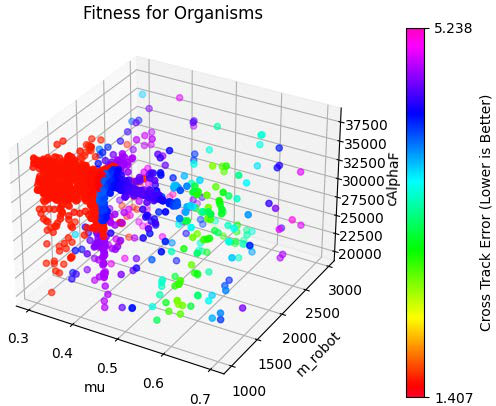}}
    \subfloat[Fitness trends.\label{fig:speedrampC}]
    {\includegraphics[width=0.32\textwidth]{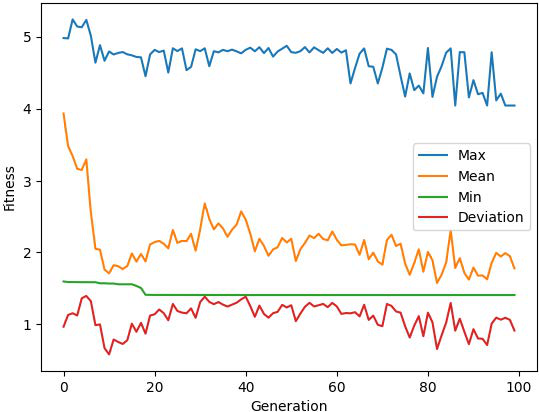}}
	\caption{Speed Ramp results}
	\label{fig:speedrampresults}
\end{figure}
\begin{figure}
    \centering
    \subfloat[Gene values across all generations.\label{fig:speedstepA}]
    {\includegraphics[width=0.32\textwidth]{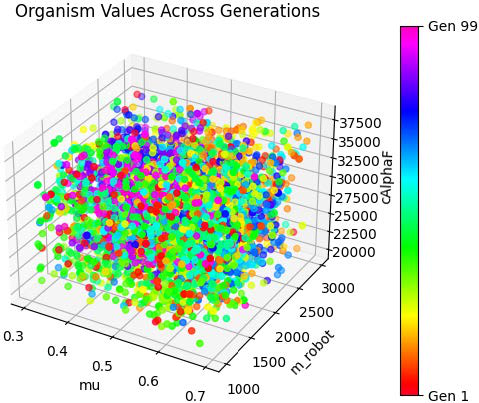}}
    \subfloat[Fitness against gene values.\label{fig:speedstepB}]
    {\includegraphics[width=0.32\textwidth]{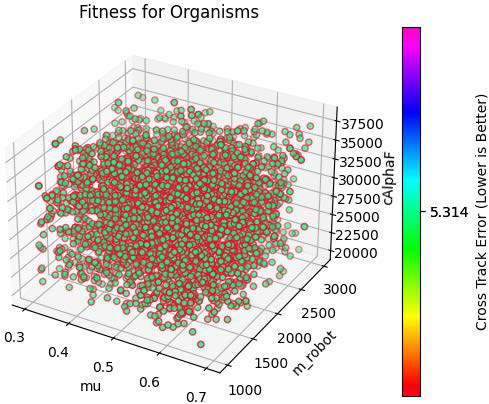}}
    \subfloat[Fitness trends.\label{fig:speedstepC}]
    {\includegraphics[width=0.32\textwidth]{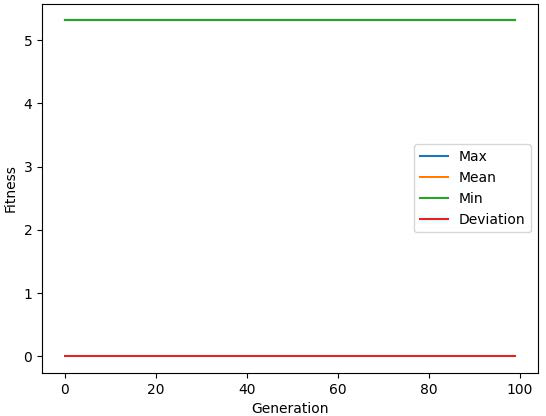}}
	\caption{Speed Step results.}
	\label{fig:speedstepresults}
\end{figure}
\begin{figure}

    \centering
    \subfloat[Gene values across all generations.\label{fig:turnrampA}]
    {\includegraphics[width=0.32\textwidth]{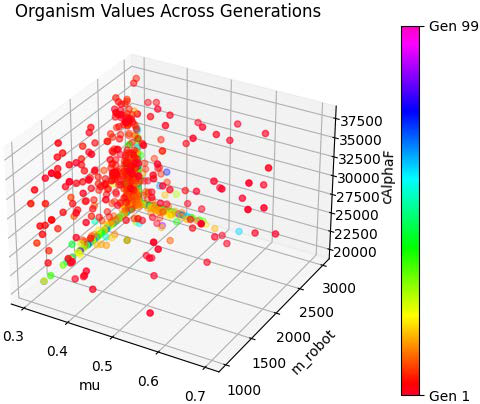}}
    \subfloat[Fitness against gene values.\label{fig:turnrampB}]
    {\includegraphics[width=0.32\textwidth]{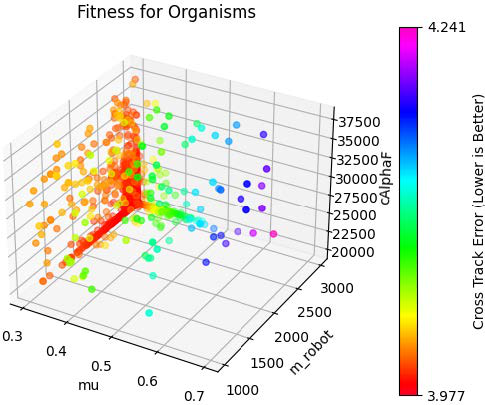}}
    \subfloat[Fitness trends.\label{fig:turnrampC}]
    {\includegraphics[width=0.32\textwidth]{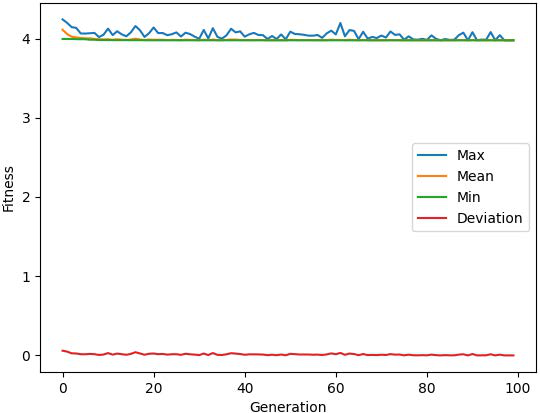}}
	\caption{Turn Ramp results.}
	\label{fig:turnrampresults}
\end{figure}

\section{Discussion and Future Work}
\label{sec:discussion}

We have provided an implementation of GAs as a basis for DSE in INTO-CPS, and evaluated this using an established case study. Some groundwork had to be laid in order to enable GA-based DSE, including the updating of existing scripts and the introduction of threading. However, the implementation appears to be viable. 

Aside from increasing the sophistication of the GA operators now implemented for INTO-CPS DSE, this opens up a range of other areas of future work.
First, CPSs such as Robotti may develop faults over time and it may be useful for the INTO-CPS toolchain to integrate Fault Space Exploration~(FSE) in addition to DSE. This could be used to find the level of fault tolerance that a system can withstand before resulting in undefined, possibly unsafe, behaviour. It is reasonable to think of this in the context of resilience modelling~\cite{Jackson&17}. Second, GAs are part of the broader class of Evolutionary Algorithms (EAs), suggesting that a wider examination of the orle of EAs in DSE would be worthwhile. Third, as a CPS may be deployed in many different environments it is not unreasonable to find that a set of parameters may be optimal for one scenario but not another, as shown in the DSE results for Robotti. Thus, it may be useful to implement an approach that can automatically produce a `compromise' set of parameter values that are optimal across all scenarios for a given CPS. This could take the results of DSEs as input and produce an `average organism' from this data.

\paragraph{Acknowledgements}

We acknowledge the European Union's support for the INTO-CPS and HUBCAP projects (Grant Agreements 644047 and 872698). Among many colleagues we are particularly grateful to Ken Pierce at Newcastle University, Casper Hansen at Aarhus University, and to AgroIntelli for permission to use Robotti data. Finally, we are grateful to the reviewers for their questions and helpful suggestions. 

\bibliographystyle{splncs03}

 \newcommand{\noop}[1]{}

\clearpage
\endgroup

\begingroup
\renewcommand\theHchapter{2-Stanley:\thechapter}
\renewcommand\theHsection{2-Stanley:\thesection}
\locallabels{2-Stanley:}
\setcounter{footnote}{0}
\setcounter{chapter}{0}
\setcounter{lstlisting}{0}
\fontfamily{ptm}\selectfont

\setlength\textfloatsep{8.0pt plus 2.0pt minus 2.0pt}
\setlength\intextsep{4.0pt plus 2.0pt minus 2.0pt}
\setlength\floatsep{4.0pt plus 2.0pt minus 2.0pt}
\setlength\abovecaptionskip{4.0pt plus 2.0pt minus 2.0pt}
\setlength\belowcaptionskip{0pt}

\lstset{basicstyle=\scriptsize,tabsize=2,frame=trBL,frameround=fttt}

\makeatletter
\def\input@path{{2-Stanley/}}
\makeatother

\graphicspath{{2-Stanley/}}

\title{	Multi-Objective Optimisation Support for Co-Simulation}
\titlerunning{}

\author{Aiden Stanley\inst{1} \and
  Ken Pierce\inst{1}
  }
\institute{School of Computing, Newcastle University, United Kingdom \\\email{A.Stanley@ncl.ac.uk,kenneth.pierce@ncl.ac.uk}
}

\maketitle
\begin{abstract}
When designing Cyber-Physical Systems (CPSs) using Model-Based Design (MBD) techniques, Design Space Exploration (DSE) can be used to discover optimal configurations. However, due to the computational complexity of co-simulations, this process can take a considerable amount of time, perhaps making it an unviable tool for some time-constrained projects. This paper reports on the integration of a Multi-Objective Optimisation (MOO) library to enhance the DSE features of INTO-CPS, a tool chain for design of CPSs based on co-simulation. The paper demonstrates that the new feature functions correctly by replicating results of a previous study, however the true benefits of MOO are not demonstrated and this paper suggests new MOO-specific case studies are needed for the INTO-CPS example repository to demonstrate these benefits.
\end{abstract}


\section{Background}
\label{sec:back}

This section provides background information on the INTO-CPS toll chain and its current Design Space Exploration (DSE) support, and introduces the concept of Multi-Objective Optimisation (MOO) and existing MOO libraries.

\subsection{INTO-CPS and Design Space Exploration (DSE) Support}
\label{sec:back:into}

INTO-CPS is an integrated tool chain for comprehensive Model-Based Design (MBD) of Cyber-Physical Systems (CPSs)''~\cite{Larsen&17a}. INTO-CPS is based around the Functional Mock-up Interface (FMI)\footnote{\url{https://fmi-standard.org/}} that allows collaborative models of CPS to be defined, configured, and co-simulated. INTO-CPS is comprised of two main components:

\begin{enumerate}
  \item A front-end called the INTO-CPS Application that allows for configuration and execution of co-simulations.
  \item A back-end program called the Co-simulation Orchestration Engine (COE) called Maestro~\cite{Thule&17}, which provides functionality for, and oversees, the actual execution of co-simulations.
\end{enumerate}

Around the core of co-simulation, INTO-CPS provides other useful functionalities such as automatic Design Space Exploration (DSE), in which multiple co-simulations are executes automatically and the results compared. This functionality is provided by a set of Python scripts which can be downloaded from within the INTO-CPS Application. The INTO-CPS Application and DSE scripts provide the ability to configure DSE using the following:

\begin{description}
  \item[Search Algorithm] The ability to choose between which optimisation algorithm should be used to perform the search. The currently available choices are `Exhaustive Search' and `Genetic'.
  \item[Experiment Parameters] The ability to define which parameters of the model should be considered part of the design space and their bounds.
  \item[Parameter Constraints] The ability to define constraints on the varying parameters e.g., ``parameter A should always have a value greater than parameter B, to be considered a feasible solution.''
  \item[External Script Objectives] The ability to point to a script that uses the results of a simulation run to calculate the model's fitness according to a custom objective.
  \item[Results Output] Once a run of DSE has completed, the current implementation outputs a `Pareto front' plot showing the optimal solutions regarding the objectives defined, and the set of input parameters required to recreate these optimal configurations.
\end{description}

The work reported in this paper aims to replicate these features using a MOO library, with the aim of providing the user with a seamless option to apply further algorithms in their exploration of design space.

\subsection{Multi-Objective Optimisation (MOO)}
\label{sec:back:moo}

Multi-Objective Optimisation (MOO) is the act of optimising a system as according to multiple specific objective criteria~\cite{gunantara2018review}. A solution is considered Pareto optimal if no aspect of that solution can be improved without making another aspect of the solution worse off. Due to the nature of this, any optimisation problem that has more than one objective will most likely have more than one Pareto-optimum solution. MOO algorithms may attempt to find a representative set of optimisations, to quantify the trade off, or pick one solution based on the preferences of the Decision Maker (DM).

The work reported in this paper links a MOO library to the INTO-CPS tool chain, thus we do not go into details regarding how various MOO algorithms work. The core concept is the same as the existing genetic DSE offered by INTO-CPS --- there is a set of possible parameters and a method which proposes solutions, and selects and refines these based on the results of co-simulation. Some MOO algorithms are based on genetic algorithms, such as the popular Non-dominated Sorting Genetic Algorithm-II (NSGA-II)~\cite{deb2002fast} and Strength Pareto Evolutionary Algorithm 2 (SPEA-2)~\cite{zitzler2001spea2} algorithms. Other MOO algorithms are not genetic, such as Particle Swarm Optimization and Simulated Annealing~\cite{suman2006survey}.

When selecting a library for the implementation, a number of options were considered. The three main options were MOEA, jMetal and pymoo. MOEA (Multi-Objective Evolutionary Algorithms) is a free and open-source framework for Java\footnote{\url{http://moeaframework.org/}}. It has an extensive list of algorithms and high-quality documentation, however is only two-years old and many of the guides are behind a paywall. jMetal\footnote{\url{http://jmetal.sourceforge.net/}} stands for Metaheuristic Algorithms in Java and is similar to MOEA. It has been established for much longer, with its fifth major release in 2015. Documentation is also freely available, although somewhat lacking for the newest version. The pymoo\footnote{\url{https://pymoo.org/}} framework offers state of the art single- and multi-objective optimization algorithms in Python. Although it has excellent documentation, it offers fewer algorithms and has not yet reached a major release. Based on this search, jMetal was selected as the most sensible option.

\section{Multi-Objective Optimisation (MOO) Support}
\label{sec:what}

\begin{figure}[tb]
\includegraphics[width=0.99\textwidth]{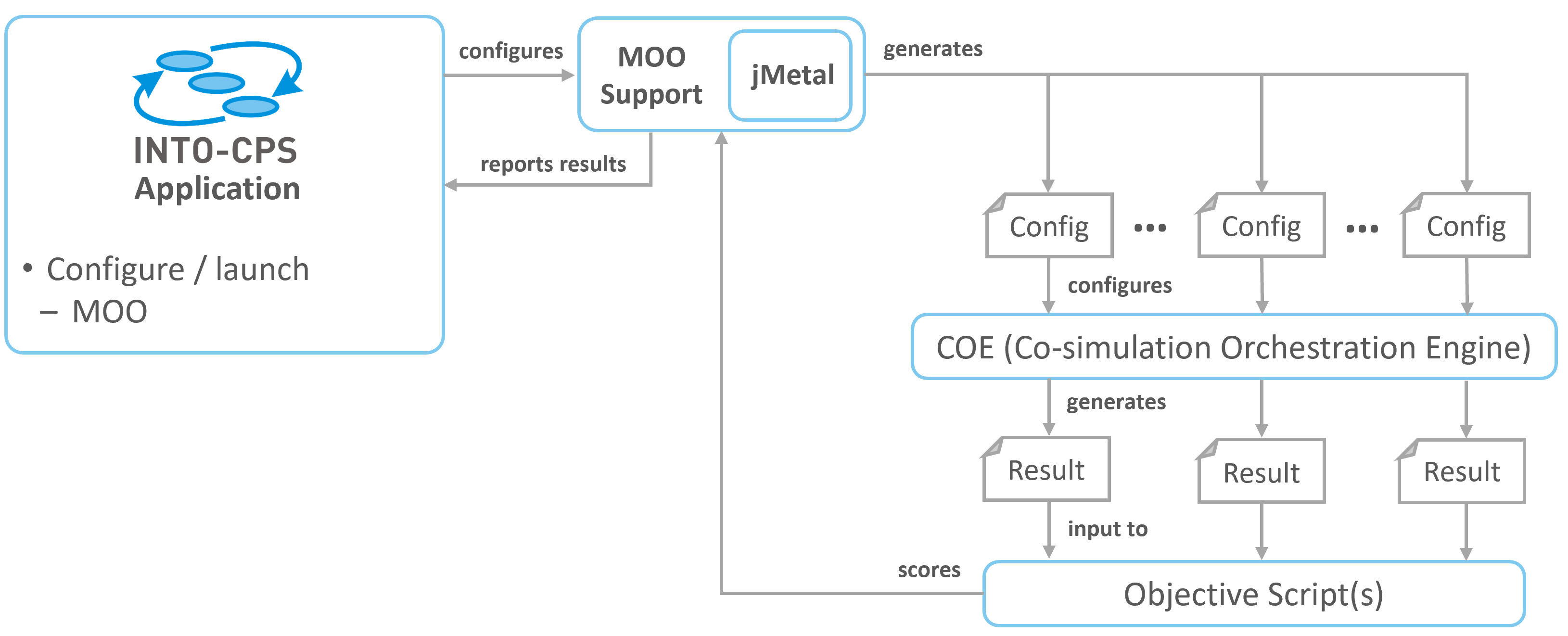}
\caption{Class diagram of the inheritance hierarchy of \texttt{Algorithm}}
\label{fig:algo}
\end{figure}

The support for Multi-Objective Optimisation (MOO) involved creating an interface between the jMetal~\cite{durillo2006jmetal} library and the Co-Simulation Orchestration Engine (COE). Figure~\ref{fig:algo} shows a diagram of the INTO-CPS Application interacting with the new MOO support tool and the COE. Here the frontend can configure a MOO problem, selecting the algorithm, parameter space, objective scripts and maximum iterations. The jMetal library then executes the selected algorithm. When testing solutions proposed by the algorithm, a co-simulation configuration is generated that configures the COE to run one or more co-simulations. The selected objective scripts are used to feed back into the jMetal library for the algorithm to generate a new set of solutions. This process repeats until a given number of iterations is reached.

The jMetal library is object-oriented. The key classes called \texttt{Problem}, representing the MOO problem to be solved, and \texttt{Solution}, representing a possible solution to the optimisation. In this case, the values of parameters of the model. The \texttt{Problem} class has a primarily method called \texttt{evaluate} which determines the fitness of a proposed \texttt{Solution}. Hence the primary effort of this implementation is to generate co-simulation configurations for \texttt{Solution} instances proposed by the MOO algorithm(s) and evaluate the \texttt{Problem} using the COE and objective scripts.

Each algorithm in JMetal comes with its own builder, which allows you to configure and initialise an algorithm. The MOO functionality was built in an object-oriented approach to wrap these algorithms in an interface. Each algorithm to be imported from jMetal is represented as its own class, which inherits from an abstract superclass called \texttt{Algorithm}. The superclass also contain an abstract method \texttt{runAlgorithm} that subclasses are required to implement. Calling this function should run the algorithm and output the results. This superclass contains empty fields for:

\begin{itemize}
  \item Each operator kind of operator, i.e.:
    \begin{itemize}
          \item Crossover
          \item Mutation
          \item Selection
    \end{itemize}
  \item Every variant of algorithm parameter, e.g.
    \begin{itemize}
          \item Crossover probability and distribution index
          \item Mutation probability and distribution index
          \item Mutation probability and distribution index
          \item Population size
          \item Maximum number of evaluations
          \item Etc.
    \end{itemize}
\end{itemize}

To allow the user or INTO-CPS Application to select an algorithm by name, all possible algorithms are enumerated in a type called \texttt{AlgorithmVariant}. A class called \texttt{RunHandler} takes a string input and attempts to resolve which subclass of \texttt{Algorithm} should construct and run. This is shown in the class diagram in Figure~\ref{fig:runhandler}.

\begin{figure}[tb]
\includegraphics[width=0.99\textwidth]{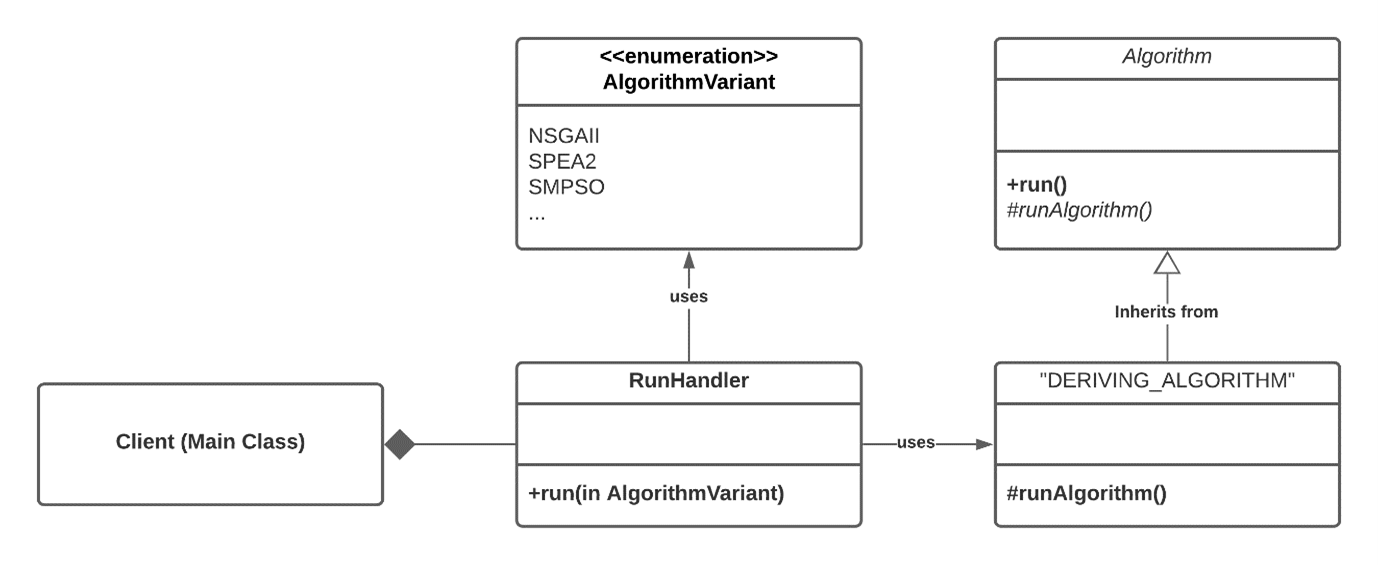}
\caption{Class diagram to showing the role of \texttt{RunHandler}}
\label{fig:runhandler}
\end{figure}

To make the integration with the INTO-CPS Application as simple as possible, the MOO implementation takes a single JSON (Javascript Object Notation) file as its only command line argument. This replicates the style of both the COE and existing DSE scripts. This means that tool integration only requires the INTO-CPS Application to fetch the required information from the user, store that information in a JSON file, then execute the MOO solution with this JSON as an argument.

\clearpage
The input JSON used should contain the following fields (based on the existing DSE JSON file), with an example is presented in Figure~\ref{fig:json}:

\begin{itemize}
  \item The path of the COE
  \item The path of the JSON configuring the base co-simulation
  \item The length of the simulation, in seconds
  \item The algorithm to use
  \item An array of the model's parameters that make up the design space, each with the following attributes:
      \begin{itemize}
        \item Parameter identifier
        \item Upper bound
        \item Lower bound
      \end{itemize}
  \item An array of the objectives, each with the following attributes:
      \begin{itemize}
        \item Type (``SCRIPT'' or ``PARAMETER'')
        \item Identifier (path to script, or parameter identifier)
        \item A Boolean value for whether the objective should be maximised (false means minimised)
      \end{itemize}
\end{itemize}

\begin{figure}[tb]
\centering
\includegraphics[width=0.75\textwidth]{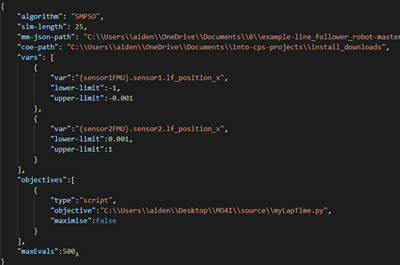}
\caption{An example of the input JSON used to configure an MOO search}
\label{fig:json}
\end{figure}

In addition to basic functionality of running each algorithm offered by jMetal, the solution also includes concurrent execution of solution evaluation and custom constraint handling through Python scripts. At the time of writing, the INTO-CPS Application is currently being updated to provide a MOO option in the existing DSE interface. A screenshot showing this is given in Figure~\ref{fig:into}. It is expected that this functionality will be available in the next release of the INTO-CPS Application expected in Q3 2021.

\begin{figure}
\centering
\includegraphics[width=0.95\textwidth]{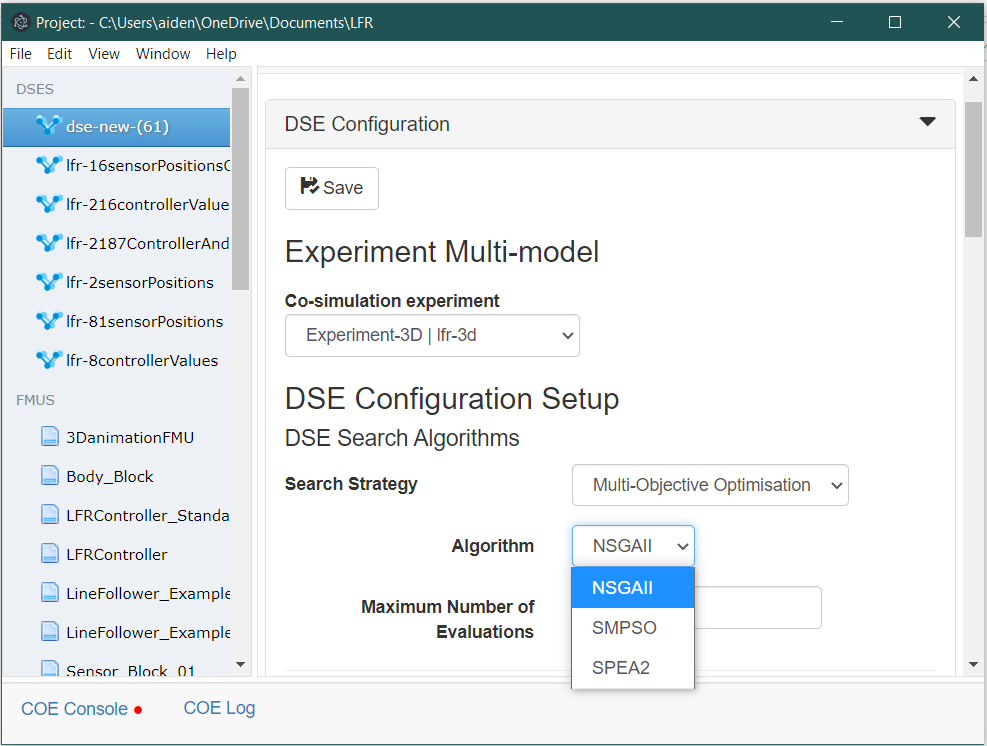}
\caption{A screenshot of the INTO-CPS Application showing MOO algorithm choices}
\label{fig:into}
\end{figure}  

\section{Case Study}
\label{sec:study}

To demonstrate the work reported in this paper against existing functionality, we use the same case study as Bogomolov et al.~\cite{bogomolov2021tuning} presented at a previous workshop. The paper reports on a pilot study to assess the viability of DSE as a tool in  improving the fidelity of a co-simulation model, based on real-world data observed in field trials of the real system. The study is based on the Robotti unmanned platform developed by Agro Intelligence (Agrointelli) for~(see Figure~\ref{fig:RealRobotti}). In the remainder of this section, the case study and previous results are discussed. Section~\ref{sec:results:time} discusses the performance of the current solution and the selection of the MOO algorithms.

The authors note that this study does not exhibit true multi-objective behaviours that require trade off to best demonstrate . We suggest in Section~\ref{sec:conc} that new case studies are required in the INTO-CPS examples repository that do show multi-objective characteristics and would better demonstrate the strengths and weaknesses of both the DSE and MOO approaches. We selected this case study because it was readily available and was previously reported at this venue, which meant that the results could be directly compared to demonstrate the functionality of the new MOO add on, if not demonstrate its full utility.

In this study, the real Robotti system carried out field trials under four scenarios with a total of 27 runs. Each scenario is a pattern of control outputs designed to assess how Robotti moves under different conditions. The control outputs and responses of the Robotti are then recorded for use in model optimisation. This data is then used to optimise the co-simulation by finding which physical parameters in the model most closely replicate those in the real Robotti. In this way, the optimisation should deliver a co-simulation that better models the actual system. The four scenarios were as follows~\cite{bogomolov2021tuning}:

\begin{description}
  \item[Speed step:] The speed of each wheel is increased incrementally in lock step, resulting in the Robotti driving straight.
  \item[Speed ramp:] The speed of each wheel is increased smoothly. This also results in a straight path as in the speed step tests.
  \item[Turn ramp:] The speed of the right wheel is increased smoothly, while the left wheel is kept constant. This results in the Robotti turning left.
  \item[Sin:] The speed of the left and right wheel is increased and decreased to produce the sinusoidal motion.
\end{description}

\begin{figure}[tb]
\centering
\includegraphics[width=0.7\columnwidth,trim=0 150 0 130, clip]{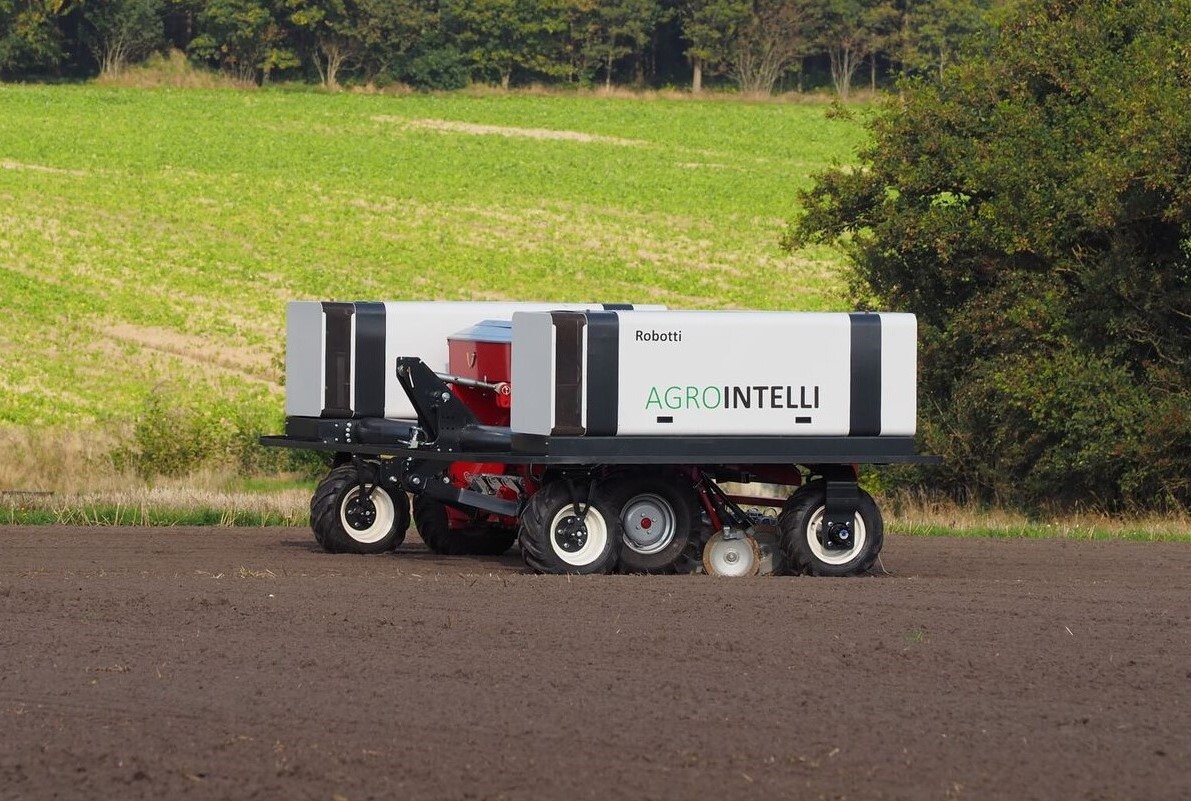}
\caption{\label{fig:RealRobotti}The Robotti unmanned platform}
\end{figure}

The measure of accuracy in this study is the mean cross-track error. Given the known path and the simulated path under a given set of parameters, the mean difference between then provides a measure of how well the parameters match the (unknown) attributes of the real Robotti. A lower mean difference therefore means a closer representation of reality.

The design space involves three parameters: \emph{m\_robot}, the mass of the Robotti; \emph{cAlphaF}, the steering stiffness; and \emph{mu}, the friction between the wheel and ground. In the study, exhaustive search is used for the DSE, with five values for each parameter. This results in 125 co-simulation per scenario, and 1250 for the ten scenarios used. Results for one each of the four scenarios are shown in Figure~\ref{fig:res_scenarios}. These show that the Sine and Turn ramp scenarios strongly suggest the same specific parameter values (in the lower right of the plots). Speed ramp shows a region of optimal solutions, and a secondary weaker region of good solutions, while Speed step suggests no overall optimal solution.

\begin{figure}[tb]
	\centering
    \subfloat[$\mathsf{sin1}$ Scenario\label{fig:res_sin1}]
	{\includegraphics[width=0.48\textwidth]{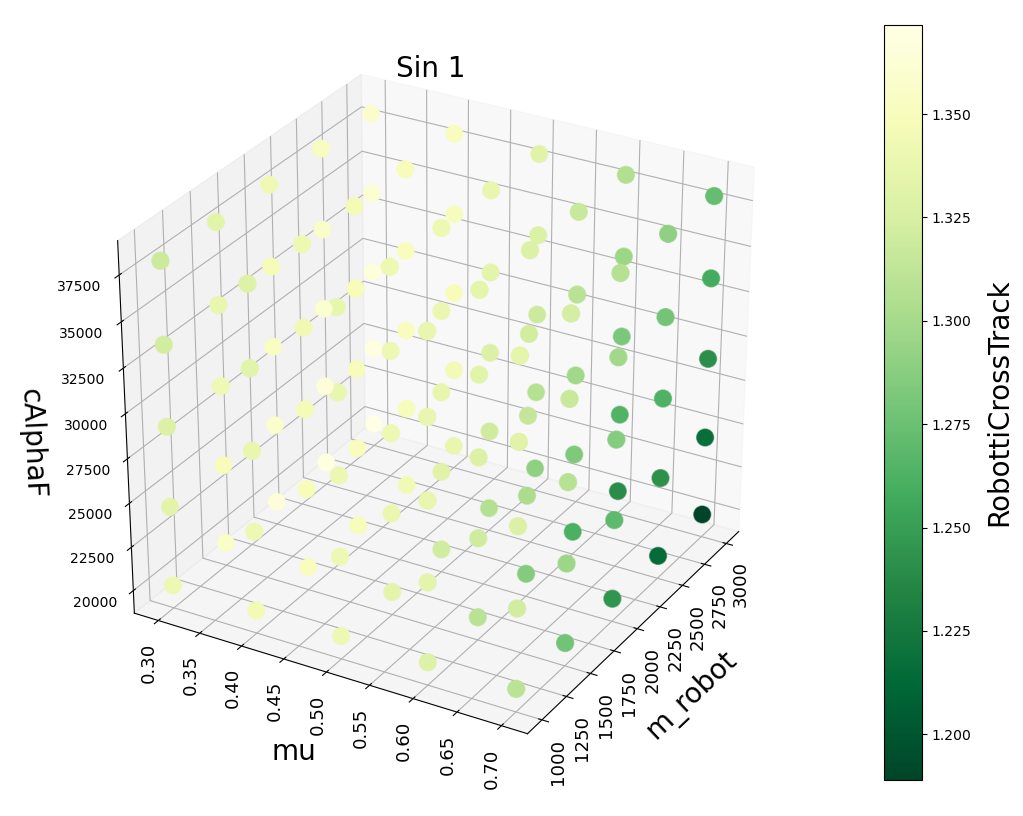}}
    \subfloat[$\mathsf{turn\_ramp1}$ Scenario\label{fig:res_turn_ramp1}]
	{\includegraphics[width=0.48\textwidth]{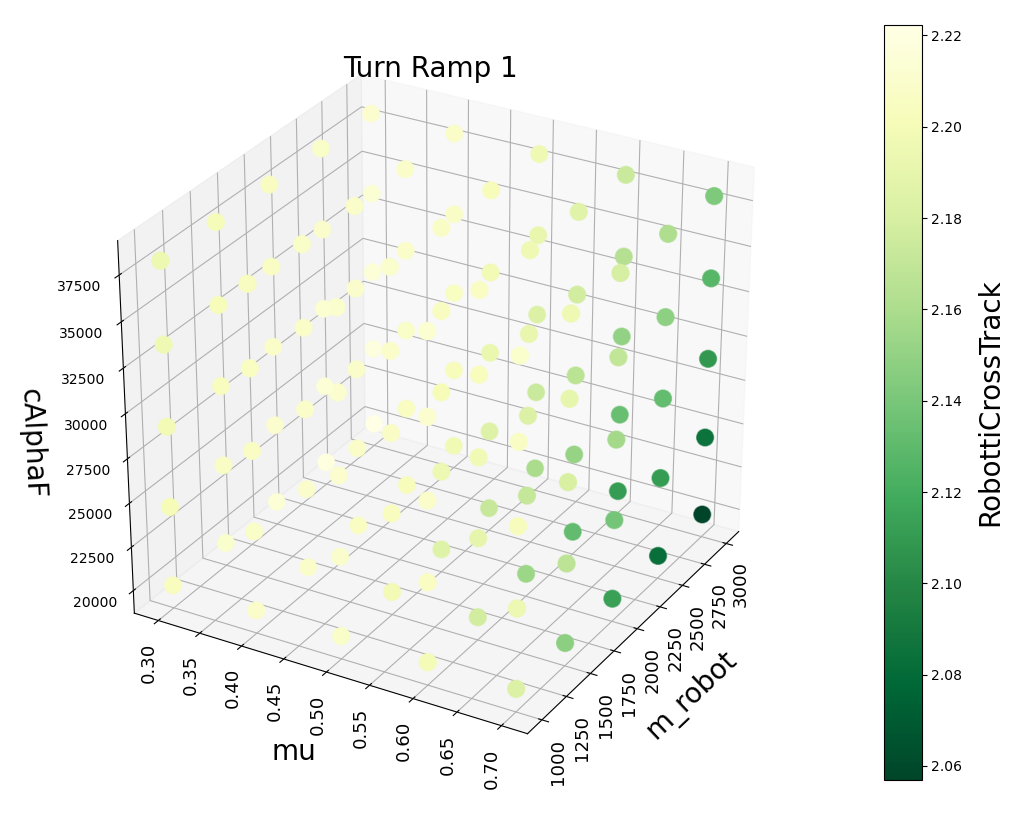}}\\
    \subfloat[$\mathsf{speed\_ramp1}$ Scenario\label{fig:res_speed_ramp1}]
	{\includegraphics[width=0.48\textwidth]{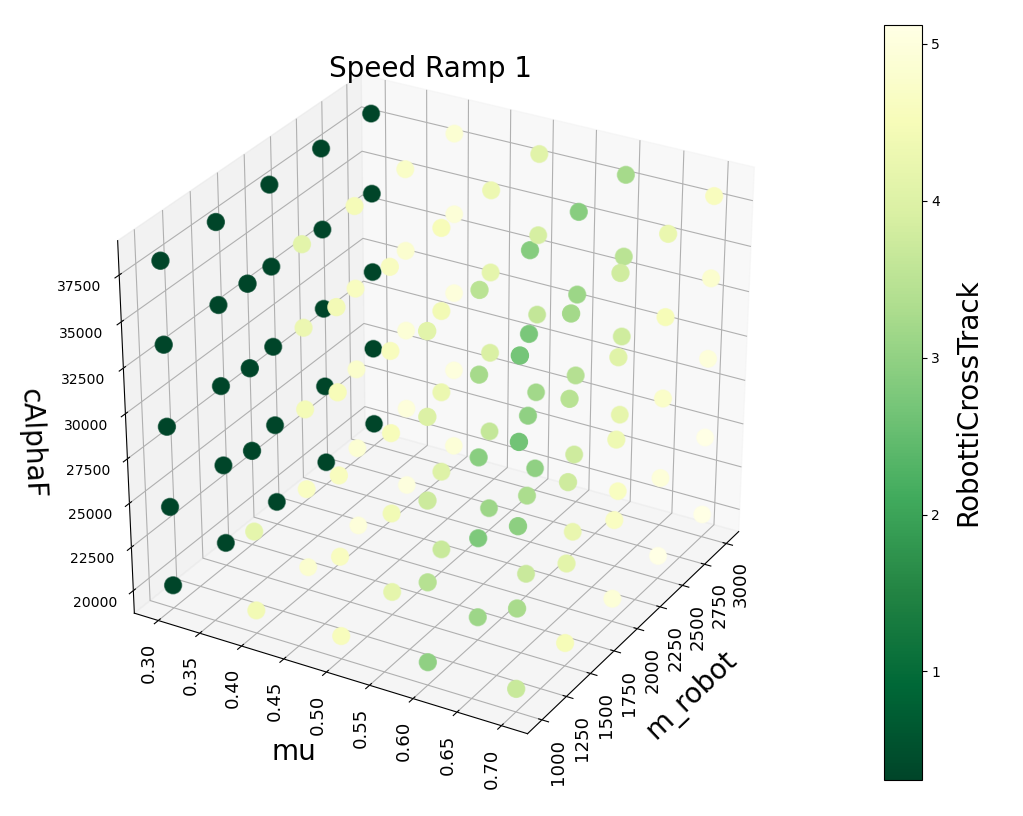}}
    \subfloat[$\mathsf{speed\_step1}$ Scenario\label{fig:res_speed_step1}]
	{\includegraphics[width=0.48\textwidth]{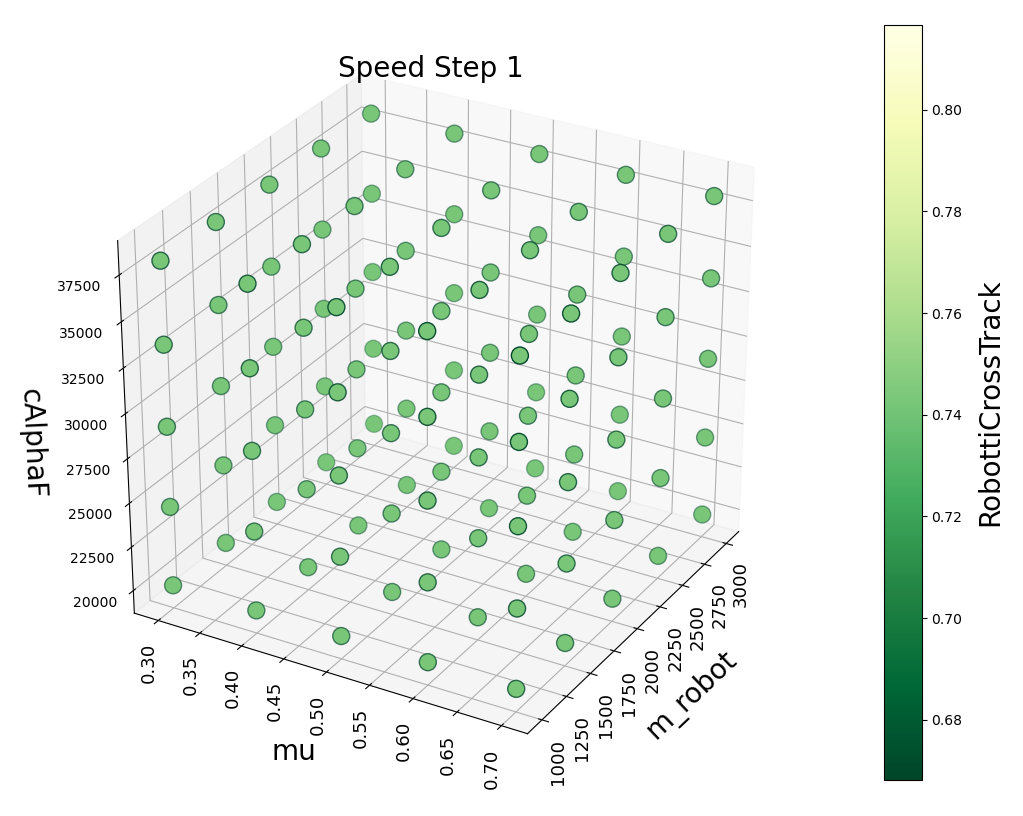}}
	\caption{Mean cross track error of four scenarios against different parameter values~\protect\cite{bogomolov2021tuning}}
\label{fig:res_scenarios}
\end{figure}


\section{Conclusions and Future Work}
\label{sec:conc}

This paper presented work on integrating support for a library of Multi-Objective Optimisations (MOO) algorithms into the INTO-CPS tool chain to provide additional options for Design Space Exploration (DSE) of models through co-simulation. It demonstrated a working solution and initial integration into the INTO-CPS Application. The solution was demonstrated by using two common MOO algorithms (NSGA-II and SMPSO) to explore the design space of the Robotti case study, comparing against results presented at a previous workshop~\cite{bogomolov2021tuning}.

The results show that the MOO algorithms were able to find similar trends to the exhaustive search presented previously, however when limited to the same 125 evaluations by co-simulation, performance was worse overall. The MOO library also introduces an overhead in execution time, which can be mitigated through parallel execution support. The case study chosen was selected given the previous work on using DSE to explore the parameter space, and provided a useful benchmark to check the correct functioning of the new MOO features.

The results show that MOO is not suitable for every use case, thus would not replace existing DSE solutions for INTO-CPS, but rather complement them. It is likely that the MOO solutions would offer an improvement to the time needed to perform DSE on CPS models when the design space in question has a high number of dimensions, in particular where multiple objectives compete and require trading off. In addition, of particular importance is to also apply this approach to case studies with a larger parameter spaces (both in the range of each parameter and the number of parameters).

One key area of future work is to develop case studies within the INTO-CPS examples repository that truly demonstrate multi-objective characteristics. This will allow a better comparison of existing DSE and MOO performance, particularly given the new multi-threading support in the existing DSE scripts. In this way, we can begin to identify guidelines for users about when and where each approach is likely to be most beneficial. 

\paragraph*{Acknowledgements.}
We are grateful to AgroIntelli for supplying Robotti data. This work was partially funded under the European Union's H2020 programme through the INTO-CPS and HUBCAP projects (Grant Agreements 644047 and 872698).

\clearpage

\endgroup

\end{document}